\RequirePackage{silence}
\documentclass[fleqn,usenatbib,useAMS]{mnras}

\usepackage{newtxtext,newtxmath}
\usepackage{fix-cm}

\usepackage[T1]{fontenc}
\usepackage{orcidlink}
\usepackage{multirow}
\usepackage{fontawesome5}
\usepackage{xcolor}
\usepackage{overpic}
\usepackage{bm}

\DeclareRobustCommand*{\matr}[1]{\mathbfss{#1}}
\DeclareRobustCommand{\dd}{\mathop{}\!\mathrm{d}}

\DeclareRobustCommand{\Mdes}{\ensuremath{M_{\mathrm{DES}}}}
\DeclareRobustCommand{\kmsMpc}{\ensuremath{\,\mathrm{km\,s^{-1}\,Mpc^{-1}}}}

\DeclareRobustCommand{\appropto}{\mathrel{\vcenter{
		\offinterlineskip\halign{\hfil##\cr
			$\propto$\cr\noalign{\kern2pt}$\sim$\cr\noalign{\kern-2pt}}}}}

\usepackage{graphicx}
\usepackage{amsmath}

\defcitealias{Planck_2020}{\emph{Planck}~2020}

\title[A homogeneous DES inverse distance ladder]{Does the inverse distance ladder rule out late-time solutions to the Hubble tension? A model-independent test using DES-Dovekie supernovae and DESI DR2 BAO}

\author[J. A. N\'ajera et al.]{Jos\'e Antonio N\'ajera$^{1}$\thanks{antonio.najera@port.ac.uk}\orcidlink{0000-0001-9738-7704}, Indranil Banik$^{1}$\thanks{indranilbanik1992@gmail.com}\orcidlink{0000-0002-4123-7325} and Harry Desmond$^{1}$\thanks{harry.desmond@port.ac.uk}\orcidlink{0000-0003-0685-9791} \\
$^{1}$Institute of Cosmology \& Gravitation, University of Portsmouth, Dennis Sciama Building, Burnaby Road, Portsmouth PO1 3FX, UK}

\date{Accepted XXX. Received YYY; in original form ZZZ}

\pubyear{\the\year{}}

\hypersetup{breaklinks=true}

\begin{document}
\label{firstpage}
\pagerange{\pageref{firstpage}--\pageref{lastpage}}
\maketitle

\begin{abstract}
It has been argued that the inverse distance ladder disfavours late-time solutions to the Hubble tension. We test this with a cosmographic expansion where $H$ is a fourth-order polynomial in $1 - a$, the order preferred by the Bayesian evidence. We combine the uncalibrated DES-Dovekie SNIa with DESI~DR2 BAO, assuming a standard sound horizon of $r_{\rm d} = 147.05 \pm 0.30$~cMpc. Using only DES SNIa, we predict that $H_0 = 76.25 \pm 2.87\kmsMpc$, which is $3.1\sigma$ above the CMB-calibrated $\Lambda$CDM prediction but consistent with SH0ES ($1.0\sigma$). Bayesian model averaging over orders 2--6 yields $H_0 = 74.77 \pm 4.83\kmsMpc$. Adding the legacy SNIa decreases the quartic $H_0$ to $67.98 \pm 0.69\kmsMpc$, $4.7\sigma$ below SH0ES. The $8.27\kmsMpc$ drop upon including the legacy SNIa cannot be attributed to expected sample variance: under a fiducial $\Lambda$CDM model, only $0.85\%$ of $10^4$ mock realisations exhibit a larger $H_0$ shift due to the changed redshift sampling in a like-for-like comparison. This points to a systematic tension between DES and legacy SNIa datasets. We next add local $H_0$ anchors (the bright siren, SNII, TRGB + SBF) and fit a $\Lambda$CDM $H(z)$ enhanced by $\Delta H(z) = A \exp (-z/z_1)$. We infer $H_0 = 70.00^{+4.78}_{-0.98}\kmsMpc$ and $z_1 < 0.086$ ($1\sigma$) using DES SNIa, but $H_0 = 67.66 \pm 1.01\kmsMpc$ and $z_1$ is unconstrained using DES + legacy. Late-time solutions thus remain possible with the homogeneous DES catalogue, but not with the legacy SNIa as published. Allowing recalibration of their absolute magnitude relative to DES, we infer an offset of $\Delta M_\mathrm{legacy} = 0.065 \pm 0.039$~mag, raising the predicted $H_0$ to $71.03 \pm 1.99\kmsMpc$. Ongoing surveys like DEBASS will provide a homogeneous SNIa sample down to redshifts accessible via the distance ladder, testing current legacy samples and hence the feasibility of late-time solutions.

\end{abstract}


\section{Introduction}
\label{Introduction}

One of the most challenging problems in modern cosmology is the tension between different estimates of the present expansion rate $H_0$. This Hubble tension describes the discrepancy between the expansion rate of the Universe inferred through local distance measurements \citep[e.g. $H_0^\mathrm{SH0ES} = 73.17 \pm 0.86\kmsMpc$ from SH0ES;][]{Breuval_2024} and the predicted expansion rate based on calibrating the parameters of the Lambda cold dark matter ($\Lambda$CDM) standard cosmological paradigm \citep*{Efstathiou_1990, Ostriker_Steinhardt_1995} using the Cosmic Microwave Background (CMB), which gives a far lower value \citep[e.g. SPT-3G + ACT~DR6 + \emph{Planck} obtained that $H_0^\mathrm{SPA} = 67.19 \pm 0.38\kmsMpc$;][]{SPT_2026}. These results are in $6.4\sigma$ tension with each other. An even higher tension has recently been claimed by the Local Distance Network based on combining several popular local determinations and taking into account their covariance \citep{H0DN_2026}. However, this does not properly model selection effects or distance priors \citep{Desmond_2026, Stiskalek_2026_Cepheids}. The three possible explanations for the tension are unaccounted-for systematics in the distance ladder, statistical inadequacy of the inferences \citep[e.g.][]{Becker_2015_impact, Feeney_2018_clarifying, Desmond_2026, Percival_2026, Stiskalek_2026_Cepheids, Stiskalek_2026_maser}, and new physics beyond the standard $\Lambda$CDM model \citep[e.g.][and references therein]{Valentino_2025}. New physics could primarily affect the Universe prior to recombination or have little impact until more recent epochs, leading to a distinction between early-time and late-time solutions.

A useful expansion probe is provided by supernovae (SNe) of Type~Ia (SNIa). These thermonuclear explosions occur when a carbon--oxygen white dwarf in a binary system reaches a mass close to $1.2 \, M_\odot$, as predicted by models like the dynamically driven double-degenerate double-detonation $\left( \mathrm{D}^6 \right)$ scenario \citep{Shen_2018a, Shen_2018b, Bauer_2021}. Since SNIa occur at nearly the same mass, they are good standardisable candles, i.e., the peak absolute magnitude $M_B$ in the rest-frame $B$ band can be estimated independently of the redshift $z$ \citep{Phillips_1993, Tripp_1998, Phillips_1999}. Due to the known $H_0$--$M_B$ degeneracy \citep{Cai_2026, Wang_2026_intercept}, SNIa require an absolute calibration to measure $H_0$. Since SNIa are very luminous ($M_B \approx -19.3$), they can be detected out to $z \approx 2$ \citep{Scolnic_2022}. However, it is difficult to access all of this redshift range using just one survey. Modern SNIa catalogues like Pantheon+ \citep{Scolnic_2022}, Union3 \citep{Rubin_2025}, and Unite \citep{Camilleri_2026_Unite} typically combine different surveys. This may create problems due to the use of different filters, selection functions, and relative calibration offsets \citep{Betoule_2014, Scolnic_2014, Scolnic_2015, Efstathiou_2025}. A homogeneous catalogue with only one survey and one selection function would be far preferable for cosmological inference.

This is provided by the Dark Energy Survey \citep[DES;][]{DES_2016} through its Dovekie catalogue \citep{Popovic_2026_Dovekie}. This single-instrument dataset with a single filter system is typically complemented by legacy SNIa from older surveys to improve the leverage on the shape of the expansion history at $z \la 0.2$, where there are very few SNIa in DES \citep[though it has SNIa down to $z = 0.073$;][]{DES_2024_SNe, Vincenzi_2024}. However, including the legacy SNIa risks the aforementioned problems with combining surveys. Assessing this legacy sample relative to the DES sample is a main aim of this paper. Our baseline analysis involves only SNIa in the DES data, which we label simply DES. We also consider an extension including the legacy SNIa (DES + legacy), which corresponds to what other authors typically mean by ``DES'' \citep[e.g.][]{DESI_2025}. This distinguishes our work from similar studies \citep[most notably][]{Popovic_2026_IDL}, which generally do not distinguish between legacy and DES SNIa.

Our main motivation is the presently open discussion on whether the legacy SNIa have been correctly recalibrated onto the DES photometric filter system, something which cannot be directly verified through further observations given the transient nature of SNIa \citep{Efstathiou_2025, Huang_2025_DESI, Ormondroyd_2026}. Since the legacy SNIa lie in $z \in [0.025, 0.087]$, they provide a critical lever on the value of $H_0$, making it crucial that they be unbiased. This motivates further analyses where we calculate the possible offset between the legacy SNIa and the main DES sample. This is also a crucial issue for the evidence for evolving dark energy, which is largely driven by the legacy SNIa; allowing an offset provides a better fit to the data in a Bayesian sense than allowing a time-variation in the equation of state of dark energy \citep{Ormondroyd_2026}. In Section~\ref{sec:Model_comparison}, we explore the extent to which the legacy SNIa affect the evidence for evolving dark energy.

There are two main routes to determine the standardised absolute SNIa magnitude $M_{\rm SNIa}$. The first is the cosmic distance ladder, in which geometric distances calibrate Cepheid variable stars \citep{Leavitt_1912, Riess_2022_comprehensive, Breuval_2024}, the tip of the red giant branch \citep[TRGB;][]{Lee_1993, Madore_1995, Freedman_2019, Hoyt_2024}, or the J-region Asymptotic Giant Branch \citep[JAGB;][]{Madore_2020, Lee_2025}. These indicators calibrate $M_B$ in nearby SNIa host galaxies, so that SNIa in the Hubble flow at $z \ga 0.023$ give $H_0$ through the slope of the Hubble diagram. Using this approach, the SH0ES collaboration derived a value of $M_B = -19.253 \pm 0.027$ \citep{Riess_2022_comprehensive, Scolnic_2022}.

The second route to $M_{\rm SNIa}$ is through the inverse distance ladder (IDL). This involves anchoring $M_{\rm SNIa}$ using our understanding of the early Universe to set the baryon acoustic oscillation (BAO) comoving sound horizon $r_{\rm d}$ at the end of the baryon drag epoch \citep[$z_{\rm d} \approx 1060$;][]{Hu_1996, Eisenstein_1998}. $r_{\rm d}$ is then propagated to intermediate redshifts using the BAO evident from galaxy clustering, which nowadays is typically obtained from the Dark Energy Spectroscopic Instrument (DESI) Data Release~2 \citep[DR2;][]{DESI_2025}. We follow the common assumption that $r_{\rm d} = 147.05 \pm 0.30$~comoving Mpc (cMpc) from \citealt{Planck_2020} (hereafter \citetalias{Planck_2020}). This explicitly adopts $\Lambda$CDM at early times, so the IDL approach implicitly assumes that the Hubble tension, if due to new physics, must be solved at late times. Under this assumption, $M_{\rm SNIa} \approx -19.30$ to $-19.40$ \citep{Camilleri_2025, Popovic_2026_IDL}, which is significantly lower than suggested by the local distance ladder. This is because a lower $H_0$ requires SNIa to be more distant at given $z$, which for given apparent magnitude requires the SNIa to be brighter. The discrepancy with the locally determined $M_B$ leads to the usual conclusion that a successful solution to the Hubble tension requires a smaller $r_{\rm d}$, which is possible only through new pre-recombination physics \citep{Cai_2026, Sabogal_2026, Tiwari_2026}.

It is also possible to get around distances altogether using the cosmic chronometer (CC) technique, which finds the elapsed time between two redshifts by comparing the relative ages of massive passive elliptical galaxies at those redshifts \citep{Jimenez_2002, Moresco_2012, Moresco_2015, Moresco_2016, Moresco_2022, Moresco_2024}. Since we can relate distance and lookback time using the speed of light $c$, we can use CC datasets to calibrate $M_B$ without assuming any value for $r_{\rm d}$, thereby obtaining an estimate for $H_0$ \citep{Gomez_2018, Gomez_2022, Favale_2023, Cogato_2024, Guo_2025}.

Early-time solutions to the Hubble tension seek to resolve it by decreasing $r_{\rm d}$, which requires modifying the redshift dependence of the sound horizon and/or the redshift of recombination \citep*{Mirpoorian_2025}. Since BAO measurements are most sensitive to the product $H_0 r_{\rm d}$, reducing $r_{\rm d}$ allows for an increased $H_0$. However, these solutions face at least seven problems \citep{Vagnozzi_2023}. In particular, they imply a reduced age of the Universe \citep{Vagnozzi_2022}, creating significant tension with the observed ages of the oldest Galactic globular clusters (GCs) and stars \citep{Montalban_2021, Valcin_2020, Valcin_2021, Cimatti_2023, Souza_2024, Valcin_2025, Valcin_2026, Banik_2026_age}. For instance, the asteroseismic age of the Methuselah star just 60~pc away is $14.2 \pm 0.4$~Gyr \citep{Lundkvist_2025}. This leaves little room to reduce the cosmic age below the 13.8~Gyr in the SPA cosmology. Moreover, currently known early-time solutions struggle with Big Bang nucleosynthesis (BBN) due to their need for a generically higher baryon density to match the CMB anisotropies, which affects primordial light element abundances \citep{Giovanetti_2026, Launders_2026, Poulin_2026}. More generally, the success of $\Lambda$CDM in fitting the CMB anisotropies by now limits the scope for substantial modifications prior to recombination \citep[\citetalias{Planck_2020};][]{Calabrese_2025, SPT_2026}, including in models where electron density fluctuations cause inhomogeneous recombination \citep{Lynch_2026}.

Late-time solutions to the Hubble tension avoid these problems by assuming standard pre-recombination physics, but such models have their own problems at $z \la 1$ \citep{Krishnan_2021_age}. In particular, \citet{Bansal_2026} find that a model modifying only the late-time expansion history cannot simultaneously fit $H_0^\mathrm{SH0ES}$, the DESI BAO data, and relative SNIa distances. A fit is possible only if $M_\mathrm{SNIa}$ takes a sharp step at very low redshift ($z \la 0.01$) or the distance-duality relation is broken \citep[see also][]{Gomez_2024}. Importantly, the analysis of \citet{Bansal_2026} rests on the Pantheon+ catalogue, in which a step in $M_B$ at low redshift is degenerate with precisely the relative calibration offsets between surveys described above. This motivates the test we present here. We note that astrophysically it is difficult to understand why SNIa might abruptly change in intrinsic brightness, with terrestrial data placing strong constraints on models with a sharp change in $G$ in the geologically recent past \citep*{Banik_2025_GSM}. Such models also face difficulties with the mass--radius relation of red dwarf stars due to their long thermal timescales \citep{Sakstein_2026}.

Our main goal is to assess if the IDL can accommodate a late-time solution to the Hubble tension, i.e., if the assumption of standard pre-recombination physics is compatible with the high $H_0$ required to match various local constraints \citep{Valentino_2025, H0DN_2026, Roy_2026}. The observational data we use for this purpose are described in Section~\ref{sec:data}. We then explain the IDL, our test of the legacy SNIa, and our Bayesian model comparison of late-time solutions to the Hubble tension (Section~\ref{sec:methodology}). We go on to present our results (Section~\ref{sec:results}) and discuss their implications (Section~\ref{sec:discussion}). We conclude in Section~\ref{sec:conclusions}. In Appendix~\ref{app:adot}, we show the reconstructed expansion history. Appendix~\ref{app:AP} presents the Alcock--Paczynski (AP) consistency test between DESI~DR2 BAO and DES-Dovekie SNIa. Appendix~\ref{app:ordersweep} reports how the inferred cosmological parameters depend on the polynomial order. Appendix~\ref{app:compCMB} details our compressed CMB likelihood. Appendix~\ref{app:triangles} contains all the posteriors from our analyses. Appendix~\ref{app:logz1} repeats our model comparison with a log-uniform prior on the \texttt{exp} model $z_1$. In Appendix~\ref{app:floor-sweep-mock}, we use $\Lambda$CDM mocks to better understand how our analysis is affected by the limited DES coverage at low redshift.

\section{Observational data}
\label{sec:data}

\subsection{DES-Dovekie SNIa}

The DES-Dovekie compilation of SNe provides 1820 spectroscopically and photometrically classified events with their full statistical + systematic covariance \citep{Popovic_2026_Dovekie}. The intrinsic scatter contribution to this covariance is calibrated so that $\Lambda$CDM yields an acceptable fit to the Hubble diagram. This is because we currently lack sufficiently large samples of SNIa with redshift-independent distances from which the intrinsic scatter can be inferred, as is usually done with other indicators like Cepheids. The total SNIa uncertainties are therefore scaled to a reference cosmology, which means that an acceptable $\chi^2$ under $\Lambda$CDM in SNIa-dominated analyses like ours does not by itself demonstrate that $\Lambda$CDM is an adequate description of the data. We consider two SN samples, which we label DES and DES + legacy throughout this paper. Our baseline is the DES sample, which uses the 1623 DES SNe alone ($\mathrm{IDSURVEY}=10$), leading to a redshift range of $z\in[0.073,1.14]$ -- though in practice there is limited constraining power at $z \la 0.2$
\citep[see figures~3 and 4 of][]{DES_2024_SNe}. We select the SNe with magnitude uncertainty $\sigma_\mu < 1$, removing 82 SNe as these are likely contaminants (e.g. other types of SN) given they are down-weighted by the Bayesian Estimation Applied to Multiple Species (BEAMS) analysis \citep{Kunz_2007}. They carry $\approx10^{-5}$ of the total statistical weight, so removing them does not significantly change the results. Although our DES sample excludes the legacy SNIa from the cosmological fit, they are still used to train the light-curve model applied to the DES SNIa, as we discuss in Section~\ref{sec:discussion_systematics_legacy}.

The DES + legacy sample adds an external low-redshift anchor from legacy surveys, which contribute a further 197 spectroscopically confirmed SNIa with $z \in [0.025, 0.087]$, lowering the redshift floor from $z\approx 0.073 \to 0.025$. It is important to stress that both DES and legacy SNIa are uncalibrated as there is no independent determination of their absolute magnitude within the DES-Dovekie catalogue. We denote this absolute magnitude $\Mdes$ to distinguish its IDL calibration from the Cepheid-based SH0ES $M_B$, although both refer to the same SALT rest-frame $B$-band magnitude (Section~\ref{sec:discussion}). The reported distance modulus $\mu(z) = m(z) - \Mdes$ assumes a fiducial $\Mdes = -19.327$, but this can be arbitrarily varied. We infer $\Mdes$ jointly with the other parameters, which implicitly imposes a calibration via the IDL. The predicted $\mu \equiv 25 + 5 \log_{10} D_{\rm L}/\mathrm{Mpc}$, with the luminosity distance $D_{\rm L}$ found as follows:
\begin{eqnarray}
    \label{eqn:DL}
    D_{\rm L} &=& \left( 1 + z_\mathrm{hel} \right) D_{\rm M} \left( z_\mathrm{HD} \right), \\
    D_{\rm M} \left( z \right) &=& c \int_0^z \frac{\mathrm{d}\tilde{z}}{H(\tilde{z})} \, ,
\end{eqnarray}
where $D_{\rm M}$ is the comoving distance, $z_\mathrm{hel}$ is the heliocentric redshift of each SNIa, and $z_\mathrm{HD}$ is its ``Hubble diagram'' redshift obtained after corrections for peculiar velocities.

To study possible differences between the main DES sample and the legacy SNIa, we also consider adding a systematic offset of $\Delta M_{\rm legacy}$ to the absolute magnitude of the legacy SNIa. This measures possible systematics when correcting the legacy SNIa for survey completeness effects or converting the observed legacy magnitudes to what would have been measured in the DES filters -- despite the lack of direct measurements in those filters. We label this as DES + legacy ($\Delta M_{\rm legacy}$) and explain it further in Section~\ref{sec:systematics-legacy}.

\subsection{DESI DR2 BAO}

The BAO is a statistical standard ruler arising from coupled matter and radiation fluctuations in the primordial Universe \citep{Eisenstein_1998}. We consider the latest DESI data release \citep[DESI~DR2;][]{DESI_2025}. They report six measurements of $D_{\rm M}/r_{\rm d}$, six of $D_{\rm H}/r_{\rm d}$ at the same redshifts, and one of $D_{\rm V}/r_{\rm d}$ at a lower redshift:
\begin{eqnarray}
    D_{\rm H} \left( z \right) &=& \frac{c}{H \left( z \right)} \, , \\
    D_{\rm V} \left( z \right) &=& \left[ z D^2_{\rm M} \left( z \right) D_{\rm H} \left( z \right) \right]^{1/3}, \\
    r_{\rm d} &=& \int_{z_{\rm d}}^\infty \frac{c_\mathrm{s} \left( \tilde{z} \right)}{H \left( \tilde{z} \right)} \, \mathrm{d}\tilde{z} \, ,
\end{eqnarray}
where $c_\mathrm{s}$ is the sound speed in the baryon--photon plasma. We use the full DESI covariance matrix, which has a block-diagonal form because measurements at different redshifts use different galaxies, making them independent.

\subsection{CC}

We use 15 CC measurements to constrain the expansion history without assumptions about $\Mdes$ or $r_{\rm d}$ \citep{Moresco_2012, Moresco_2015, Moresco_2016}. CC measurements constrain the Hubble--Lemaitre parameter via the following relation
\begin{eqnarray}
    H(z) ~=~ -\frac{\dot{z}}{1+z} \, ,
\end{eqnarray}
where an overdot denotes a time derivative. $\dot{z}$ is measured through the differential age evolution of early-type galaxies \citep{Jimenez_2002}. We use the full covariance reported in \citet{Moresco_2020}. The errors quoted in CC compilations like \citet{Moresco_2022} omit the stellar-population modelling systematic. We find that this is crucial as it can significantly change the cosmological parameter posteriors. In particular,
the error on $r_{\rm d}$ roughly doubles when considering the full covariance rather than only the diagonal terms. Since the covariance with other CC measurements is not known, we do not use the other 18 of the 33 available CC measurements \citep{Moresco_2020, Wang_2026_CC}.

We use the CC data for an extended analysis in which a direct comparison between differential age $H(z)$ measurements and $D_{\rm H}/r_{\rm d}$ BAO measurements can independently calibrate $r_{\rm d}$ (Section~\ref{sec:results-params}). This affords an IDL independent of the CMB.

\section{Methodology}
\label{sec:methodology}

\subsection{Model setup}
\label{sec:setup}

The cosmic expansion history $a(t)$ can be approximated as a Taylor series around the present epoch, where $a$ is the cosmic scale factor normalised to unity today and $t$ is the time since the Big Bang. The use of a Taylor series minimises assumptions about the expansion history, making the results model-independent. The series expansion is commonly based on redshift, but since we consider data out to $z = 2.33$, we instead expand in the lookback scale factor $\widetilde{a} = 1 - a$, limiting the range to 0--1 \citep{Cattoen_2007, Busti_2015, Yang_2020}. We therefore use the following series expansion:
\begin{eqnarray}
    E \left( \widetilde{a} \right) ~=~ 1 + \sum_{k = 1}^{k_{\max}} C_k \, \widetilde{a}^k \, , \qquad \widetilde{a} ~\equiv~ \frac{z}{1 + z} ~=~ 1 - a,
    \label{eqn:cosmographic-expansion}
\end{eqnarray}
where $E(\widetilde{a}) \equiv H(\widetilde{a})/H_0$ is the normalised expansion rate and $k_{\max}$ is the order of the polynomial with coefficients $C_k$. Expanding in $\widetilde{a}$ rather than $z$ is important here because a Taylor series in $z$ converges only for $\lvert z \rvert < 1$ \citep{Cattoen_2007}, whereas DESI reaches out to $z = 2.33$. Fitting a series expansion in $z$ to data at $z > 1$ would thus amount to a model-dependent polynomial fit, whose parameters $H_0, q_0, j_0, \dots$ would no longer be the standard cosmographic parameters. This problem does not arise in our setup because $\widetilde{a} \leq 0.70$. Since the $z \ll 1$ correspondence is $\widetilde{a} = z - z^2 + \mathcal{O}(z^3)$, the deceleration parameter can be found from $q_0 = C_1 - 1$.

\begin{figure}
    \centering
    \includegraphics[width=\linewidth]{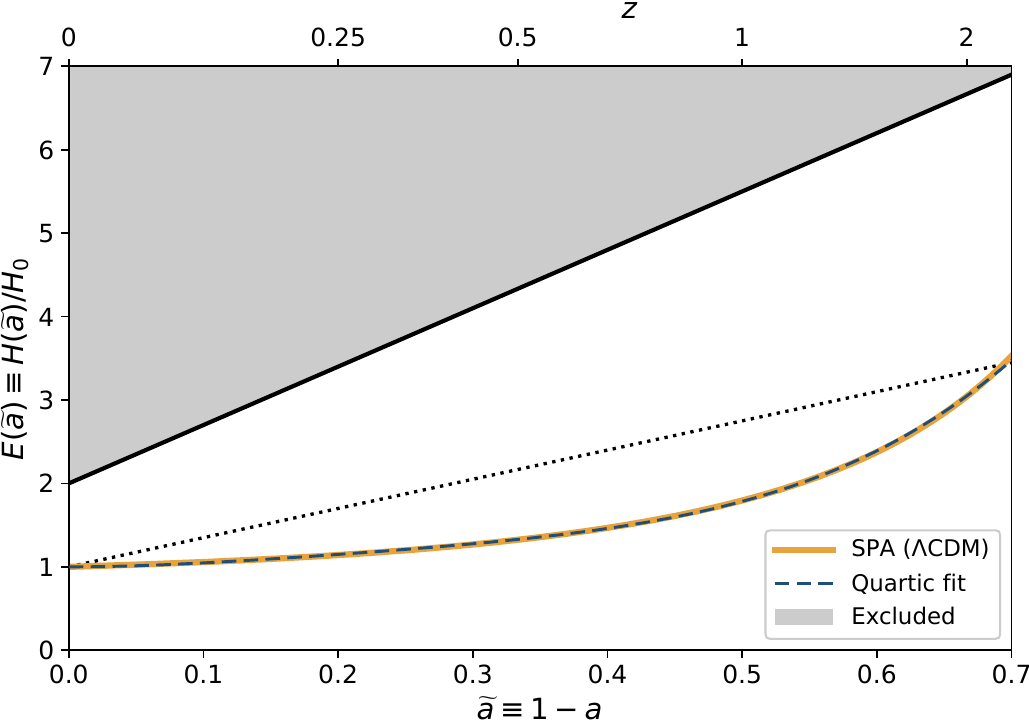}
    \caption{Illustration of our prior on the shape of the expansion history $E(\widetilde{a}) \equiv H(\widetilde{a})/H_0$ (Eq.~\ref{eqn:cosmographic-expansion}). The thick orange line shows flat $\Lambda$CDM with $\Omega_{\rm m} = 0.3175$, which has negligible errors for the purpose of this plot \citep{SPT_2026}. The dashed blue line shows the best quartic fit to the $\Lambda$CDM prediction. The dotted black line at $E = 1 + 3.5\widetilde{a}$ approximately connects the fixed point at $\widetilde{a} = 0$ to the $\Lambda$CDM prediction at $\widetilde{a}=0.7$. The solid black line at twice this level ($2 + 7\widetilde{a}$) shows our prior upper limit on $E$, above which the grey shading indicates the excluded region. We also require $E \geq 0$ to enforce an expanding universe. Our limits apply only to the range $0 \leq \widetilde{a} \leq 0.70$ as we do not use data outside this range. For the AP test in Appendix~\ref{app:AP}, we further restrict to $0 \leq \widetilde{a} \leq 0.53$ for the same reason. We adopt a uniform prior on $E$ at $k_{\max}$ control points in $\widetilde{a}$ and then check that the unique polynomial through these points with $E \equiv 1$ today remains within the above bounds.}
    \label{fig:E_prior}
\end{figure}

A crucial aspect for a cosmographic analysis is the prior on the polynomial coefficients $C_k$, but it is difficult to estimate an appropriate prior range. Instead of directly bounding each $C_k$ separately, we set a prior on the shape of the expansion history by requiring the following:
\begin{eqnarray}
    0 ~\leq~ E(\widetilde{a}) ~\leq~ 2 + 7 \, \widetilde{a} , \qquad 0 \leq \widetilde{a} \leq \widetilde{a}_{\max} \, ,
    \label{eqn:envelope}
\end{eqnarray}
where $\widetilde{a}_{\max} = 0.70$ is around the largest lookback scale factor considered by the data (the DESI~DR2 BAO point at $z = 2.33$). The lower limit imposes an expanding universe. The upper limit is twice our $\Lambda$CDM-inspired linear model of $1 + 3.5\widetilde{a}$, which approximately joins the two ends of the flat $\Lambda$CDM $E(\widetilde{a})$ relation over this range with the SPA parameters (Figure~\ref{fig:E_prior}). This means the upper limit lies between $\approx 2\times$ and $3.3\times$ the $\Lambda$CDM expansion rate at every $\widetilde{a}$ in this range. The broad allowed range of $E$ permits any plausible late-time departure from $\Lambda$CDM. We sample our $E$ prior (Eq.~\ref{eqn:envelope}) using a uniform prior on $E$ over the allowed range at each of $k_{\max}$ control points, which we place at the Chebyshev--Lobatto nodes $\widetilde{a}_j = \widetilde{a}_{\max} \left[ 1 - \cos \left( j \pi / k_{\max} \right) \right] / 2$, where $j = 1 \ldots k_{\max}$. Given also that $E \equiv 1$ today, this defines a unique polynomial of order $k_{\max}$ through a fixed linear transformation dependent only on $k_{\max}$. We then reject the polynomial if it goes outside our prior limits anywhere in the interval $0 \leq \widetilde{a} \leq \widetilde{a}_{\max}$. This implies a joint prior which is uniform in the $C_k$ over the region allowed by Eq.~\ref{eqn:envelope}, even though the marginal prior on each $C_k$ is not uniform. We show in Appendix~\ref{app:prior} that this condition does not truncate our posteriors, which give relatively narrow ranges across which the priors are essentially flat.

Table~\ref{tab:priors} shows the prior on every parameter considered in this paper. We sample the posterior with the No-U-Turn Sampler (NUTS) as implemented in \texttt{numpyro} \citep{Hoffman_2014, Bingham_2019, Phan_2019}, ensuring convergence with the Gelman-Rubin diagnostic $R - 1 < 0.02$ \citep{Gelman_1992, Brooks_1998}. We obtain similar results using least-squares fits, which need no explicit prior ranges and assume that $\chi^2$ varies quadratically with the model parameters near its minimum (Section~\ref{sec:robustness_tests}). We therefore use least squares for many of our analysis variants to reduce the computational cost.

\begin{table}
    \centering
    \begin{tabular}{ll}
    \hline
    Parameter & Prior\\
    \hline
    \multicolumn{2}{l}{\textit{Cosmographic expansion (Eq.~\ref{eqn:cosmographic-expansion})}} \\
    $H_0$ & $\mathcal{U}(50,\,90)\kmsMpc$ \\
    $E(\widetilde{a})$, $0 \leq \widetilde{a} \leq 0.70$ & $\mathcal{U}(0,\,2 + 7\widetilde{a})$ \\
    $\Mdes$ & $\mathcal{U}(-21,\,-18)$~mag \\
    $r_{\rm d}$ (configuration~A) & $\mathcal{N}(147.05,\,0.30)$~cMpc \\
    $r_{\rm d}$ (configuration~B) & $\mathcal{U}(120,\,175)$~cMpc \\
    $\Delta M_\mathrm{legacy}$ & $\mathcal{U}(-1,\,+1)$~mag \\
    \multicolumn{2}{l}{\textit{Flat $\Lambda$CDM (Eq.~\ref{eqn:lcdm-expansion})}} \\
    $H_0$ & $\mathcal{U}(50,\,90)\kmsMpc$ \\
    $\Omega_{\rm m}$ & $\mathcal{U}(0.05,\,0.70)$ \\
    \multicolumn{2}{l}{\textit{Model comparison (Sections~\ref{sec:evidence} and \ref{sec:Model_comparison})}} \\
    $H_0$ & $\mathcal{U}(40,\,100)\kmsMpc$ \\
    $\omega_{\rm b}$ & $\mathcal{U}(0.020,\,0.025)$ \\
    $\Omega_{\rm bc}$ & $\mathcal{U}(0.05,\,0.70)$ \\
    $A$ & $\mathcal{U}(-10,\,+10)\kmsMpc$ \\
    $z_1$ & $\mathcal{U}(0.001,\,0.50)$ \\
    $w_0$ & $\mathcal{U}(-2,\,0)$ \\
    $w_a$ & $\mathcal{U}(-3,\,2)$ \\
    \hline
    \end{tabular}
    \caption{Priors on the parameters considered in this paper. The shape of the expansion history is required to lie within the bounds specified in Eq.~\ref{eqn:envelope}, with $E$ assigned a uniform prior over its allowed range at each of $k_{\max}$ control points in $\tilde{a}$. This is followed by a check that the resulting polynomial (Eq.~\ref{eqn:cosmographic-expansion}) remains within bounds for all $0 \leq \widetilde{a} \leq 0.70$ (see the text). In analyses with a more rigorous CMB comparison, $\omega_{\rm b} \equiv \Omega_{\rm b} h^2$ is the physical baryon density and $\Omega_{\rm bc}$ is the present density parameter of baryons + CDM, from which we derive $r_{\rm d}$ (Section~\ref{sec:evidence}).}
    \label{tab:priors}
\end{table}

We also fit the flat $\Lambda$CDM model for comparison. This involves replacing the cosmographic expansion of Eq.~\ref{eqn:cosmographic-expansion} with the following:
\begin{eqnarray}
    H \left( z \right) ~=~ H_0 \sqrt{\Omega_{\rm m} \left( 1 + z \right)^3 + \left( 1 - \Omega_{\rm m} \right)} \, ,
    \label{eqn:lcdm-expansion}
\end{eqnarray}
where $\Omega_{\rm m}$ is the matter density parameter. All other aspects of the analysis remain unchanged, but the free parameters are now $\Omega_{\rm m}$, $H_0$, $\Mdes$, and $r_{\rm d}$. The deceleration parameter is the derived quantity $q_0 = \tfrac{3}{2}\Omega_{\rm m} - 1$. Since Eq.~\ref{eqn:lcdm-expansion} has much less flexibility in the shape of $H(z)$, comparing it to our cosmographic fit shows how conclusions like the predicted $H_0$ and any calibration offset in the legacy SNIa (Section~\ref{sec:systematics-legacy}) depend on this flexibility.

Either expansion history enters every likelihood: the CC directly through $H(z)$, the SNIa through the luminosity distance $D_{\rm L}$ (Eq.~\ref{eqn:DL}), and the BAO through $D_{\rm M}$, $D_{\rm H}$, and $D_{\rm V}$. The $C_k$ coefficients (or $\Omega_\mathrm{m}$ for $\Lambda$CDM), $\Mdes$, and $r_{\rm d}$ are sampled jointly, so the three types of observation are coupled through a common expansion history and cannot drift apart. Since we consider both luminosity distances to SNIa and angular diameter distances for the BAO, our results assume the Etherington distance-duality relation $D_{\rm L} = (1 + z) D_{\rm M}$ \citep{Etherington_1933}.

Throughout the paper, our baseline truncates Eq.~\ref{eqn:cosmographic-expansion} at $k_{\max} = 4$, the order with the highest Bayesian evidence for DES. We test how the results depend on this choice by repeating the analysis for $k_{\max} = 2$--6 and computing the Bayesian evidence $Z(k_{\max})$ of each order. We use the nested sampling algorithm \texttt{dynesty} \citep{Skilling_2004, Skilling_2006, Speagle_2020, Koposov_2024} with random slice sampling, 1000 live points, and a stopping criterion of $\Delta \ln \hat{Z} = 0.1$, where $\hat{Z}$ denotes the approximate remaining Bayesian evidence. Assigning every order the same prior probability, the weighted posterior probability of order $k_{\max}$ is as follows:
\begin{eqnarray}
    w \left( k_{\max} \right) ~=~ Z \left( k_{\max} \right) \div \displaystyle\sum_{k = 2}^{6} Z \left( k \right).
    \label{eqn:order_weight}
\end{eqnarray}
We present the results using different $k_{\max}$ as a robustness test of our baseline in Section~\ref{sec:robustness_tests}, where we then average the results for different $k_{\max}$ using the above weights to get the Bayesian model average (BMA; see Appendix~\ref{app:ordersweep}).

The basic idea behind the IDL is to calibrate datasets assuming the validity of $\Lambda$CDM prior to recombination. In our setup, the total $\chi^2$ is therefore
\begin{eqnarray}
    \chi^2_{\rm total} ~=~ \chi^2_{\rm{SNIa}} + \chi^2_{\rm BAO} + \begin{cases} \chi^2_{\mathrm{CMB}} & \text{(configuration A),} \\ \chi^2_{\rm CC} & \text{(configuration B),} \end{cases}
    \label{eqn:chisq_tot}
\end{eqnarray}
where $\chi^2_{\rm{SNIa}}$ is the contribution of SNIa from DES (possibly supplemented by the legacy SNIa), $\chi^2_{\rm BAO}$ is the contribution of DESI~DR2 BAO measurements, $\chi^2_{\mathrm{CMB}}$ is the CMB constraint imposed via a Gaussian prior of $r_{\rm d} = 147.05 \pm 0.30$~cMpc \citepalias{Planck_2020}, and $\chi^2_{\rm CC}$ is the CC contribution. We use either the CMB constraint on $r_{\rm d}$ assuming $\Lambda$CDM prior to recombination, or we use CC and impose only a broad uninformative prior on $r_{\rm d}$. We break the well-known $H_0$--$\Mdes$ degeneracy in the SNIa by requiring consistency with BAO, but the BAO data itself suffers from a $D_{\rm M,H,V}$--$r_{\rm d}$ degeneracy. This is because the observables are the BAO angular scale and redshift depth, which measure distances relative to $r_{\rm d}$.

We obtain an independent calibration of $r_{\rm d}$ using two different approaches:
\begin{enumerate}
    \item \textbf{Baseline (configuration A):} Assuming standard pre-recombination physics, which requires $r_{\rm d} = 147.05 \pm 0.30$~cMpc \citepalias[TT, TE, EE + low-E modes in the CMB;][]{Planck_2020}; or
    \item \textbf{Extension (configuration B):} Comparing CC measurements of $H(z)$ to BAO measurements of $D_{\rm H}/r_{\rm d}$, with a wide uniform $r_{\rm d}$ prior covering 120--175~cMpc. This approach does not assume $\Lambda$CDM prior to recombination, instead relying on stellar physics.
\end{enumerate}
Our baseline assumes $\Lambda$CDM behaviour at high redshift. The extension relaxes this assumption, reducing the statistical constraining power of the analysis. Since the extension constrains $r_{\rm d}$ independently of the CMB, it tests whether early-time solutions to the Hubble tension can be accommodated by late-time observables, albeit without considering the CMB as this will depend on details of the new physics. Such models leave the BAO constraint on $H_0 r_{\rm d}$ unchanged and therefore require a low $r_{\rm d}$, which we compare with the $r_{\rm d}$ inferred in configuration~B (Section~\ref{sec:discussion}).

\subsection{Redshift-floor sweep}
\label{sec:z_floor_method}

To check if our results are driven by SNIa at especially low redshifts, we conduct a sweep of the redshift floor $z_\mathrm{floor}$ below which SNIa are neglected, recalibrating the IDL each time. Some studies report a descending trend in $H_0$ with the redshift of the data used to measure it \citep{Krishnan_2020, Krishnan_2021_H0z, Lopez_2025, Jia_2025a, Jia_2026}. Those studies infer $H_0$ from data in a narrow redshift range centred on $z$, whereas we use a redshift floor, so we denote our results as $H_0(z_\mathrm{floor})$. A clear trend in $H_0(z_\mathrm{floor})$ which becomes asymptotically flat would indicate that the lowest redshift SNIa carry the cosmological parameters away from the more robust result from higher redshift SNIa. On the other hand, a flat trend in $H_0(z_\mathrm{floor})$ at the low-redshift end would indicate that different SNIa subsets prefer much the same present expansion rate, though error bars should widen at high $z_\mathrm{floor}$ as more data is excluded, placing greater reliance on the extrapolation to $z \to 0$.

\subsection{Alcock--Paczynski consistency}
\label{sec:AP_consistency}

We subject the IDL to the AP consistency test \citep{Alcock_1979}. This is based on the AP ratio $F_\mathrm{AP} \equiv D_{\rm M}/D_{\rm H}$ (Appendix~\ref{app:AP}). Using SNIa alone and assuming smoothness of the distance--redshift relation combined with the Etherington relation, we predict $F_\mathrm{AP}$ at three redshifts where particularly precise BAO measurements of $F_\mathrm{AP}$ are available from DESI~DR2. Taking the ratio causes $r_{\rm d}$ to cancel out, so the AP test is independent of the absolute calibration of the BAO sound horizon. It is therefore also independent of the SNIa absolute magnitude -- and thus $H_0$. The AP test instead probes the \emph{shape} of the reconstructed distance--redshift relation, or equivalently of our cosmographic $E(1-a)$ expansion.

\subsection{Systematics with the legacy SNIa}
\label{sec:systematics-legacy}

The DES-Dovekie compilation \citep{Popovic_2026_Dovekie} combines the homogeneous, single-instrument DES catalogue of SNe ($z\in[0.073,1.14]$) with an external low-redshift anchor assembled from several legacy surveys ($z\in[0.025,0.087]$). The legacy SNIa come almost entirely from the Foundation sample (117 SNIa) or from the Center for Astrophysics CfA3 and CfA4 samples (72), with a further 8 from the Carnegie Supernova Project (CSP-I). We denote the CfA and CSP samples together as CfA/CSP. All 197 legacy SNIa lie at $z \la 0.1$, where DES has practically no SNIa and BAO data are scant \citep[though see table~1 of][and references therein]{Banik_2025_BAO}. This makes the legacy SNIa an important lever on the $z \to 0$ region. We therefore test whether they follow the same distance--redshift relation as DES, and what this means for $H_0$.

We stress that the redshift distributions of the legacy SNIa and the main DES sample have very little overlap \citep[see figure~3 of][]{DES_2024_SNe}. This means our test relies on a downward extrapolation of the DES sample, not a comparison of SNIa apparent magnitudes at the same redshift between the two subsamples. The legacy SNIa span $z\in[0.025, 0.087]$ while DES begins at $z=0.073$, so only 2 DES SNe and 10 legacy SNIa lie in the overlap region. Given the large number of legacy SNIa with modest error bars, the uncertainty in $\Delta M_\mathrm{legacy}$ is dominated by extrapolation uncertainty, i.e., how much freedom the DES reference has below its own floor. The main uncertainty in $\Delta M_\mathrm{legacy}$ is thus not from the legacy SNIa themselves -- at least if we take their uncertainties at face value.

Our test for systematics in the legacy SNIa is designed to require no absolute calibration. We proceed in three steps:
\begin{enumerate}
  \item We fit the shape of the DES Hubble diagram (using only DES SNIa) with a rigid fourth-order polynomial as the baseline model (Eq.~\ref{eqn:cosmographic-expansion}), the same expansion additionally constrained by the DESI result for $F_{\rm AP} \equiv D_{\rm M}/D_{\rm H}$ (which is independent of $r_{\rm d}$; see Section~\ref{sec:AP_consistency}), rigid $\Lambda$CDM with free $\Omega_{\rm m}$ (Eq.~\ref{eqn:lcdm-expansion}), and rigid $\Lambda$CDM likewise constrained by $F_{\rm AP}$.
  \item We extrapolate the fitted relation below the DES floor ($z = 0.073$) down to the redshifts of the legacy SNIa, allowing us to calculate the residuals $r_i \equiv m_{{\rm legacy},i} - m_{\rm pred}(z_i)$ between observed legacy SNIa and our extrapolated predictions.
  \item We take $\Delta M_\mathrm{legacy}$ to be the weighted mean of the $r_i$ under the full statistical + systematic covariance of the compilation, including the legacy--DES cross-covariance.
\end{enumerate}

It is important to highlight why this approach is calibration-free, since this property is what makes $\Delta M_\mathrm{legacy}$ comparable across the four parametrisations. Rescaling $H \to \lambda H$ maps $D_{\rm M} \to D_{\rm M}/\lambda$ and hence $m \to m - 5\log_{10}\lambda$, so that $(H_0 \to \lambda H_0, \, M \to M + 5\log_{10}\lambda)$ is an exact flat direction of the fit. We therefore fix $H_0 = 70\kmsMpc$ and let $\Mdes$ float freely. Every predicted apparent magnitude is invariant along this direction, making the fitted $\Mdes$ meaningless by construction. This means that no absolute scale enters $\Delta M_\mathrm{legacy}$, making it independent of the sound horizon and the local distance ladder. In particular, $\Delta M_\mathrm{legacy}$ inherits no part of the calibration chain that later produces $H_0$.

To constrain $H_0$, we cannot simply add the $\Delta M_\mathrm{legacy}$ obtained above to the legacy SNIa because $H_0$ and $\Delta M_\mathrm{legacy}$ are correlated. We therefore conduct a separate analysis where we sample $\Delta M_\mathrm{legacy}$ alongside the other parameters. The constraints we apply are SNIa + BAO + \citetalias{Planck_2020} $r_{\rm d}$. We consider the cosmographic expansion as well as $\Lambda$CDM. For comparison, we also report results for the two cases that do not have $\Delta M_\mathrm{legacy}$ as a model parameter: excluding the legacy SNIa altogether, or including them at face value ($\Delta M_\mathrm{legacy} = 0$). These results highlight the role of the legacy SNIa in $H_0$ determinations via the IDL.

\subsection{Bayesian evidence for a late-time solution to the Hubble tension}
\label{sec:evidence}

\subsubsection{Alternative models}
\label{sec:Alternatives}

To assess if the data prefers a late-time solution to the Hubble tension over $\Lambda$CDM, we consider the following phenomenological expansion history \citep{Najera_2026}:
\begin{eqnarray}
    && H \left( z \right) ~=~ H_{0, \Lambda \mathrm{CDM}} \sqrt{\Omega_{\rm m} \left( 1 + z \right)^3 + \left( 1 - \Omega_{\rm m} \right)} + \Delta H \left( z; A, z_1 \right), \nonumber \\
    && \Delta H(z; A, z_1) ~=~ A \exp \left( -z/z_1 \right),
    \label{eqn:H_z_extra}
\end{eqnarray}
where $A$ is the amplitude and $z_1$ is the redshift scale of $\Delta H$, an extra contribution to $H(z)$. A solution to the Hubble tension should have $A \approx 6\kmsMpc$, while we expect $z_1 \la 0.2$ \citep{Najera_2026}. This \texttt{exp} model represents a flexible late-time solution to the Hubble tension in which $H_0 = H_{0, \Lambda \mathrm{CDM}} + A$. We adopt $A \sim \mathcal{U}(-10, 10)\kmsMpc$ (Table~\ref{tab:priors}). This range comfortably covers the $A \approx 6\kmsMpc$ needed to solve the Hubble tension. The prior is symmetric about zero, so it does not favour positive $A$.

We also consider the Chevallier-Polarski-Linder \citep*[CPL;][]{Chevallier_2001, Linder_2003, Linder_2008} model in which the dark energy equation of state is linear in $a$ ($w = w_0 + \widetilde{a} w_a$), so the expansion history is as follows:
\begin{eqnarray}
    H \left (z \right) &=& H_0 \sqrt{\Omega_{\rm m} \left( 1 + z \right)^3 + \left( 1 - \Omega_{\rm m} \right) f(z)} \, , \\
    f \left( z \right) &=& \left( 1 + z \right)^{3 \left( 1 + w_0 + w_a \right)} \exp \left( -3 \widetilde{a} w_a \right).
    \label{eq:CPL}
\end{eqnarray}
We can recover $\Lambda$CDM behaviour by setting $w_0 = -1$ and $w_a = 0$, in which case $w = -1$ and $f(z) = 1 \forall z$, indicating a constant dark energy contribution.

The Bayesian evidence of model $\mathcal{M}_i$ is defined as follows:
\begin{eqnarray}
    Z \left( \mathcal{M}_i \right) ~\equiv~ \int \dd\theta \, \pi \left( \theta \vert \mathcal{M}_i \right) \mathcal{L} \left(D \vert \theta, \mathcal{M}_i \right),
\end{eqnarray}
where $\pi$ is the prior and $\mathcal{L}$ is the likelihood of the data $D$ when the vector of model parameters is $\theta$. We compare the \texttt{exp} and CPL models to $\Lambda$CDM through the Bayes factor
\begin{eqnarray}
    B_{10} ~\equiv~ \frac{Z \left( \mathcal{M}_1 \right)}{Z \left( \mathcal{M}_0 \right)} \, ,
\end{eqnarray}
where $\mathcal{M}_0$ is $\Lambda$CDM and $\mathcal{M}_1$ is an alternative, which in our case is either the \texttt{exp} model (Eq.~\ref{eqn:H_z_extra}) or the CPL model (Eq.~\ref{eq:CPL}). Both models are phenomenological, with the \texttt{exp} model aimed at the Hubble tension \citep{Najera_2026} and the CPL model aimed at the preference of DESI~DR2 BAO for evolving dark energy \citep{DESI_2025}. The Bayesian evidence automatically applies an Occam penalty for the two extra parameters that these models have compared to $\Lambda$CDM. Since both models have two extra parameters, differences in evidence between them do not reflect parameter counting, but the Occam penalty still depends on how much of the prior range the data exclude, and therefore on the adopted prior ranges (Table~\ref{tab:priors}).

To compare these models on an equal footing, we must use the same dataset in all cases. We use the SNIa, BAO, and CC data described in Section~\ref{sec:data}, possibly supplemented by local anchors on $H_0$. As described below, we supplement these with additional constraints from the early Universe.
\begin{eqnarray}
    \chi^2_\mathrm{total} ~=~ \chi^2_{\rm SNIa} + \chi^2_{\rm BAO} + \chi^2_{\rm{CMB}} + \chi^2_{\rm{BBN}} + \chi^2_{\rm CC} \, \left( + \chi^2_{\rm anchor} \right),
    \label{eqn:evidence_chi2-new}
\end{eqnarray}
with the anchor term entering only where stated (Section~\ref{sec:H0_anchors}). Since we need the Bayesian evidence, we again use \texttt{dynesty} \citep{ Skilling_2004, Skilling_2006, Feroz_2009, Speagle_2020, Koposov_2024} with random slice sampling, 800 live points, and a stopping criterion of $\Delta \ln \hat{Z} = 0.01$.

\subsubsection{Compressed CMB likelihood and BBN}
\label{sec:comp_CMB}

In the rest of this paper, the CMB is included only via the \emph{Planck} prior on $r_{\rm d}$ because the cosmographic expansion cannot be extended to high redshifts. But since this is possible for both the \texttt{exp} and CPL models, we can implement the CMB constraint in a more rigorous manner. We calculate the CMB term in Eq.~\ref{eqn:evidence_chi2-new} following the compressed CMB likelihood approach of \citet{Banik_2026_z0}, whose protocol we summarise here. We no longer impose a prior on $r_{\rm d}$ because it must now be found self-consistently using the physics of sound waves in the baryon--photon plasma prior to decoupling. This requires both the baryon and CDM densities, so we sample the present density parameter of baryons + CDM ($\Omega_{\rm bc}$) and the physical baryon density
$\omega_{\rm b} \equiv \Omega_{\rm b} h^2$ from their prior ranges, where $h \equiv H_0/(100\kmsMpc)$ and
$\omega_x \equiv \Omega_x h^2$ for any component $x$ (Table~\ref{tab:priors}). In the \texttt{exp} model, these densities refer to the underlying $\Lambda$CDM background, so $h$ is set by $H_{0, \Lambda \mathrm{CDM}}$ rather than $H_0$. This is because $\Delta H$ (Eq.~\ref{eqn:H_z_extra}) is completely negligible long before recombination for any $z_1$ in our prior range, so it has no effect on the expansion history around the time of recombination. However, the $\Delta H$ term is important when calculating the comoving distance to recombination, which affects the angular scale of its anisotropies.

Within this framework, the CMB constraints are expressed through the parameter vector $\bm{x} \equiv (100 \, \theta_\star, \omega_{\rm b}, \omega_{\rm bc})^{\rm T}$, where $\theta_\star \equiv r_\star/D_{\rm M}(z_\star)$ is the angular size of the comoving sound horizon $r_\star$ at recombination, $D_{\rm M}(z_\star)$ is the comoving distance to recombination, and $\star$ subscripts denote quantities evaluated at that epoch \citep{Lemos_2023}. CMB observations indicate that $\bm{x}$ has a mean value of $\bm{\mu}_{\rm CMB}$ and covariance matrix $\matr{C}_{\rm CMB}$, both of which were obtained by \citet{Banik_2026_z0} from the chains of \citet{SPT_2026}. Their analysis combines their SPT-3G results with \citetalias{Planck_2020} and ACT~DR6 \citep{Louis_2025}. We find that this SPA combination can be summarised as follows:
\begin{eqnarray}
    \bm{\mu}_{\rm CMB} &=& \left( 1.04161, \, 0.02240, \, 0.14265 \right)^{\rm T}, \\
    \matr{C}_{\rm CMB} &=& 10^{-9} \times {\setlength{\arraycolsep}{3pt}
    \begin{bmatrix}
    53.8482 & 1.10723 & -28.7005 \\
    1.10723 & 9.20447 & -17.6016 \\
    -28.7005 & -17.6016 & 843.132
    \end{bmatrix}}. \nonumber
    \label{eqn:CMB_compressed}
\end{eqnarray}
The CMB term is then $\chi^2_{\rm CMB} = (\boldsymbol{x} - \boldsymbol{\mu}_{\rm CMB})^{\rm T} \matr{C}_{\rm CMB}^{-1} (\boldsymbol{x} - \boldsymbol{\mu}_{\rm CMB})$, with $\boldsymbol{x}$ predicted by each model.

We obtain $D_{\rm M}(z_\star)$ by integrating $c/H(z)$ for each model all the way to recombination. This means that a late-time departure from $\Lambda$CDM changes the predicted $\theta_\star$ despite leaving the early universe untouched, which is precisely how the CMB constrains late-time solutions to the Hubble tension. In particular, any extra expansion at low redshift has to be compensated elsewhere in the expansion history in order to preserve the comoving distance to recombination. Since the integral extends to $z_\star \approx 1090$, the expansion history now includes photons and neutrinos, which we model following \citet{Banik_2026_z0}. We compute $r_\star$ and $r_{\rm d}$ from $\omega_{\rm b}$ and $\Omega_{\rm bc}$ using the analytic result of \citet{Hu_1996}, with the photon temperatures at recombination and at the end of the drag epoch taken from fits to a grid of \textsc{class} runs \citep{CLASS}. We give the details in Appendix~\ref{app:compCMB}.

In flat $\Lambda$CDM, this compressed likelihood recovers the $H_0$ and its uncertainty inferred by \citet{SPT_2026} from the full CMB power spectra to within $0.01\kmsMpc$, demonstrating that it retains the information relevant to the background expansion \citep[see appendix~C of][]{Banik_2026_z0}. Following their study, we also add the BBN constraint to Eq.~\ref{eqn:evidence_chi2-new} through a Gaussian prior on
$100 \, \omega_{\rm b} = 2.1915 \pm 0.0215$ \citep{Giovanetti_2025_BBN}.

\subsubsection{Local \texorpdfstring{$H_0$}{H\textsubscript{0}} anchors}
\label{sec:H0_anchors}

To assess the feasibility of a late-time solution to the Hubble tension, we consider the case with and without anchors on $H_0$ from the standard distance ladder. We use either the SH0ES constraint of $H_0 = 73.17 \pm 0.86\kmsMpc$ from the traditional Cepheid--SNIa route \citep{Leavitt_1912, Breuval_2024}, or we use three anchors which are independent of Cepheid and SNIa absolute magnitudes, thereby making them independent of some of the main contributors to the Hubble tension.
We assume all three are independent of each other because they come from non-overlapping datasets, as follows:
\begin{enumerate}
    \item The TRGB \citep*{Lee_1993, Madore_1995} is used to calibrate distances to galaxies with measured surface brightness fluctuations \citep[SBF;][]{Tonry_1988}. Our adopted TRGB + SBF value is $H_0 = 73.8 \pm 0.7\,({\rm stat}) \pm 2.3\,({\rm sys})\kmsMpc$ \citep{Jensen_2025}. We combine their statistical and systematic errors in quadrature to obtain a single uncertainty of $2.4\kmsMpc$.
    \item The GW170817 standard siren \citep{Abbott_2017, LIGO_Virgo_2017} is used to obtain a luminosity distance. This is combined with a peculiar velocity correction and the very long baseline interferometry (VLBI) constraint on the inclination to obtain an estimated $H_0=68.3^{+4.6}_{-4.5}\kmsMpc$ \citep{Mukherjee_2021}, as listed in table~2 of \citet{Alfradique_2026}. We adopt the larger uncertainty of $4.6\kmsMpc$ to be conservative. We do not use the headline result of \citet{Alfradique_2026} as this also combines the single bright siren with 17 dark sirens, which would significantly complicate our analysis while only slightly improving the constraining power.
    \item Type~II SNe (SNII) with the expanding photosphere method \citep[EPM;][]{Vogl_2025}. Their study gives an estimated $H_0 = 74.9 \pm 2.7\kmsMpc$ if we assume the statistical and systematic errors are comparable, which the authors indicate is a reasonable conservative assumption.
\end{enumerate}

We stress that adding SH0ES and legacy SNIa to the same analysis risks overlapping data. This is the reason why we lead with the non-SH0ES anchors, which we can use even if we also include the legacy SNIa. However, if the latter are excluded, there is negligible overlap between the SNIa used by DES and by SH0ES \citep{Riess_2022_comprehensive}.

There is an important caveat that we must take into account for the \texttt{exp} model. The above anchors obtain $H_0$ by fitting their luminosity distances with a second-order kinematic expansion in $z$ that assumes $q_0 = -0.55$ and a jerk parameter $j_0 = 1$. This expansion is then extrapolated down to $z \to 0$. If $z_1 \la z_\mathrm{max}$, the maximum redshift of the data considered, the \texttt{exp} model $H_0$ does not correspond to the anchor measurement, so we do not compare the two directly. We instead calculate the \texttt{exp} model distance moduli at the redshifts of the real data and fit them with the same kinematic expansion, giving each object equal weight. We can then read off $H_0$ \citep[similarly to section~4.1.1 of][]{Najera_2026}. This calculates the value of $H_0$ that would have been reported by observers from distance measurements to anchors with the same redshift distribution as the real data, if the universe followed the phenomenological \texttt{exp} cosmology. Since SH0ES obtains $H_0$ from SNIa at $z = 0.023$--0.15, a model whose excess expansion rate is concentrated at $z \la 0.05$ predicts a lower SH0ES measurement than its actual $H_0$ because the observers would fit a smooth Hubble law across a window in which most of the excess has already decayed. We therefore also quote $H_0^{\rm win}$, the value that the SH0ES analysis would infer from each \texttt{exp} and CPL posterior, given the SH0ES redshift window (Table~\ref{tab:evidence}). In other words, $H_0$ remains the expansion rate today, while $H_0^{\rm win}$ decides whether a model reproduces the SH0ES measurement.

\section{Results}
\label{sec:results}

\subsection{Baseline IDL constraints and the impact of the legacy SNIa}
\label{sec:results-params}

\begin{figure}
    \centering
    \includegraphics[width=1.0\linewidth]{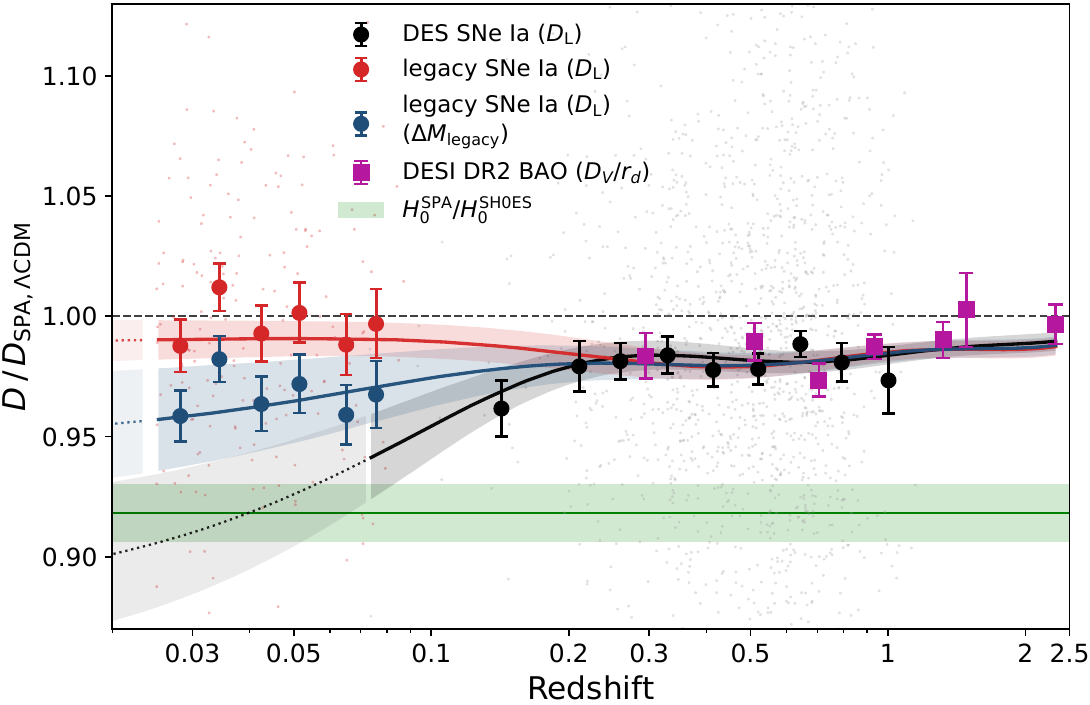}
    \caption{Observed distances divided by the flat $\Lambda$CDM prediction with the SPA parameters \citep[$H_0 = 67.19\kmsMpc$ and $\Omega_{\rm m} = 0.3175$;][]{SPT_2026} and $r_{\rm d} = 147.05$~cMpc \citepalias{Planck_2020}. We neglect the small uncertainties in these parameters and in $\Mdes$, which is fixed at the best fit of each analysis. Faint grey (magenta) points show individual DES (legacy) SNIa at their Hubble diagram redshift. Circles show inverse variance-weighted mean SNIa luminosity distances relative to SPA using the full DES-Dovekie covariance for DES (black), legacy at face value (red), and legacy recalibrated by the jointly inferred $\Delta M_\mathrm{legacy}$ (blue; Section~\ref{sec:systematics-legacy}), which lowers them by $\approx 3\%$. We use uniform $\ln z$ bins for the SNIa. Magenta squares show DESI~DR2 $D_\mathrm{V}/r_{\rm d}$, combining $D_{\rm M}$ and $D_{\rm H}$ using their published covariance. Lines in the same colour as the data show the best-fitting $k_{\max} = 4$ cosmography (Eq.~\ref{eqn:cosmographic-expansion}) up to $z = 2.33$, with $1\sigma$ Gaussian-equivalent bands. Results are extrapolated below the lowest SNIa of each sample, as marked by the lines becoming dotted and a short gap in the band. The green band is $H_0^\mathrm{SPA}/H_0^\mathrm{SH0ES}$, the level that a late-time solution to the Hubble tension must reach as $z \to 0$.}
    \label{fig:dz_ratio}
\end{figure}

To gain a first impression of whether the Hubble tension might plausibly be solved at late times, we can consider the ratio between predicted distances in the flat $\Lambda$CDM cosmology that best fits the CMB \citep[the SPA cosmology;][]{SPT_2026} and observed distances from SNIa and BAO. The results are shown in Figure~\ref{fig:dz_ratio}, which plots DESI~DR2 BAO measurements of $D_\mathrm{V}/r_{\rm d}$ and binned $D_{\rm L}$ from SNIa, using uniform bins in $\ln z$. The SNIa are calibrated according to the best-fit value of $\Mdes$, which has very little uncertainty due to the need for consistency with the BAO. This is evident in the excellent agreement between SNIa and BAO measurements in the regime where they overlap ($z = 0.3$--1). In particular, the DES SNIa have a weighted mean ratio of $D_{\rm L}$ relative to the SPA prediction of $0.981 \pm 0.003$, while the corresponding result for DESI~DR2 BAO measurements of $D_{\rm V}/r_{\rm d}$ is $0.987 \pm 0.003$. This agreement is inevitable in the IDL approach since it is how we calibrate $\Mdes$. However, it does not guarantee consistency over the wide redshift range where there is an overlap between SNIa and BAO. We revisit this issue in Section~\ref{sec:AP_consistency} using the AP test, which does not rely on an assumed $r_{\rm d}$.

As we move towards lower redshift, Figure~\ref{fig:dz_ratio} shows a drastic difference between the DES SNIa, the legacy SNIa at face value, and the legacy SNIa recalibrated by the value of $\Delta M_{\rm legacy}$ that we infer jointly with the other parameters (Section~\ref{sec:systematics-legacy}). The legacy SNIa cover the crucial range of $z = 0.025$--0.087, but they sit systematically above where we would expect them to sit based on the DES SNIa. Recalibrating the legacy SNIa naturally reduces this discrepancy using an offset of $\Delta M_\mathrm{legacy} = +0.065$~mag. Since $H_0$ is inversely related to the distance scale in the $z \to 0$ limit, these differences can be expected to significantly affect our conclusions regarding the IDL prediction for $H_0$. To highlight this, we add a green horizontal band at $H_0^\mathrm{SPA}/H_0^\mathrm{SH0ES} = 0.918 \pm 0.012$, which is the level that must be reached as $z \to 0$ in order to solve the Hubble tension. Our results suggest that this is quite likely if we consider SNIa from only the homogeneous DES catalogue, although this relies on extrapolating below its floor of $z = 0.073$. A solution is marginally possible if we include the legacy SNIa after recalibrating them to line up with the extrapolated relation constrained by BAO + DES, but becomes impossible if we take the legacy SNIa at face value. The rest of this paper elaborates this basic picture without much altering it.

The corresponding expansion history $\dot{a}(z) = H(z)/(1 + z)$ is shown in Figure~\ref{fig:adot-reconstruction}. It is higher than the SPA expectation at $z \approx 0.4$ and rises sharply towards lower $z$ at the very lowest redshifts probed by the SNIa, most strongly with DES alone and least with DES + legacy. We note that our cosmographic analyses include the CMB only through $r_{\rm d}$. Preserving the comoving distance to recombination would require a compensating reduction of a few percent in the overall normalisation of $\dot{a}$, which would not qualitatively change our conclusions because this distance is barely affected by the $z \la 0.1$ regime crucial to the Hubble tension. A full CMB comparison is possible in the \texttt{exp} and CPL models (Section~\ref{sec:comp_CMB}).

\begin{figure}
    \centering
    \includegraphics[width=1.0\linewidth]{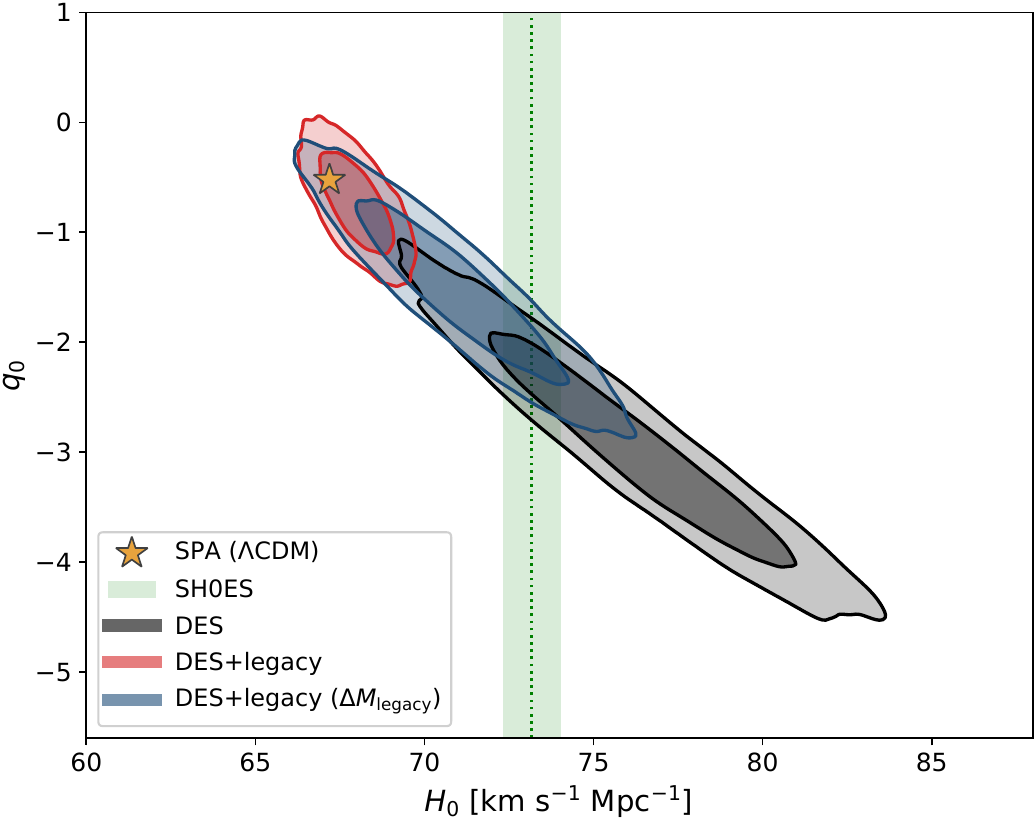}
    \caption{The joint inference on $H_0$ and $q_0$ from our $k_{\max} = 4$ cosmographic expansion using SNIa from DES (black), DES + legacy (red), and DES + legacy with free $\Delta M_\mathrm{legacy}$ (blue). In all cases, the darker shading shows the $1\sigma$ region and the lighter shading shows the $2\sigma$ region. The orange star shows the $\Lambda$CDM expectation based on the latest CMB observations \citep{SPT_2026}. The vertical green band shows $H_0^\mathrm{SH0ES}$ \citep{Breuval_2024}.}
    \label{fig:h0_q0}
\end{figure}

The reconstructed expansion history thus deviates moderately from the SPA cosmology at $z \la 0.8$, even though we adopt the \citetalias{Planck_2020} prior of $r_{\rm d} = 147.05 \pm 0.30$~cMpc and BAO data show at most modest deviations from $\Lambda$CDM predictions calibrated using the CMB \citep{Banik_2026_z0}. These predictions are evidently accurate at high redshift, so any deviations at low redshift should be apparent in both the Hubble and deceleration parameters. We therefore use Figure~\ref{fig:h0_q0} to show the joint posterior of $H_0$ and $q_0 = C_1 - 1$. The black contours show that without the legacy SNIa, the parameters are indeed significantly different to the $\Lambda$CDM expectation, which we show using an orange star \citep[neglecting the very small uncertainties;][]{SPT_2026}. Remarkably, the predicted $H_0$ with our DES-only cosmographic IDL is very much in line with the local $H_0$ anchors mentioned in Section~\ref{sec:evidence}, even slightly exceeding the central value of $H_0^\mathrm{SH0ES}$ \citep{Breuval_2024}. The high $H_0$ is related to a very negative inferred $q_0$, which implies the expansion rate accelerates more than expected at late times.

Adding the legacy SNIa tightens the constraints considerably, but it also shifts both $H_0$ and $q_0$ very close to the CMB-calibrated $\Lambda$CDM prediction (Figure~\ref{fig:h0_q0}). The $2\sigma$ constraints with and without the legacy SNIa barely overlap in the $(H_0, q_0)$ plane, which is unusual given the latter is strictly a subset of the former. This suggests that the estimated DES filter system apparent magnitudes of the legacy SNIa might have unrecognised systematics. If we allow for this possibility using our analysis with free $\Delta M_\mathrm{legacy}$ (Section~\ref{sec:systematics-results}), we get intermediate results.

\renewcommand{\arraystretch}{1.2}
\begin{table*}
    \centering
    \begin{tabular}{llccccc}
        \hline
        Configuration & Sample & $H_0 \; [\mathrm{km\,s^{-1}\,Mpc^{-1}}]$ & $r_{\rm d}$ [cMpc] & $\Mdes$ & $q_0$ & $\chi^2/{\rm dof}$\\
        \hline
        \multicolumn{7}{l}{(A) SNIa + BAO + \emph{Planck} $r_{\rm d}$}\\
        \hline
        \multicolumn{7}{l}{\textit{Cosmography, $k_{\max} = 4$}}\\
        & DES        & $76.25\pm2.87$ & $147.05\pm0.30$ (prior) & $-19.380\pm0.012$ & $-3.13^{+0.84}_{-0.60}$ & $0.921$\\
        & DES + legacy & $67.98\pm0.69$ & $147.05\pm0.30$ (prior) & $-19.378\pm0.012$ & $-0.73\pm0.30$ & $0.942$\\
        \noalign{\smallskip}
        \multicolumn{7}{l}{\textit{Flat $\Lambda$CDM}}\\
        & DES        & $68.98\pm0.48$ & $147.05\pm0.30$ (prior) & $-19.377\pm0.012$ & $-0.55\pm0.01$ & $0.920$\\
        & DES + legacy & $68.60\pm0.46$ & $147.05\pm0.30$ (prior) & $-19.379\pm0.012$ & $-0.54\pm0.01$ & $0.942$\\
        \hline
        \multicolumn{7}{l}{(B) SNIa + BAO + CC, free $r_{\rm d}$ (no CMB)}\\
        \hline
        \multicolumn{7}{l}{\textit{Cosmography, $k_{\max} = 4$}}\\
        & DES        & $75.7\pm5.3$ & $147.89\pm8.59$ & $-19.39\pm0.13$ & $-2.84\pm0.72$ & $0.920$\\
        & DES + legacy & $67.6\pm3.6$ & $148.31\pm7.77$ & $-19.39\pm0.11$ & $-0.71\pm0.29$ & $0.939$\\
        \noalign{\smallskip}
        \multicolumn{7}{l}{\textit{Flat $\Lambda$CDM}}\\
        & DES        & $69.0^{+3.4}_{-4.6}$ & $148.48\pm8.39$ & $-19.40\pm0.12$ & $-0.55\pm0.01$ & $0.917$\\
        & DES + legacy & $68.1\pm4.0$ & $148.51\pm8.68$ & $-19.40\pm0.13$ & $-0.54\pm0.01$ & $0.938$\\
        \hline
    \end{tabular}
    \caption{Summary of the most important cosmological parameters in different setups. Configuration~A (SNIa + BAO + \emph{Planck} $r_{\rm d}$) pins the sound horizon to the \citetalias{Planck_2020} prior ($r_{\rm d} = 147.05\pm0.30$~cMpc). Configuration~B (SNIa + BAO + CC) replaces the $r_{\rm d}$ prior with CC datasets, so $r_{\rm d}$ is derived without CMB input. In the first block for each configuration, we obtain the deceleration parameter $q_0 = C_1 - 1$ in our cosmographic IDL analysis with $k_{\max} = 4$ (Eq.~\ref{eqn:cosmographic-expansion}). The second block of each configuration repeats the IDL assuming flat $\Lambda$CDM (Eq.~\ref{eqn:lcdm-expansion}), for which we obtain $q_0 = \tfrac{3}{2}\Omega_{\rm m} - 1$. The final column shows the reduced $\chi^2$ using the posterior mean value of each parameter. We quote the posterior mean and standard deviation~$\sigma$, except for markedly non-Gaussian posteriors in which the mode and mean differ by more than $0.3\sigma$ or the distances from the mode to the two ends of the $1\sigma$ highest posterior density interval differ by more than $0.3\sigma$. In these cases, we quote the mode and the $1\sigma$ interval.}
    \label{tab:baseline}
\end{table*}
\renewcommand{\arraystretch}{1.0}

Our main results are presented in Table~\ref{tab:baseline}. Our baseline (cosmographic configuration~A) result is that the predicted $H_0 = 76.25 \pm 2.87 \kmsMpc$ for DES. This is only $1.03\sigma$ above SH0ES \citep{Breuval_2024}, but it lies $3.13\sigma$ above the latest CMB-constrained flat $\Lambda$CDM prediction \citep{SPT_2026}. Our mock data reveal an upward bias in our estimator, but only 0.84\% of mocks in the SPA cosmology reach this $H_0$ in a like-for-like comparison (Section~\ref{sec:robustness_tests}). Importantly, the tension with SPA depends on our choice of $k_{\max} = 4$, because DES lacks effective coverage at $z \la 0.2$ and so requires an extrapolation that depends on the order (Section~\ref{sec:robustness_tests}). We can reduce this extrapolation error if we extend the low-redshift coverage by adding the legacy SNIa, in which case the predicted $H_0$ using our cosmographic IDL drops to $67.98 \pm 0.69\kmsMpc$. This is only $1.00\sigma$ above $H_0^\mathrm{SPA}$ \citep{SPT_2026}, but lies $4.69\sigma$ below SH0ES \citep{Breuval_2024}. Adding the legacy SNIa therefore completely changes the result.

To reduce sensitivity to the assumption of $\Lambda$CDM prior to recombination, we can drop the \textit{Planck} prior on $r_{\rm d}$ and set the absolute scale of the expansion history using CC datasets. This is our configuration~B, which yields $H_0 = 75.7 \pm 5.3\kmsMpc$ for DES and $67.6 \pm 3.6 \kmsMpc$ for DES + legacy. The inferred $r_{\rm d}$ lies within $0.10\sigma$ ($0.16\sigma$) of the standard \textit{Planck} value for DES (DES + legacy). In flat $\Lambda$CDM, configuration~B gives $r_{\rm d} = 148.48 \pm 8.39$ ($148.51 \pm 8.68$)~cMpc for DES (DES + legacy), also within $0.2\sigma$ of the \citetalias{Planck_2020} value. We note that our results without this prior are not as precise as in \citet{Guo_2025}, who used a somewhat different approach and a larger number of CC measurements, combined as here with uncalibrated BAO and SNIa. Interestingly, their predicted $H_0 = 68.4^{+1.0}_{-0.8}\kmsMpc$ lies $3.6\sigma$ below the central SH0ES value. A clearer picture will emerge with further improvements to CC measurements, especially at low redshift where leverage from DES remains limited \citep{Favale_2026, Wang_2026_CC}.

Our constraints rely on an absolute calibration of DES SNIa, for which we use DESI~DR2 BAO calibrated assuming a standard $r_{\rm d}$. In all cases, the results in Table~\ref{tab:baseline} are consistent with an absolute SNIa calibration of $\Mdes \approx -19.38$, with a narrow range of $-19.380 \pm 0.012$ for DES alone to $-19.378 \pm 0.012$ for DES + legacy. These results are in $4.30\sigma$ and $4.23\sigma$ tension, respectively, with the SH0ES Cepheid-based result of $M_B = -19.253 \pm 0.027$ \citep{Riess_2022_comprehensive}. We use this value because \citet{Breuval_2024} do not directly publish $M_B$. Figure~\ref{fig:dz_ratio} shows that our result for $\Mdes$ is largely set by BAO data at $z = 0.51$--0.93 combined with the \emph{Planck} prior on $r_{\rm d}$, so $\Mdes$ is not dependent on how we extrapolate below the DES floor. This explains why $\Mdes$ and its uncertainty are barely affected by inclusion of the legacy SNIa, even though these significantly extend the coverage towards lower redshift.

Flat $\Lambda$CDM gives the same calibration to within 0.02~mag (second block of each configuration in Table~\ref{tab:baseline}), so the tension with the SH0ES $M_B$ is $4.2$--$4.3\sigma$ in every configuration~A analysis but only $1.1$--$1.2\sigma$ in configuration~B, where the CC constrain $r_{\rm d}$ far less tightly than the CMB. We discuss possible origins of the difference between $\Mdes$ and $M_B$ in Section~\ref{sec:discussion_pantheon_DES}. The quality of fit to the data remains the same regardless of whether we use flat $\Lambda$CDM or our cosmographic expansion, with reduced $\chi^2 = 0.917$--0.942. Given $N_\mathrm{data} = 1555$ (1752) data points without (with) the legacy SNIa, we expect the reduced $\chi^2$ to follow approximately a Gaussian distribution centred on 1 with dispersion 0.036 (0.034). The reduced $\chi^2$ drops from 0.938--0.942 to 0.917--0.921 upon excluding the legacy SNIa in all analysis variants, indicating a slightly better fit quality without the legacy SNIa.

\subsection{Redshift-floor sweep}
\label{sec:z_floor_result}

\begin{figure}
    \centering
    \includegraphics[width=1.0\linewidth]{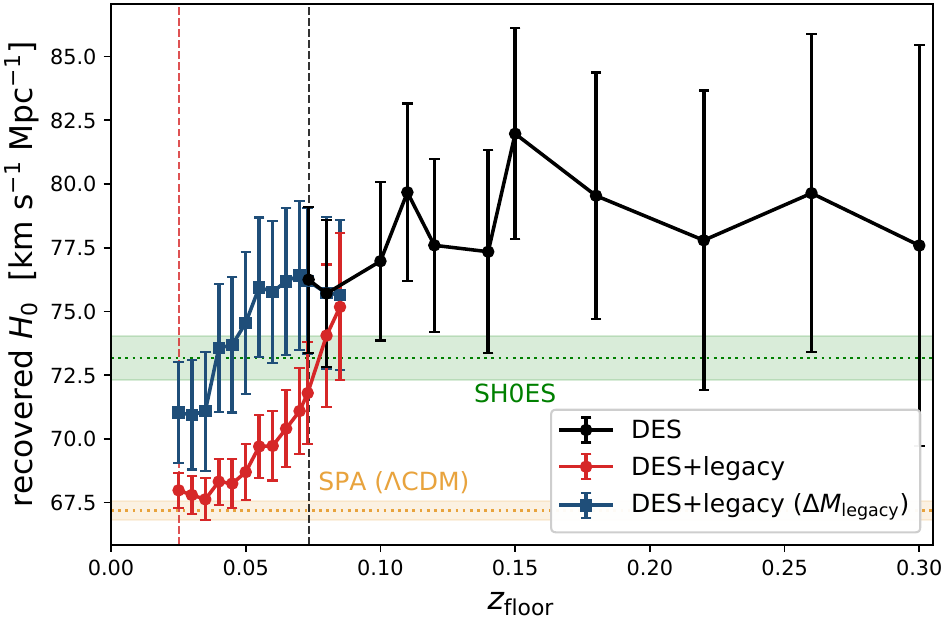}\\
    \includegraphics[width=1.0\linewidth]{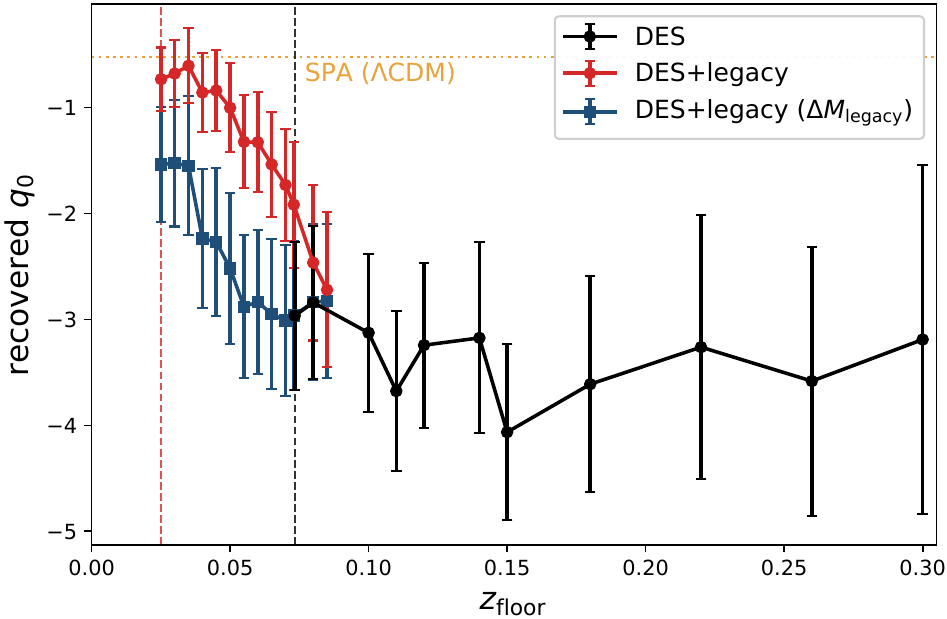}
    \caption{The solid lines show the recovered $H_0$ (\emph{top}) and $q_0$ (\emph{bottom}) in our $k_{\max} = 4$ cosmographic IDL constrained by DES-Dovekie SNIa + DESI~DR2 BAO + \emph{Planck} $r_{\rm d}$, shown as a function of the redshift floor $z_{\rm floor}$ below which SNIa are excluded and the sample re-fitted. In both panels, we present results with DES (black), DES + legacy (red), and DES + legacy with free recalibration of the legacy SNIa by $\Delta M_{\rm legacy}$ (blue; Eq.~\ref{eqn:legacy_offset}). The dashed vertical lines show the lower limit of each dataset without truncation. The upper limits of $z = 0.087$ for the legacy SNIa and $z = 1.14$ for DES are not shown here. At redshift floors which reject all legacy SNIa, we show results only for DES. In the top panel, the horizontal dotted lines with $1\sigma$ error bands show $H_0^\mathrm{SH0ES}$ \citep{Breuval_2024} and $H_0^\mathrm{SPA}$ \citep{SPT_2026}, as indicated. In the bottom panel, the horizontal dotted orange line shows $q_0^{\rm SPA}$.}
    \label{fig:floorsweep-real-new}
\end{figure}

The top panel of Figure~\ref{fig:floorsweep-real-new} shows how the predicted $H_0$ depends on $z_\mathrm{floor}$. The black line shows no obvious trend with only DES SNIa -- all results are $\approx 2.5$--$3.5\sigma$ above $H_0^\mathrm{SPA}$ until we get to high $z_\mathrm{floor}$, when the error bars become large due to the small number of SNIa. The predicted $H_0$ is always above $H_0^\mathrm{SH0ES}$ \citep{Breuval_2024}, but the level of tension is always $<2.1\sigma$ for $z_{\rm floor} = 0.073 - 0.3$.

The bottom panel of Figure~\ref{fig:floorsweep-real-new} shows that the high $H_0$ is associated with a very low deceleration parameter of $q_0 \approx -3$. This is expected since the expansion history needs to be similar to that in the \emph{Planck} cosmology at high redshift, so we can get elevated $H_0$ only with extra acceleration at late times. Similarly to the results for $H_0$, the black line shows no obvious trend in $q_0$ with $z_\mathrm{floor}$ when considering SNIa from only the homogeneous DES catalogue.

These results change substantially upon including the legacy SNIa at low redshift, as shown by the red curve in each panel of Figure~\ref{fig:floorsweep-real-new}. As we start to include legacy SNIa by decreasing $z_\mathrm{floor}$, the predicted $H_0$ drops sharply until it becomes consistent with $H_0^\mathrm{SPA}$. This behaviour is mirrored for $q_0$, which is far below the SPA value of $\tfrac{3}{2}\Omega_{\rm m} - 1 = -0.52$ at high $z_\mathrm{floor}$, but rises sharply towards and ultimately becomes consistent with it as $z_\mathrm{floor} \to 0$. Adding the legacy SNIa considerably reduces the formal uncertainties, so the predicted $H_0$ at the lowest $z_\mathrm{floor}$ values lies well below SH0ES and other local $H_0$ anchors. This tension is clearly driven by the legacy SNIa. We note that when we set $z_\mathrm{floor} = 0.073$ in our DES + legacy analysis to include all DES SNIa while minimising the number of legacy SNIa, we end up including 10 legacy SNIa. Adding these 10 legacy SNIa decreases $H_0$ by $4.45\kmsMpc$, which is $1.6\times$ the uncertainty prior to adding them. This suggests a tension between DES and the legacy SNIa where the two samples meet, although only 2 DES SNe lie in this overlap region, so the legacy SNIa are compared with the DES relation extrapolated slightly below its floor rather than with DES SNe at the same redshift.

Such large parameter shifts are not expected on the basis of the stable DES-only results. This is apparent from repeating our redshift-floor sweep test on the $10^4$ mock datasets presented in Appendix~\ref{app:floor-sweep-mock}, which we generate assuming correctness of the published BAO and SNIa covariance matrices and analyse similarly to the real data. Despite the scatter between the $H_0(z_\mathrm{floor})$ curves and a gradual upwards drift due to our cosmographic expansion not being exact, it is difficult to find a mock realisation of $\Lambda$CDM where the behaviour resembles that of the real data (Figure~\ref{fig:floorsweep-mock}). The sharp feature in the DES + legacy result for $H_0(z_\mathrm{floor})$ at precisely the redshift where the catalogue transitions from legacy SNIa to DES suggests a systematic affecting the legacy SNIa rather than a genuine transition at low redshift.

\subsection{Systematics with the legacy SNIa}
\label{sec:systematics-results}

Adding the legacy SNIa to the homogeneous DES sample shifts the inferred cosmological parameters in a manner that would not be expected from the latter (Figure~\ref{fig:h0_q0}). This might be due to systematics with the legacy SNIa. We test this possibility by adding an offset parameter $\Delta M_\mathrm{legacy} \sim \mathcal{U} \left(-1, 1 \right)$~mag to the absolute magnitude of the legacy SNIa, which therefore becomes as follows:
\begin{eqnarray}
    M_\mathrm{legacy} ~=~ \Mdes + \Delta M_\mathrm{legacy} \, .
    \label{eqn:legacy_offset}
\end{eqnarray}
A higher $\Delta M_\mathrm{legacy}$ corresponds to intrinsically fainter legacy SNIa, which for fixed apparent magnitude implies that they must be closer to us. Since the redshifts are spectroscopically confirmed, this would lead to a higher $H_0$.

\begin{figure}
  \centering
  \includegraphics[width=\columnwidth]{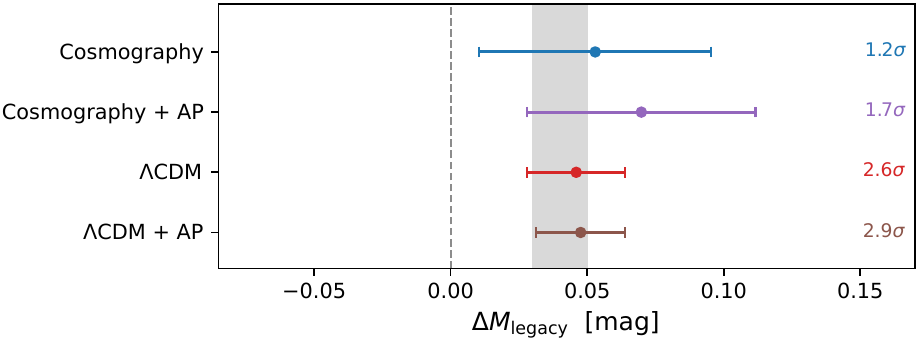}
  \caption{The relative offset $\Delta M_\mathrm{legacy}$ (Eq.~\ref{eqn:legacy_offset}) of the legacy SNIa from the DES-Dovekie relation extrapolated to $z \to 0$, for the four analyses described in Section~\ref{sec:systematics-legacy}. The vertical grey band shows the $\approx 0.04$~mag offset independently reported by \citet{Efstathiou_2025}.}
  \label{fig:offset_result}
\end{figure}

Our results for $\Delta M_\mathrm{legacy}$ are shown in Figure~\ref{fig:offset_result} for the four considered cases discussed in Section~\ref{sec:systematics-legacy}, namely cosmography, cosmography + AP, $\Lambda$CDM, and $\Lambda$CDM + AP. We consistently infer an offset of $\Delta M_\mathrm{legacy} = +0.05$ to $+0.07$~mag, with the detection significance ranging from $1.2\sigma$ for the more flexible cosmographic expansion, to $2.9\sigma$ for the less flexible $\Lambda$CDM parametrisation with the AP consistency requirement. In our baseline analysis where we use a cosmographic IDL and infer $\Delta M_\mathrm{legacy}$ alongside the other parameters, we find that $\Delta M_\mathrm{legacy} = +0.065 \pm 0.039$~mag, a $1.65\sigma$ offset (cf. Figure~\ref{fig:floorsweep-real-new}).

\begin{figure}
  \centering
  \includegraphics[width=\columnwidth]{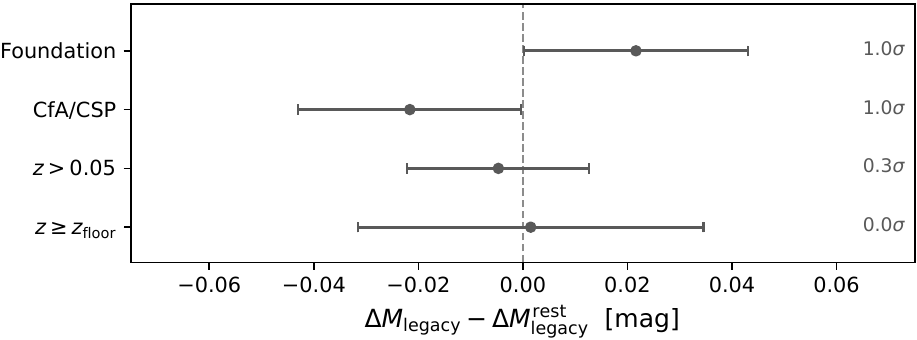}
  \caption{Test of whether the legacy offset depends on which subset of the legacy SNIa we consider, based on splitting the legacy SNIa into two subsamples according to the distinguishing feature indicated at the left (e.g., the first row splits off the Foundation SNIa from the remaining legacy SNIa). The values show $\Delta M_\mathrm{legacy} - \Delta M_\mathrm{legacy}^\mathrm{rest}$, the inferred offset for the subsample of interest and the rest of the legacy SNIa, respectively. We jointly infer both offset parameters. Note this does not test whether the legacy SNIa are offset from the main DES sample (which is shown in Figure~\ref{fig:offset_result}), but rather whether the offset depends on subsample within the legacy set.}
  \label{fig:offset_split}
\end{figure}

To gain further insight into how a calibration offset might have arisen, we split the legacy SNIa into two subsamples and infer a separate $\Delta M_\mathrm{legacy}$ for each. Figure~\ref{fig:offset_split} shows $\Delta M_\mathrm{legacy} - \Delta M_\mathrm{legacy}^\mathrm{rest}$ for a few possible splits, where $\Delta M_\mathrm{legacy}^\mathrm{rest}$ is the offset that we obtain for the remaining legacy SNIa. The legacy SNIa largely come from Foundation or from CfA. The former have a positive $\Delta M_\mathrm{legacy} - \Delta M_\mathrm{legacy}^\mathrm{rest}$, which is mirrored by an almost equal and opposite value for the CfA SNIa because `rest' then largely means Foundation, but the difference between them is significant at only $1.0\sigma$.

We also check for any redshift-dependent trend in $\Delta M_\mathrm{legacy}$ by splitting the legacy SNIa in redshift. A trend here might arise due to the greater importance of peculiar velocity at low redshift, while survey incompleteness may be an issue towards the high-redshift end of the $z = 0.025$--0.087 range covered by the legacy SNIa. We therefore split the legacy SNIa into subsamples below and above a redshift threshold of 0.05 or 0.073, the lowest redshift SNIa in the main DES sample. Figure~\ref{fig:offset_split} shows that in both cases, $\Delta M_\mathrm{legacy} - \Delta M_\mathrm{legacy}^\mathrm{rest}$ is very consistent with 0. We therefore conclude that the difference between the main DES sample and the legacy SNIa is not driven by any particular redshift bin.

\begin{figure}
  \centering
  \includegraphics[width=\columnwidth]{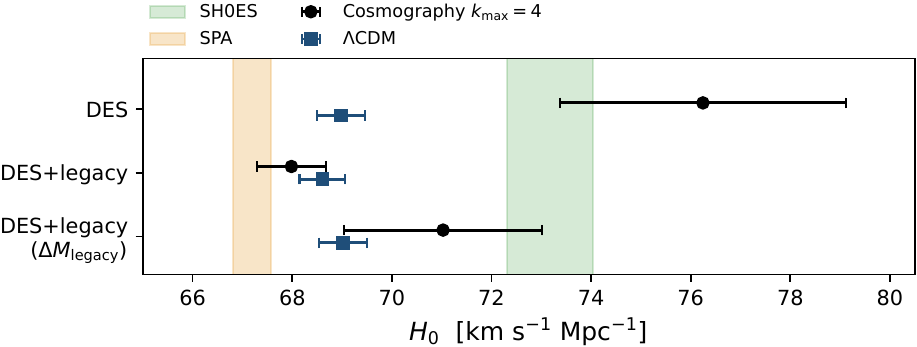}
  \includegraphics[width=\columnwidth]{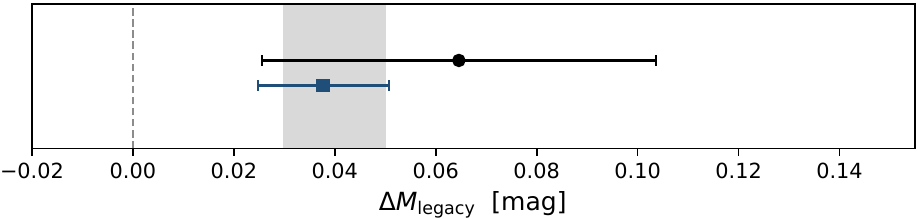}
  \caption{\emph{Top}: $H_0$ inferred from the IDL. Circles show results from our cosmographic analysis with $k_{\max} = 4$ (Eq.~\ref{eqn:cosmographic-expansion}). Squares show results using flat $\Lambda$CDM (Eq.~\ref{eqn:lcdm-expansion}). The vertical shaded bands show $H_0^\mathrm{SPA}$ \citep{SPT_2026} and $H_0^\mathrm{SH0ES}$ \citep{Breuval_2024}, as indicated in the legend. The top row excludes legacy SNIa, while the middle row includes them without correction ($\Delta M_\mathrm{legacy} = 0$). The bottom row includes the legacy SNIa allowing free $\Delta M_\mathrm{legacy}$. \emph{Bottom}: Our inferred $\Delta M_\mathrm{legacy}$ when we sample it jointly with $r_{\rm d}$ and $\Mdes$, assuming our cosmographic IDL. The vertical grey band shows the range reported by \citet{Efstathiou_2025}. Both panels adopt the \citetalias{Planck_2020} prior on $r_{\rm d}$.}
  \label{fig:legacy-h0}
\end{figure}

The legacy SNIa calibration offset is directly related to $H_0$ because intrinsically fainter SNIa must be closer to have the same apparent magnitude, leading to a higher $H_0$. We explore what impact this might have on the IDL using the approach described in Section~\ref{sec:systematics-legacy}. The results are presented in Figure~\ref{fig:legacy-h0} and summarised in Table~\ref{tab:legacy-h0}. For our cosmographic expansion, allowing $\Delta M_\mathrm{legacy} \neq 0$ increases $H_0$ from $67.98 \pm 0.69\kmsMpc$ to $71.03 \pm 1.99\kmsMpc$, which is now only $0.99\sigma$ below SH0ES. Allowing the Foundation and CfA/CSP legacy SNIa to have different offsets leaves this result almost unchanged, with $H_0 = 70.91 \pm 2.00 \kmsMpc$. The full IDL thus gains little from allowing separate offset parameters for Foundation and CfA/CSP.

\begin{table}
    \centering
    \begin{tabular}{lcc}
        \hline
        & \multicolumn{2}{c}{$H_0$ [km s$^{-1}$ Mpc$^{-1}$] using ...} \\
        Case & cosmography & $\Lambda$CDM \\ \hline
        Legacy excluded (DES only) & $76.25 \pm 2.87$ & $68.98 \pm 0.48$ \\
        Legacy included uncorrected & $67.98 \pm 0.69$ & $68.60 \pm 0.46$ \\ \hline
        Legacy zero-point sampled & $71.03 \pm 1.99$ & $69.02 \pm 0.48$ \\
        \multicolumn{1}{c}{$\Delta M_\mathrm{legacy}$ [mag]} & $+0.065 \pm 0.039$ & $+0.038 \pm 0.013$ \\ \hline
    \end{tabular}
    \caption{Inferred $H_0$ in our baseline analysis (SNIa + DESI~DR2 BAO + \citetalias{Planck_2020} $r_{\rm d}$) with the legacy SNIa excluded, included without correction, or included after correction by $\Delta M_\mathrm{legacy}$ to align them with the DES relation extrapolated to $z \to 0$. The final row shows $\Delta M_\mathrm{legacy}$.}
    \label{tab:legacy-h0}
\end{table}

The preference for $H_0 > H_0^\mathrm{SPA}$ is still discernible in flat $\Lambda$CDM, but it is far weaker due to the less flexible expansion history (Eq.~\ref{eqn:lcdm-expansion}). In this case, $H_0$ rises from $68.60 \pm 0.46\kmsMpc$ without $\Delta M_\mathrm{legacy}$ to $69.02 \pm 0.48 \kmsMpc$ with free $\Delta M_\mathrm{legacy}$ (Table~\ref{tab:legacy-h0}). The very small difference arises because fixing both the shape of the expansion history and its normalisation assuming $\Lambda$CDM necessarily implies a \emph{Planck}-like $H_0$ regardless of whether the legacy SNIa can be recalibrated. However, assuming $\Lambda$CDM at late times automatically excludes the possibility of solving the Hubble tension at late times. Interestingly, we still find that $H_0 > H_0^{\mathrm{SPA}}$ at $2.38\sigma$ ($2.98\sigma$) without (with) free $\Delta M_\mathrm{legacy}$. The latter result is similar to that without the legacy SNIa ($2.93\sigma$).

Our results suggest that a calibration offset with the legacy SNIa contributes $\approx 3\kmsMpc$ to the $H_0$ gap between local anchors and the IDL including legacy SNIa. Remarkably, the $H_0$ value recovered by the IDL using DES without legacy SNIa is very similar to the standard distance ladder result obtained by anchoring the DESI Fundamental Plane using the distance to the Coma Cluster \citep{Scolnic_2025} or with a broader range of anchors \citep{Carr_2026}. This agreement survives and in the latter case even improves upon averaging over the cosmographic order $k_{\max}$ using BMA, which gives $H_0 = 74.77 \pm 4.83 \kmsMpc$ (Table~\ref{tab:ordersweep-ls-A}). The Bayesian evidence assigns 20\% of the probability to $k_{\max} = 5$, which gives a lower and more uncertain prediction for $H_0$, as discussed next.

\subsection{Robustness tests}
\label{sec:robustness_tests}

The most striking feature of the results in Table~\ref{tab:baseline} is the substantial difference between the $H_0$ predictions of our cosmographic IDL constrained by DES and by DES + legacy, which we denote $H_0^\mathrm{DES}$ and $H_0^\mathrm{DES+legacy}$, respectively. The poorer low-redshift coverage with SNIa from DES alone forces our cosmographic expansion to extrapolate across the $z \la 0.2$ regime, though DES has SNIa down to $z = 0.073$. Since our cosmographic expansion does not exactly match the true $H(1-a)$ for any finite $k_{\max}$, it inevitably introduces some systematic error into the predicted $H_0$, even in the absence of measurement errors.

This can be investigated using mock data. The size of the effect depends on the unknown true expansion history; for illustration we quantify it in the flat $\Lambda$CDM cosmology that best fits the CMB \citep{SPT_2026}. We set up $10^4$ mock datasets assuming this SPA cosmology, with the mock SNIa having the same redshift distribution and magnitude uncertainties as the real SNIa in DES or DES + legacy (further details are provided in Appendix~\ref{app:floor-sweep-mock}). The difference $\Delta H_0 \equiv H_0^\mathrm{DES} - H_0^\mathrm{DES+legacy}$ with $k_{\max} = 4$ is $8.27\kmsMpc$ using our posterior means. However, we fit each mock by least squares to reduce computational costs, so a like-for-like comparison requires the least-squares best fit to the real data. This gives $H_0^\mathrm{DES} = 76.70\kmsMpc$ and $H_0^\mathrm{DES+legacy} = 68.00\kmsMpc$, widening the gap to $\Delta H_0 = 8.70\kmsMpc$. Only 0.85\% of the SPA mocks exceed this $\Delta H_0$ (Figure~\ref{fig:dh0-hist}). The observed $\Delta H_0$ is therefore unlikely to arise from random fluctuations in a $\Lambda$CDM cosmology coupled with extrapolation error due to the limited DES coverage at $z \la 0.2$.

However, our mocks suggest that these issues can provide a partial explanation given their tendency to find $\Delta H_0 > 0$. This bias is related to a generic overestimation of $H_0$ by $\approx 1\sigma$, even if the DES coverage at low redshift is further limited by a redshift floor (Figure~\ref{fig:floorsweep-mock}). As shown by our noise-free mock, the bias is a truncation error arising from differences between the assumed $\Lambda$CDM expansion history and our fourth-order cosmographic expansion. We do not know how accurately this can represent the real cosmological model, especially as a late-time solution to the Hubble tension must by definition differ from $\Lambda$CDM. Even so, our mocks quantify the effect in the SPA cosmology. The scatter of $H_0^\mathrm{DES}$ across mocks ($2.86\kmsMpc$) matches our stated uncertainty ($2.87\kmsMpc$), so the error bar captures the statistical uncertainty. However, the mean $H_0$ exceeds the input by $2.76\kmsMpc$, so the $\pm 1\sigma$ ($\pm 2\sigma$) interval contains the true value in only 49\% (85\%) of the mocks. Calibrated on the mocks, the least-squares $H_0^\mathrm{DES} = 76.70\kmsMpc$ of the real data is reached in 84 of the $10^4$ realisations (0.84\%), so the tension with SPA is weaker than the $3.13\sigma$ implied by the formal uncertainty. The least-squares $q_0 = -3.10$ is less unusual on its own, with a lower value in 3.30\% of the realisations. However, $H_0$ and $q_0$ are strongly anti-correlated across the mocks (correlation coefficient $-0.96$), so 83 of the 84 realisations with $H_0 \geq 76.70\kmsMpc$ also have $q_0 \leq -3.10$. Both thresholds are therefore crossed in 0.83\% of the mocks. If we subtract the bias, $H_0^\mathrm{DES}$ becomes $73.49 \pm 2.87\kmsMpc$. This is $2.18\sigma$ above SPA and in excellent agreement with SH0ES ($0.11\sigma$). We note that this correction assumes the SPA cosmology, so it illustrates the size of the bias rather than correcting for it in general.

\begin{figure}
    \centering
    \includegraphics[width=1.\linewidth]{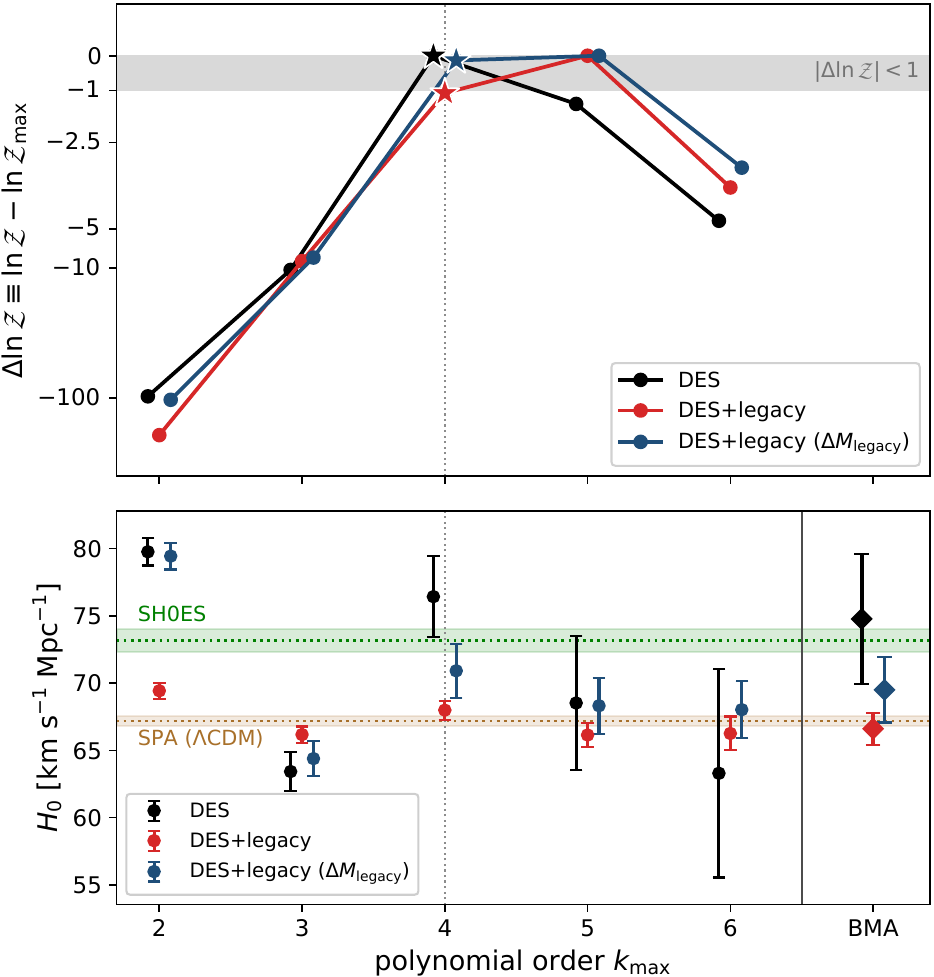}
    \caption{Polynomial order sweep for our cosmographic IDL in configuration~A using DES (black), DES + legacy (red), and DES + legacy with the legacy zero point $\Delta M_\mathrm{legacy}$ free (blue). \emph{Top}: Bayesian evidence relative to the most probable order of each sample, $\Delta \ln Z \equiv \ln Z - \ln Z_{\max}$. The grey band marks $\lvert \Delta \ln Z \rvert < 1$, inside which two orders are not distinguished \citep{Trotta_2008}. Stars and the vertical dotted line mark our baseline choice of $k_{\max} = 4$. \emph{Bottom}: The posterior on $H_0$ at each order, with the Bayesian model average over $k_{\max} = 2$--6 shown as diamonds in the column labelled ``BMA'' (Appendix~\ref{app:blend}). The horizontal green and orange bands show $H_0^\mathrm{SH0ES}$ \citep{Breuval_2024} and $H_0^\mathrm{SPA}$ \citep{SPT_2026}, respectively.}
    \label{fig:polynomial-order}
\end{figure}

The truncation error can be reduced with a higher-order expansion, but this makes the model more flexible, which inflates the uncertainties and can lower the Bayesian evidence. We therefore test how our results depend on the order by repeating the analysis for $k_{\max} = 2$--6. The results are shown in Figure~\ref{fig:polynomial-order}, with further details in Appendix~\ref{app:ordersweep}. Although $H_0$ changes significantly between orders, $k_{\max} = 4$ is the lowest order that provides an adequate fit to the data. Compared to this order, $k_{\max} = 2$ and 3 raise the DES $\chi^2_{\min}$ by 210 and 30, respectively, because they cannot describe the shape of the Hubble diagram across the DES redshift range. With $k_{\max} = 5$, we instead obtain $H_0 = 68.52 \pm 4.99 \kmsMpc$ for DES and $H_0 = 66.14 \pm 0.90 \kmsMpc$ for DES + legacy. This means that DES with $k_{\max} = 5$ or 6 gives results that are consistent with both SH0ES and SPA because of their larger uncertainties, whereas $k_{\max} = 4$ is consistent only with SH0ES. Even so, the higher orders remain consistent with the $k_{\max} = 4$ result within the $1.58\sigma$ confidence level (C.L.) for DES and the $1.61\sigma$~C.L. for DES + legacy.

The Bayesian evidence of each order (Eq.~\ref{eqn:order_weight}) quantifies how strongly the data support our baseline. The first result to highlight is that $k_{\max} = 4$ is the most probable order for DES, carrying 80\% of the posterior probability in configuration~A and 83\% in configuration~B. The evidence at this order exceeds that at $k_{\max} = 2$ and 3 by an overwhelming 97 and 10~nats, respectively. Beyond $k_{\max} = 4$, the Bayesian evidence falls again, by 1.4 at $k_{\max} = 5$ and by 4.8 at $k_{\max} = 6$. This is because the fifth coefficient improves $\chi^2_{\min}$ by only 2.4, which does not compensate for the extra prior volume. This is different for DES + legacy, where $k_{\max} = 5$ improves $\chi^2_{\min}$ by 9.3 and is preferred over $k_{\max} = 4$ by $\Delta \ln Z = 1.1$, but even this preference is only weak evidence on the Jeffreys scale \citep{Trotta_2008}. Once $\Delta M_\mathrm{legacy}$ is free, the two orders are indistinguishable, with $\Delta \ln Z = 0.1$. Thus, the data support our baseline choice of $k_{\max} = 4$ for DES and do not significantly disfavour it for DES + legacy, regardless of whether we allow recalibration of the legacy SNIa. This allows us to keep the two samples on an equal footing by fixing $k_{\max}$. However, the evidence assigns 20\% of the probability to the higher orders for DES, while for DES + legacy it slightly prefers $k_{\max} = 5$, so we cannot treat the fourth order as certain in either case.

One way to deal with the uncertain polynomial order $k_{\max}$ is to combine the results at different orders by averaging over the considered $k_{\max} = 2$--6 range using BMA (Eq.~\ref{eqn:bma}). We present this in Appendix~\ref{app:blend}. For DES alone, it gives $H_0 = 74.77 \pm 4.83 \kmsMpc$, compared with $76.43 \pm 3.01 \kmsMpc$ at $k_{\max} = 4$ from the same nested sampling. The central value moves little because $k_{\max} = 4$ carries 80\% of the probability. However, the uncertainty grows $1.6\times$ because the average includes the spread between orders, which in this case includes the $7.9\kmsMpc$ lower $H_0$ at $k_{\max} = 5$. As a result, the tension with SPA falls from $3.13\sigma$ to $1.56\sigma$. Once we include the legacy SNIa, $k_{\max} = 5$ carries 73\% of the probability and the BMA gives $H_0 = 66.61 \pm 1.18 \kmsMpc$. This is $1.37\kmsMpc$ below our $k_{\max} = 4$ result, but it remains $4.50\sigma$ below SH0ES, so our conclusions on the legacy SNIa remain unchanged. After recalibrating the legacy SNIa, the BMA gives $H_0 = 69.50 \pm 2.43 \kmsMpc$, which is only $1.43\sigma$ below SH0ES. Our fiducial results focus on the same polynomial order (4\textsuperscript{th}) for all SNIa configurations to help isolate the effect of including the legacy sample without simultaneously changing the flexibility of the reconstruction, as would effectively occur if we had used BMA due to the different weights it would inevitably assign to the different orders.

We do not repeat every test in this paper with the BMA because it does not change our conclusions on the legacy SNIa. The BMA fit quality is the same as at $k_{\max} = 4$, with $\chi^2/{\rm dof} = 0.920$ for DES and 0.938 for DES + legacy (cf. Table~\ref{tab:baseline}). $\Mdes$ is affected only by 2~mmag, so the tension with the SH0ES $M_B$ is unchanged. Moreover, the redshift-floor sweep of Section~\ref{sec:z_floor_result} behaves in the same way with the BMA. To show this, we fit each order by least squares and approximate its evidence with the Laplace approximation, which reproduces the nested-sampling BMA $H_0$ to within $0.1\sigma$ in all three of our configuration~A analyses. With DES alone, the averaged $H_0(z_\mathrm{floor})$ then shows no trend, staying within $72.7$--$80.7\kmsMpc$ for every $z_\mathrm{floor} \leq 0.3$. In contrast, the 10 legacy SNIa at $z \geq 0.073$ lower the predicted $H_0$ by $6.52\kmsMpc$, which is $1.4\times$ its uncertainty prior to adding them. The main effect of the BMA is thus to widen the DES-only uncertainty, which reduces the tension with SPA. We note that the BMA is very computationally intensive as it needs the posterior at every order, which is affordable for the analyses in Appendix~\ref{app:ordersweep} but much more costly for the $\Delta H_0$ test and the redshift-floor sweep on our $\Lambda$CDM mock datasets (Appendix~\ref{app:floor-sweep-mock}), the AP test (Appendix~\ref{app:AP}), and the offset of the legacy SNIa relative to the DES relation (Section~\ref{sec:systematics-results}). Repeating each of these at five different orders is thus not practical. Since $k_{\max} = 4$ is the most probable order for DES and carries 80\% of the total weight in the BMA, we perform these tests only at $k_{\max} = 4$.

It is also important to consider how our results are affected by the prior, which could matter for the DES-alone case because DES does not contain SNIa at $z < 0.073$. Our prior bounds only the expansion history itself (Eq.~\ref{eqn:envelope}), with no posterior approaching the prior limits (Section~\ref{sec:setup}). In particular, the best-fitting expansion history for DES stays at least 0.92 in $E$ from both limits, so prior truncation effects should be negligible. A least-squares fit in the $C_k$, which approximates $\chi^2$ as quadratic near its minimum and needs no prior ranges, gives $H_0 = 76.70 \pm 2.97 \kmsMpc$ for DES and $68.00 \pm 0.71 \kmsMpc$ for DES + legacy, fully consistent with our NUTS inference. Moreover, the marginal prior on each $C_k$ is $24$--$73\times$ wider than the corresponding posterior, while other priors give virtually the same $H_0$ (Appendix~\ref{app:prior}).

Finally, the AP test checks the shape of the expansion history independently of $r_{\rm d}$ (Section~\ref{sec:AP_consistency} and Appendix~\ref{app:AP}). We find consistency for both DES and DES + legacy, with the $p$-value (likelihood of a higher $\chi^2$) being 9.7\% and 10.5\%, respectively. There is a mild $\approx 2\sigma$ deficit in both cases at $z = 0.510$, while at $z = 0.934$, there is a mild excess of $2.0\sigma$ for DES and $1.8\sigma$ for DES + legacy. However, there is excellent agreement at $z = 0.706$. This justifies our approach of combining SNIa and BAO datasets assuming the Etherington relation \citep[cf.][]{Das_2026}. We note that independently of this relation, the DESI~DR2 BAO measurements at $z = 0.706$ are in some tension with previous BAO results at a similar redshift \citep[section~3.6 of][]{Banik_2026_z0}.

\subsection{Bayesian model comparison of late-time solutions}
\label{sec:Model_comparison}

\renewcommand{\arraystretch}{1.2}
\newlength{\ExcSNIaAdjustment}
\setlength{\ExcSNIaAdjustment}{-0.5ex}
\begin{table*}
    \centering
    \setlength{\tabcolsep}{4pt}
    \begin{tabular}{llcccccccc}
        \hline
        Sample & Anchor & Model & $\ln B_{10}$ & $\Omega_{\rm m}$ & $w_0/A$ & $w_a/z_1$ & $H_0$ & $H_0^{\rm win}$ & $\Delta M_\mathrm{legacy}$ \\
        \hline
        DES & \multirow{2}{*}{None} & \texttt{exp} & $-1.08 \pm 0.19$ & $0.310 \pm 0.008$ & $1.26 \pm 2.40$ & Unc. & $68.83 \pm 2.02$ & $68.64 \pm 1.00$ & --- \\
        &  & CPL & $-1.79 \pm 0.20$ & $0.311 \pm 0.009$ & $-0.845 \pm 0.098$ & $-0.639 \pm 0.315$ & $67.71 \pm 0.93$ & $67.98 \pm 0.78$ & --- \\
        \noalign{\smallskip}
        & \multirow{2}{*}{\raisebox{\ExcSNIaAdjustment}{\shortstack{Excl.\\SNIa}}} & \texttt{exp} & $+1.54 \pm 0.19$ & $0.308^{+0.006}_{-0.005}$ & $4.14^{+3.47}_{-1.81}$ & $<0.086$ & $70.00^{+4.78}_{-0.98}$ & $70.23 \pm 0.99$ & --- \\
        & & CPL & $-1.19 \pm 0.20$ & $0.299^{+0.008}_{-0.006}$ & $-0.943 \pm 0.085$ & $-0.349^{+0.229}_{-0.332}$ & $68.96^{+0.69}_{-0.94}$ & $68.96 \pm 0.68$ & --- \\
        \noalign{\smallskip}
        & \multirow{2}{*}{SH0ES} & \texttt{exp} & $+10.61 \pm 0.20$ & $0.310 \pm 0.005$ & $9.29^{+0.63}_{-2.09}$ & $0.031^{+0.017}_{-0.011}$ & $76.70^{+0.83}_{-2.10}$ & $71.93 \pm 0.67$ & --- \\
        &  & CPL & $+4.06 \pm 0.20$ & $0.287 \pm 0.006$ & $-1.053 \pm 0.076$ & $-0.174 \pm 0.258$ & $70.58^{+0.54}_{-0.87}$ & $70.29 \pm 0.58$ & --- \\
        \noalign{\smallskip}
        \hline
        DES  & \multirow{2}{*}{None} & \texttt{exp} & $+0.07 \pm 0.19$ & $0.302 \pm 0.005$ & $-2.17 \pm 1.65$ & $<0.194$ & $66.87^{+0.89}_{-0.98}$ & $67.18 \pm 0.46$ & --- \\
        + legacy &  & CPL & $+1.70 \pm 0.20$ & $0.314 \pm 0.005$ & $-0.802 \pm 0.056$ & $-0.770 \pm 0.220$ & $67.37 \pm 0.55$ & $67.72 \pm 0.49$ & --- \\
        \noalign{\smallskip}
        & \multirow{2}{*}{\raisebox{\ExcSNIaAdjustment}{\shortstack{Excl.\\SNIa}}} & \texttt{exp} & $-1.67 \pm 0.19$ & $0.303 \pm 0.006$ & $-0.61 \pm 1.26$ & Unc. & $67.66 \pm 1.01$ & $67.76 \pm 0.46$ & --- \\
        & & CPL & $+0.87 \pm 0.20$ & $0.310^{+0.004}_{-0.006}$ & $-0.829 \pm 0.056$ & $-0.742 \pm 0.223$ & $67.92 \pm 0.53$ & $68.22 \pm 0.48$ & --- \\
        \noalign{\smallskip}
        & \multirow{2}{*}{SH0ES} & \texttt{exp} & $-1.78 \pm 0.19$ & $0.304^{+0.010}_{-0.005}$ & $1.22 \pm 1.28$ & $>0.288$ & $68.84 \pm 0.70$ & $68.74 \pm 0.48$ & --- \\
        &  & CPL & $+3.41 \pm 0.20$ & $0.301 \pm 0.005$ & $-0.858 \pm 0.055$ & $-0.766 \pm 0.224$ & $68.87 \pm 0.48$ & $69.10 \pm 0.43$ & --- \\
        \noalign{\smallskip}
        \hline
        DES & \multirow{2}{*}{None} & \texttt{exp} & $-1.22 \pm 0.19$ & $0.307^{+0.008}_{-0.007}$ & $1.35 \pm 1.99$ & Unc. & $68.86 \pm 1.56$ & $68.59 \pm 0.81$ & $0.045 \pm 0.022$ \\
        + legacy &  & CPL & $-1.39 \pm 0.20$ & $0.311 \pm 0.008$ & $-0.845 \pm 0.091$ & $-0.653 \pm 0.295$ & $67.78 \pm 0.85$ & $68.05 \pm 0.72$ & $0.013 \pm 0.021$ \\
        \noalign{\smallskip}
        ($\Delta M_\mathrm{legacy}$) & \multirow{2}{*}{\raisebox{\ExcSNIaAdjustment}{\shortstack{Excl.\\SNIa}}} & \texttt{exp} & $+0.54 \pm 0.19$ & $0.309^{+0.012}_{-0.007}$ & $3.47 \pm 1.86$ & Unc. & $69.75^{+1.30}_{-0.98}$ & $69.61 \pm 0.76$ & $0.068 \pm 0.023$ \\
        & & CPL & $-0.84 \pm 0.20$ & $0.300 \pm 0.007$ & $-0.945 \pm 0.085$ & $-0.416 \pm 0.277$ & $68.98 \pm 0.81$ & $69.06 \pm 0.68$ & $0.036 \pm 0.020$ \\
        \noalign{\smallskip}
        & \multirow{2}{*}{SH0ES} & \texttt{exp} & $+7.87 \pm 0.19$ & $0.311^{+0.012}_{-0.008}$ & $5.61 \pm 1.43$ & $<0.225$ & $71.24^{+2.17}_{-0.96}$ & $71.05 \pm 0.75$ & $0.104 \pm 0.028$ \\
        &  & CPL & $+4.60 \pm 0.20$ & $0.287 \pm 0.006$ & $-1.050 \pm 0.077$ & $-0.191 \pm 0.257$ & $70.44 \pm 0.70$ & $70.31 \pm 0.59$ & $0.062 \pm 0.019$ \\
        \noalign{\smallskip}
        \hline
    \end{tabular}
    \caption{Parameter constraints for the \texttt{exp} (Eq.~\ref{eqn:H_z_extra}) and CPL (Eq.~\ref{eq:CPL}) models and results of Bayesian model comparison against $\Lambda$CDM. A model is preferred over $\Lambda$CDM if its $\ln B_{10} > 0$. All relative $\ln B$ uncertainties include a systematic contribution of 0.13 from the $\Lambda$CDM analysis. We use the DES-Dovekie + DESI~DR2 BAO + CC + CMB + BBN constraint in all cases (Section~\ref{sec:evidence}). We also consider the SH0ES anchor \citep{Breuval_2024} or the three $H_0$ anchors that are independent of the SNIa absolute magnitude, imposed jointly and labelled ``Excl. SNIa''. We use three different cases for the DES-Dovekie sample: DES alone, DES + legacy, and DES + legacy with a magnitude offset of $\Delta M_\mathrm{legacy}$ on the legacy SNIa (Eq.~\ref{eqn:legacy_offset}). The $w_0/A$ column shows $w_0$ for CPL and $A$ for the \texttt{exp} model, while the $w_a/z_1$ column shows $w_a$ for CPL and $z_1$ for the \texttt{exp} model. Missing model parameters are marked with a dash, while ``Unc.'' indicates parameters that are largely unconstrained by the data, with a posterior spanning essentially the whole prior range (Table~\ref{tab:priors}). Uncertainties follow the convention of Table~\ref{tab:baseline}, with one-sided limits enclosing 68\% of the posterior. $H_0$, $H_0^{\rm win}$, and $A$ are given in $\kmsMpc$, while $\Delta M_\mathrm{legacy}$ is given in mag. $H_0^{\rm win}$ is the $H_0$ that SH0ES would infer if the Universe followed each posterior, i.e. the kinematic fit of Section~\ref{sec:H0_anchors} applied over the redshifts of the SH0ES Hubble-flow SNIa ($z = 0.023$--0.149), for which we quote the posterior mean and standard deviation. The triangle plots corresponding to these results are shown in Appendix~\ref{app:triangles}.}
    \label{tab:evidence}
\end{table*}
\renewcommand{\arraystretch}{1.0}

The Bayesian evidence of each considered model (\texttt{exp} and CPL) is shown in Table~\ref{tab:evidence} relative to the $\Lambda$CDM control, whose uncertainty contributes a systematic component of 0.13 to all the stated $\ln B_{10}$ uncertainties. Without relying on any local $H_0$ anchor,
$\lvert \ln B_{10} \rvert \leq 1.22$ for the \texttt{exp} model, implying at most weak evidence for or against it -- though it is generally disfavoured in this case.

The results for the CPL model are critically dependent on whether we include the legacy SNIa. Without them, CPL is disfavoured by 1.79 natural units (nats). Including them at face value leads to a 1.70~nat preference for CPL, but including them and allowing an arbitrary recalibration reverts this to a 1.39~nat preference against CPL. These results clarify that without local $H_0$ anchors, the Bayesian evidence for CPL largely comes from adding the legacy SNIa at face value, which we have argued elsewhere is most likely not appropriate \citep[cf.][]{Ormondroyd_2026}.

The predicted $H_0$ in the CPL model (without $H_0$ anchors) is always similar to $H_0^\mathrm{SPA}$, but this is true in the \texttt{exp} model only if we include the legacy SNIa at face value. If we neglect them altogether or add them with arbitrary recalibration, the predicted $H_0$ is almost $69\kmsMpc$, which given the uncertainties is marginally compatible with $H_0^\mathrm{SH0ES}$ (e.g. DES alone gives $H_0 = 68.83 \pm 2.02\kmsMpc$, which is within $2\sigma$ of SH0ES). It is therefore possible to combine local $H_0$ anchors with the other data only for the \texttt{exp} model and even then only if the legacy SNIa are excluded or recalibrated. We cannot add local $H_0$ anchors if we use the CPL model, regardless of how we treat the legacy SNIa. If we take these at face value, we cannot add local $H_0$ anchors for either model -- or indeed for any model with standard $r_{\rm d}$ (Figure~\ref{fig:dz_ratio}).

\begin{figure}
    \centering
    \includegraphics[width=1.0\linewidth]{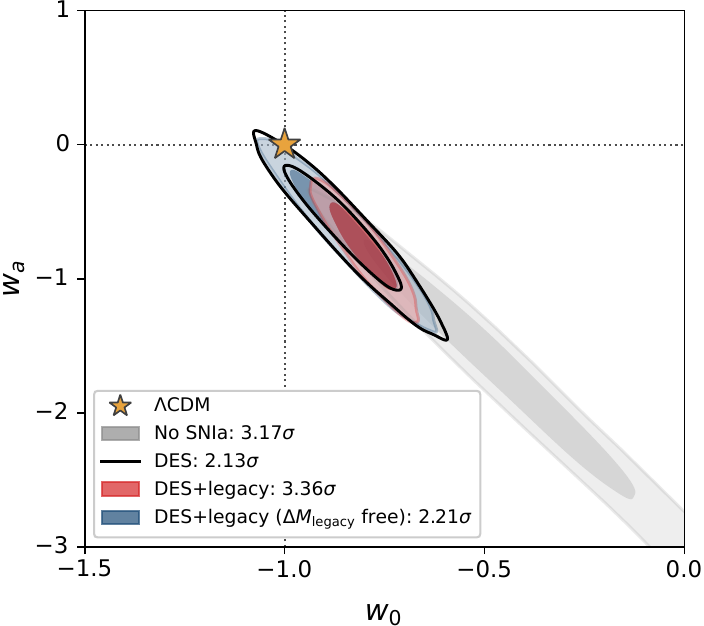}
    \caption{The contours show the $1\sigma$ and $2\sigma$ allowed regions of the CPL parameters ($w_0, w_a$) with SNIa only from DES (unfilled black) or with the legacy SNIa included, either taking them at face value (filled red) or allowing them to be recalibrated relative to the main DES sample (filled blue). The faint filled grey contours in the background show results without SNIa. The values in the legend indicate the tension based on the $\chi^2$ reduction once we free $w_0$ and $w_a$ from the $\Lambda$CDM prediction (orange star).}
    \label{fig:w0wa_contours}
\end{figure}

The CPL model was not proposed as a solution to the Hubble tension, but rather as a way to better fit available BAO and uncalibrated SNIa data consistently with the CMB. We find that when constraining the CPL model with DES + legacy, $(w_0, w_a)$ is in $3.36\sigma$ tension with the $\Lambda$CDM limit, which we obtain from the improvement in the fit when $w_0$ and $w_a$ are freed ($\Delta \chi^2 = 14.32$), converted to a Gaussian significance for two additional parameters. However, this $w_0$--$w_a$ tension gets reduced to $2.13\sigma$ upon removing the legacy SNIa, while allowing an arbitrary recalibration of them relative to DES leads to $2.21\sigma$ tension. Both cases are plausibly consistent with $\Lambda$CDM. The departure from $\Lambda$CDM expectations claimed by \citet{DESI_2025} is therefore reliant on the legacy SNIa \citep{Ormondroyd_2026}. Indeed, the actual BAO observables (angular scale and redshift depth) can be fit jointly with the CMB at an acceptable total $\chi^2$ if SNIa are neglected \citep{Banik_2026_z0}. On the other hand, their work found that $\Lambda$CDM faces a $3.07\sigma$ CMB--BAO tension in the $(H_0 r_{\rm d}, \Omega_{\rm m})$ parameter space without considering SNIa. This is similar to our result that the $w_0$--$w_a$ tension is $3.17\sigma$ without SNIa \citep[see also][]{Torres_2026}, though the parameters are quite uncertain in this case (Figure~\ref{fig:w0wa_contours}). Combined with our result in Figure~\ref{fig:dz_ratio}, this is suggestive of a growing anomaly between CMB-calibrated $\Lambda$CDM predictions and observations as we move towards lower redshift. However, CPL need not be the correct explanation for this anomaly. This model is actually one of the least sensitive to any genuine evolution of the dark energy density such as might be expected in theoretical models \citep{Colgain_2021}. Given the CPL preference for $w_0 > -1$ implies a decreasing dark energy density towards lower redshift, it is inevitable that the CPL model struggles to solve the Hubble tension \citep{Lee_2022_H0}, as discussed in the appendix to \citet{Colgain_2026}.

When adding the SNIa-independent $H_0$ anchors, the \texttt{exp} model is disfavoured compared to $\Lambda$CDM if we add the legacy SNIa at face value, but this dataset combination is not valid because the legacy SNIa cannot be reconciled with local $H_0$ measurements in any plausible model with standard $r_{\rm d}$. In the other cases, the non-SH0ES anchors on $H_0$ lead to a preference for the \texttt{exp} model, especially for DES alone (+1.54~nats). In this case, the inferred $H_0 = 70.00^{+4.78}_{-0.98}\kmsMpc$, which is seemingly consistent with a full solution to the Hubble tension. However, the excess expansion rate in this model decays on a scale $z_1 < 0.086$, so much of it lies below the Hubble-flow window of SH0ES. The $H_0$ that SH0ES would measure in this model is therefore only $H_0^{\rm win} = 70.23 \pm 0.99\kmsMpc$, which is $2.25\sigma$ below $H_0^{\mathrm{SH0ES}}$ (Table~\ref{tab:evidence}). This is substantially better than the $6.4\sigma$ tension with SPA, but it does not completely remove the Hubble tension. With DES + legacy allowing recalibration, the free $\Delta M_\mathrm{legacy}$ raises $H_0$ to $69.75^{+1.30}_{-0.98}\kmsMpc$, which is still short of a full solution as SH0ES would measure only $H_0^{\rm win} = 69.61 \pm 0.76\kmsMpc$ in this model, $3.10\sigma$ below $H_0^{\mathrm{SH0ES}}$. The DES-only case gives a low inferred $z_1 < 0.086$ at the $1\sigma$ C.L., which suggests that the solution is localised at quite low redshift \citep[as also found by][]{Najera_2026}. We discuss this further in Section~\ref{sec:logz1}.

Only when we add the SH0ES anchor on $H_0$ do we get a strong preference for the \texttt{exp} model, with $\ln B_{10} = +10.61$ for DES and +7.87 for DES + legacy with free $\Delta M_\mathrm{legacy}$. In these cases, the model comes close to reproducing the SH0ES measurement itself, with $H_0^{\rm win} = 71.93 \pm 0.67\kmsMpc$ for DES and $71.05 \pm 0.75\kmsMpc$ for DES + legacy with free $\Delta M_\mathrm{legacy}$, i.e. $1.14\sigma$ and $1.86\sigma$ below SH0ES. Their posterior mean $H_0$ is higher ($75.49 \pm 1.57$ and $72.30 \pm 1.58\kmsMpc$, respectively) because the excess expansion rate peaks at $z = 0$ and has largely decayed across the SH0ES window. We stress that the anchor is now part of the fit, so these values are not predictions. But even with the strong SH0ES anchor, the DES + legacy case (without $\Delta M_\mathrm{legacy}$) still disfavours the \texttt{exp} model relative to $\Lambda$CDM by 1.78~nats and predicts a low $H_0 = 68.84 \pm 0.70\kmsMpc$.

The CPL model is at most weakly favoured by the Bayesian evidence if we add the non-SH0ES local anchors on $H_0$, with the best case being for DES + legacy without recalibration (0.87~nat preference for CPL). Adding the SH0ES anchor leads to a clear preference for the CPL model,
which is 4.06~nats for DES alone. This is far weaker than the 10.61~nat preference for the \texttt{exp} model over $\Lambda$CDM with the same dataset combination. Moreover, the inferred $H_0$ in the CPL + SH0ES case barely exceeds $70\kmsMpc$ and has a small uncertainty, which still leads to considerable tension with SH0ES if we discard the legacy SNIa or try to rectify potential problems with them by allowing an arbitrary recalibration. The low $H_0$ problem is worsened if we add the legacy SNIa at face value, since then the inferred $H_0 = 68.87 \pm 0.48\kmsMpc$, far below $H_0^{\mathrm{SH0ES}}$. We attribute this to any deviation from $\Lambda$CDM in the CPL model declining too gradually with redshift -- the linear dependence of $w$ upon $a$ prevents a rapid convergence to $-1$. The CPL model also has a theoretical problem in that matching the local $H_0$ anchors requires $w < -1$ today and/or in the recent past. This implies a phantom fluid, violating the weak energy condition \citep{Sen_2008, Lewis_2025}.

\subsubsection{Log-uniform prior on the \texorpdfstring{\texttt{exp}}{exp} model \texorpdfstring{$z_1$}{z\textsubscript{1}}}
\label{sec:logz1}

The Hubble tension might be solved by a change to the background expansion history of a homogeneous universe. In this case, since it is not known a priori when deviations from $\Lambda$CDM would start to set in, a uniform prior on the \texttt{exp} model $z_1$ is appropriate. But if the Hubble tension is actually caused by a local underdensity or void \citep{Haslbauer_2020, Banik_2026_z0, Banik_2026_void}, then $z_1$ provides an indirect measure of the void size, since the gravitational impact of a void would persist even after the density has returned to the cosmic mean \citep{Banik_2025_BAO}. Even so, a smaller void can generally be expected to show up in our \texttt{exp} model analysis as a lower $z_1$. Because we do not know a priori how large a void we might be in, a scale-free prior is arguably a better choice. We therefore revisit the above \texttt{exp} model results using a uniform prior on $\ln z_1$, while keeping the limits unchanged.

We present the full results in Appendix~\ref{app:logz1}. Our overall conclusions regarding the \texttt{exp} model do not change much depending on whether we use a uniform prior on $z_1$ or on $\ln z_1$. A late-time solution with $A \approx 6\kmsMpc$ is preferred only when we impose a local $H_0$ anchor and do not consider the legacy SNIa, though there is a weak preference if we include them allowing recalibration. Without any local $H_0$ anchor, the data are consistent with a late-time solution to the Hubble tension, but do not prefer the high required $A$. Using a log-uniform prior on $z_1$ increases $\ln B_{10}$ in nearly all cases (typically by about 0.7) by increasing the prior weight on low values of $z_1$, which are preferred even for a uniform prior. As a result, the DES-only case with SNIa-independent anchors shifts from $\ln B_{10} = +1.54$ to $+2.53$, which is now a clear preference for the \texttt{exp} model. Even so, SH0ES would still measure only $H_0^{\rm win} = 70.21 \pm 0.97\kmsMpc$ in this case (Table~\ref{tab:logz1}).

With the strong SH0ES anchor and SNIa only from DES, we infer that $A \approx 8\kmsMpc$ and $z_1 \approx 0.03$ with either $z_1$ prior. However, we stress that our prior on $A$ bounds both values. Since most of the SH0ES Hubble-flow SNIa lie at $z > z_1$, the distances they measure respond mainly to the integrated excess $\int \Delta H \, \mathrm{d}z = A z_1$ rather than to $A$ and $z_1$ separately. The likelihood therefore has a ridge along which $A$ rises as $z_1$ falls, with $A z_1 = 0.29^{+0.12}_{-0.08}\kmsMpc$. The upper edge of our prior at $A = 10\kmsMpc$ truncates this ridge. Since 31\% of the posterior lies at $A > 9\kmsMpc$ (36\% with the log-uniform prior on $z_1$), a wider prior would raise $A$ and lower $z_1$. It would also change $\ln B_{10}$, both through the larger prior volume and through the additional likelihood along the ridge, which have competing effects.

\section{Discussion}
\label{sec:discussion}

\subsection{Reliability of the homogeneous DES IDL}
\label{sec:discussion_DES}

Our cosmographic IDL predicts an elevated $H_0 = 76.25 \pm 2.87\kmsMpc$ when SNIa are taken only from DES, a modern homogeneous survey with well-understood selection function \citep{DES_2016, DES_2024_SNe, Vincenzi_2024}. We treat this as our baseline result because the legacy SNIa normally added to the DES-Dovekie catalogue appear to have a systematic offset (Section~\ref{sec:systematics-results}), possibly for the reasons that we discuss in Section~\ref{sec:discussion_systematics_legacy}. If the legacy SNIa turn out to be reliable, discarding them would cause a loss of precision, but this should not change our overall conclusion. On the other hand, if there are indeed significant biases with the legacy SNIa, excluding them prevents a significant bias in the IDL prediction for $H_0$.

Restricting to SNIa in DES avoids issues with the legacy SNIa, but there can still be biases internally to DES. Since its SNIa cover $z = 0.073$--1.14, there are appreciable changes in the SNIa fluxes relative to the detection threshold and in the rest-frame wavelengths probed by the DECam filters, which become increasingly ultraviolet at high redshifts. Even so, we provide three reasons to be confident that the elevated $H_0$ prediction ($3.13\sigma$ above $H_0^{\rm SPA}$) is not driven by redshift-dependent biases:
\begin{enumerate}
    \item The shape of the DES SNIa distance--redshift relation agrees with DESI~DR2 BAO where they overlap, assuming the Etherington relation. The SNIa predict the BAO $F_\mathrm{AP}$ with $\chi^2 = 6.33$ for three redshifts ($P = 9.7\%$; Appendix~\ref{app:AP-result}). This test is independent of $r_{\rm d}$ and $\Mdes$, but it only reaches down to $z = 0.51$, the lowest redshift at which DESI~DR2 measures $F_\mathrm{AP}$;
    \item The high $H_0$ is not driven by the lowest redshift DES bin in Figure~\ref{fig:dz_ratio}. Raising the redshift floor from 0.073 to 0.3 leaves $H_0$ between 75.7 and $82.0\kmsMpc$ at every step (Figure~\ref{fig:floorsweep-real-new}). Even at $z_\mathrm{floor} = 0.3$, the predicted $H_0 = 77.6 \pm 7.9\kmsMpc$. The central value is practically identical to our baseline result, though the uncertainty is nearly tripled due to the loss of so much data at low redshifts. The high $H_0$ remains apparent at $z_\mathrm{floor} = 0.3$ because the distance deficit relative to the SPA cosmology does not appear suddenly at low redshift. The binned DES distances are already $\approx 2\%$ below the SPA prediction at $z \approx 0.2$--0.5. The deficit then grows to $3.8 \pm 1.2\%$ in the lowest bin at $z \approx 0.14$, which is thus a reasonable result given higher redshift data. The BAO show a similar deficit where they overlap with the SNIa: the lowest redshift DESI~DR2 measurement at $z = 0.295$ is $1.6 \pm 0.9\%$ below the SPA prediction. However, DESI~BAO measurements do not yet reach the redshifts at which the deficit is largest;
    \item Residual selection effects would act against the signal because at the low redshift end of DES, it would be able to detect fainter SNIa that would be missed further out. These faint SNIa would lead to an overestimated distance, since $\Mdes$ is set mainly by more distant SNIa that overlap with BAO measurements at $z > 0.3$. However, SNIa at the low redshift end of DES appear closer than predicted by the SPA cosmology.
\end{enumerate}

\subsection{Cosmological implications}
\label{sec:discussion_implications}

The first result to highlight is that the IDL verdict on late-time solutions to the Hubble tension depends on how the low-redshift legacy SNIa are treated. Using only the homogeneous DES sample, our baseline predicts $H_0^\mathrm{DES} = 76.25 \pm 2.87\kmsMpc$, which agrees with SH0ES \citep{Breuval_2024} but lies $3.13\sigma$ above $H_0^\mathrm{SPA}$ \citep{SPT_2026}. Adding the 197 legacy SNIa gives instead $H_0^\mathrm{DES+legacy} = 67.98 \pm 0.69\kmsMpc$, which lies $4.69\sigma$ below SH0ES. We stress that the first of these tensions is not the usual one between the local and early Universe. This is because $H_0^\mathrm{DES}$ inherits its absolute scale from the CMB through the \citetalias{Planck_2020} prior on $r_{\rm d}$, so its tension with $H_0^\mathrm{SPA}$ compares two CMB-calibrated predictions that differ only in whether $\Lambda$CDM holds after recombination. It is therefore a measure of departures from the $\Lambda$CDM expansion history at late times.

However, the DES-only result is less firm than its formal uncertainty suggests. DES has very few SNIa at $z \la 0.2$, so $H_0^\mathrm{DES}$ relies on extrapolating our cosmographic expansion below the DES floor of $z = 0.073$. This extrapolation depends on the polynomial order. Averaging over $k_{\max} = 2$--6 using BMA predicts that $H_0 = 74.77 \pm 4.83\kmsMpc$, which reduces the tension with SPA to $1.56\sigma$ (Section~\ref{sec:robustness_tests}). Moreover, our mocks show that the fourth-order expansion overestimates $H_0$ by $\approx 1\sigma$ in the SPA cosmology, so the tension falls to around $2\sigma$ once calibrated with them. Thus, DES alone favours a relatively high $H_0$ and is consistent with a late-time solution to the Hubble tension, but it does not determine $H_0$ precisely enough to require one. Even so, this indicates that the answer to the paper title is ``no''.

The next important result is that the $8.27\kmsMpc$ gap between $H_0^\mathrm{DES}$ and $H_0^\mathrm{DES+legacy}$ is unlikely to arise from sample variance and extrapolation error alone, since only 0.85\% of our $10^4$ $\Lambda$CDM mocks show a larger gap in a like-for-like comparison. Furthermore, the drop in $H_0$ occurs sharply at $z \approx 0.08$, where the catalogue switches from DES to legacy SNIa. The DES-only results and the mocks vary smoothly with $z_\mathrm{floor}$ (Figures~\ref{fig:floorsweep-real-new} and \ref{fig:floorsweep-mock}). We point out that the mocks assume an SPA cosmology and $k_{\max} = 4$, so they quantify how unusual the gap would be in $\Lambda$CDM -- results may differ in other expansion histories. Even so, the location of the feature points towards systematics with the legacy SNIa rather than a genuine transition in the expansion history at low redshift.

The Bayesian model comparison tells the same story in terms of particular models (Table~\ref{tab:evidence}). Without local $H_0$ anchors, the data neither favour nor disfavour the \texttt{exp} model ($\lvert \ln B_{10} \rvert \leq 1.22$), so the IDL alone cannot decide whether a late-time solution is required. The SNIa-independent anchors give a preference for the \texttt{exp} model with DES ($\ln B_{10} = +1.54$). This becomes strong with the SH0ES anchor ($+10.61$), whereas DES + legacy disfavours the model with either anchor. We stress that a high $H_0$ does not necessarily imply a solution to the Hubble tension. For DES with the SNIa-independent anchors, SH0ES would only measure $H_0^{\rm win} = 70.23 \pm 0.99 \kmsMpc$ in this model, which is $2.25\sigma$ below the published value, since most of the excess expansion rate is below the SH0ES Hubble-flow window. The \texttt{exp} model therefore reduces the Hubble tension substantially, but does not remove it. In contrast, the CPL model never reaches the required $H_0$ because its departure from $\Lambda$CDM decays too gradually with redshift. Its preference over $\Lambda$CDM without anchors also relies on taking the legacy SNIa at face value, since the $(w_0, w_a)$ tension with $\Lambda$CDM decreases from $3.36\sigma$ to $2.13\sigma$ once they are removed. This reduces the Bayesian preference for CPL over $\Lambda$CDM from $+1.70$~nats down to $-1.79$~nats, indicating $\Lambda$CDM is preferred over CPL without the legacy SNIa (Section~\ref{sec:Model_comparison}).

We highlight that the \texttt{exp} model prefers a very low $z_1$. This brings fine-tuning problems, namely why the expansion history would follow $\Lambda$CDM for many Gyr and only very recently depart from it. One possibility is outflow from a local underdensity or void \citep*{Haslbauer_2020, Banik_2025_BAO, Banik_2026_z0}, which needs enhanced structure growth on $\ga 100$~Mpc scales \citep[for a review, see][]{Banik_2026_void}. In this framework, a lower $z_1$ indicates a smaller void, which is less fine-tuned because it requires a more modest adjustment to $\Lambda$CDM structure growth. This motivates us to consider a log-uniform prior on $z_1$, which has little effect on our conclusions (Section~\ref{sec:logz1}). However, the prior on $A$ truncates the posterior along the ridge of constant $A z_1$, so the preferred $z_1 \approx 0.03$ with the SH0ES anchor cannot be considered a measurement of the void size.

Configuration~B replaces the \citetalias{Planck_2020} prior on $r_{\rm d}$ with CC, relaxing the assumption of $\Lambda$CDM before recombination. With DES SNIa and flat $\Lambda$CDM at late-times, this gives $r_{\rm d} = 148.48 \pm 8.39$~cMpc and $H_0 = 69.0^{+3.4}_{-4.6}\kmsMpc$, which is $1.2\sigma$ below SH0ES (Table~\ref{tab:baseline}). An early-time solution must preserve the BAO constraint $H_0 r_{\rm d} = 10154 \pm 73~\mathrm{km\,s^{-1}}$ \citep{DESI_2025}, so matching SH0ES requires $r_{\rm d} = 138.8 \pm 1.9$~cMpc. Our result lies only $1.1\sigma$ above this, with little change upon including the legacy SNIa. Thus, SNIa + BAO + CC cannot yet distinguish between early and late-time solutions due to the presently limited constraining power of CC. Even so, early-time solutions face the other difficulties discussed in Section~\ref{Introduction}.

The IDL calibrates the DES SNIa to $\Mdes = -19.380 \pm 0.012$, which is $4.30\sigma$ brighter than the SH0ES constraint $M_B = -19.253 \pm 0.027$ \citep{Riess_2022_comprehensive}. Although DES observes in the DECam passbands \citep[$g, r, i, z$;][]{Flaugher_2015, Burke_2018}, both DES-Dovekie and Pantheon+ use SALT light-curve models to standardise their SNIa to a peak absolute magnitude in the rest-frame $B$ band \citep{Guy_2007, Kenworthy_2021, Scolnic_2022}. Both estimates thus refer to the same quantity, with the actually used passbands entering only through the conversion to rest-frame $B$. A 0.13~mag difference between two calibrations of the same standardised SNIa would be an order of magnitude larger than typical photometric cross-calibration offsets, which making a purely photometric origin very unlikely \citep{Stubbs_2006, Brout_2022_supercal}. We note that $\Mdes$ is set mainly by the BAO at $z = 0.51$--0.93 and is almost identical with and without the legacy SNIa (Table~\ref{tab:baseline}), so this comparison does not test the legacy SNIa. Importantly, it also does not depend on the late-time expansion history because $\Mdes$ comes from SNIa at the redshifts of the BAO, while $M_B$ comes from SNIa in hosts with Cepheid distances. A late-time solution can therefore only move this tension into a zero-point difference between the nearby SNIa and the more distant DES SNIa, which is the kind of difference that $\Delta M_\mathrm{legacy}$ measures. Recovering the full SH0ES $H_0$ in this way would need the total 0.13~mag, of which $\Delta M_\mathrm{legacy} = 0.065 \pm 0.039$~mag accounts for about half. \citet{Efstathiou_2026_late_dynamical} makes a similar claim. In Section~\ref{sec:discussion_pantheon_DES}, we discuss whether $\Mdes = -19.38$ can be reconciled with $M_B = -19.25$.

The discrepancy between $\Mdes$ and $M_B$ is closely related to the conventional Hubble tension, since the inferred $M_{\rm SNIa}$ and $H_0$ are tightly linked when the shape of the expansion history is fixed (there is practically no uncertainty on $M_{\rm SNIa} - 5 \log_{10} H_0$). However, our cosmographic analysis demonstrates that these constraints are not equivalent when the shape of the late-time expansion history is allowed to vary. Although the DES-only IDL can yield a value of $H_0$ consistent with SH0ES, its absolute magnitude calibration remains discrepant at $\approx 4.3\sigma$ despite operating in similar passbands. Thus, agreement in $H_0$ alone does not establish a complete resolution of the tension: a successful late-time solution must also account for the difference between the absolute magnitudes of nearby and more distant SNIa on the respective instruments used to characterise them. We discuss possible reasons for this in Sections~\ref{sec:discussion_systematics_legacy} and \ref{sec:discussion_pantheon_DES}.

This issue is avoided by early-time solutions, as their revised $r_{\rm d}$ would alter $\Mdes$, potentially aligning it with $M_B$. However, the problem then becomes the predicted 1~Gyr reduction in $A_{\rm U}$, the age of the Universe \citep{Vagnozzi_2022, Cimatti_2023, Banik_2026_age, Valcin_2026}. This is in tension with the ages of the oldest stars and GCs in our own Galaxy. Other probes of the expansion history at intermediate epochs also argue against the expansion rate being 9\% faster than in the SPA cosmology, as would be required in an early-time solution \citep{Banik_2025_cosmology}. We therefore consider it more likely that the gap between $\Mdes$ and $M_B$ reflects differences between the nearby and distant SNIa, either intrinsically and/or in the way they are analysed.

This leads to a three-way tension between the high local $H_0$, the lack of a corresponding reduction in $A_{\rm U}$ in old stars and GCs, and the ``shape wall'' argument of \citet{Sabogal_2026} that $E(a)$ has not deviated much from $\Lambda$CDM based on combining BAO and SNIa (see also Section~\ref{sec:discussion-literature}). All three arguments cannot be simultaneously correct, since a standard $A_{\rm U} \approx 13.8$~Gyr and high $H_0$ imply a non-standard shape to the expansion history. The solution must involve at least one of the following:
\begin{enumerate}
    \item The Hubble tension is caused by systematics in the local distance ladder biasing $M_B$ towards fainter values and having a similar effect on many other distance indicators that also give a high $H_0$ \citep{Valentino_2025, Cai_2026, H0DN_2026};
    \item The shape wall does not exist, allowing a late-time solution. As shown above, this also requires a zero-point difference between nearby and more distant SNIa;
    \item The oldest stars and GCs are almost 1~Gyr younger than observations indicate, accommodating an early-time solution. This is because BAO and SNIa fix the late-time shape of the expansion history, so $A_{\rm U} \appropto 1/H_0$. Increasing $H_0$ to $H_0^\mathrm{SH0ES}$ thus reduces $A_{\rm U}$ from $\approx 13.8$~Gyr \citepalias{Planck_2020} to $\approx 12.7$~Gyr, but observations favour the former \citep{Vagnozzi_2022, Cimatti_2023, Valcin_2025, Banik_2026_age}.
\end{enumerate}

A possible path out of this trilemma is to combine the first two options. If the legacy offset is real, the shape wall is lower than usually assumed. In particular, allowing a legacy offset raises the IDL prediction to $H_0 = 71.03 \pm 1.99\kmsMpc$ (Table~\ref{tab:legacy-h0}). The recalibrated legacy SNIa, which come from the same surveys as much of the SH0ES Hubble-flow sample, then have $\Mdes + \Delta M_\mathrm{legacy} \approx -19.31$. This would correspond to $H_0 \approx 71 - 72\kmsMpc$, which is supported by several distance ladder analyses \citep{Anand_2022, Dhawan_2023, Uddin_2024, Hogas_2025, Li_2026, Stiskalek_2026_Cepheids, Stiskalek_2026_Cepheids_err, Stiskalek_2026_maser, Stiskalek_2026_TRGB}. The 0.13~mag difference between $\Mdes$ and $M_B$ would then be shared roughly equally between the legacy offset and the local distance ladder, with no change to pre-recombination physics and hence only a modest change to the predicted $A_{\rm U}$. This would limit the extent to which we have to question the observations underpinning any single side to the trilemma.

\subsection{Systematics with the legacy SNIa}
\label{sec:discussion_systematics_legacy}

The main open question is why the legacy SNIa disagree with the DES relation. We infer an offset of $\Delta M_\mathrm{legacy} = 0.065 \pm 0.039$~mag (Section~\ref{sec:systematics-results}), which is only a $1.65\sigma$ detection. It shows no dependence on redshift and at most a weak dependence on which legacy survey the SNIa come from (Figure~\ref{fig:offset_split}). We stress that this offset is not an established calibration error but is only suggestive of one. This is mostly based on extrapolating the DES relation below the DES floor (Section~\ref{sec:systematics-legacy}). Even so, the 10 legacy SNIa at $z = 0.073$--0.087 already lower $H_0$ by $4.45\kmsMpc$ (Section~\ref{sec:z_floor_result}). SNIa at the same redshift must trace the same expansion history, so a late-time departure from $\Lambda$CDM would not make these SNIa disagree with DES, whereas a difference in calibration would. However, we note that there are only 2 DES SNIa in this range, so this comparison is mainly with the DES relation constrained at higher redshift rather than with DES SNIa at the same redshift. Even so, the large difference in $H_0$ caused by so few legacy SNIa leads us to consider a systematic in the legacy SNIa as the more plausible explanation. Such a systematic could arise in three broad ways:
\begin{enumerate}
    \item Photometric cross-calibration: Since SNIa are transients, the legacy SNIa cannot be re-observed in the DECam passbands, so their recalibration onto these passbands relies on white dwarfs and other standard stars observed in both systems. Unrecognised shifts in filter transmission curves or colour terms routinely introduce $\mathcal{O}(0.01)$~mag offsets \citep{Stubbs_2006, Brout_2022_supercal}, while zero-point standards carry their own cross-calibration uncertainties between telescopes \citep{Oke_1983, Bohlin_2014}. Moreover, the spectral features of SNIa interact with the filter edges in a way that the nearly featureless spectrum of a white dwarf does not, so nm-scale shifts in filter centring might matter more for SNIa. Standard stars bright enough for legacy instruments can also saturate DECam, while shorter exposures introduce shutter-timing uncertainties and older CCDs may respond non-linearly. A calibration valid for bright standards might thus not hold for legacy SNIa at the $<0.1$~mag level, especially as small calibration changes can have a $6\times$ larger impact on SNIa distances due to the impact on training of the light-curve fitting algorithm and the way in which SNIa are standardised \citep{Popovic_2026_systematics};
    \item Corrections for survey selection and other observational effects: Each survey needs its own completeness and selection-bias corrections, so an error in modelling the selection function of a legacy survey would shift its SNIa coherently. This would naturally produce the survey-specific behaviour hinted at by our split between Foundation and CfA/CSP. Accurately correcting for selection is harder for the legacy surveys than for DES because most legacy SNIa were discovered by independent searches, several of which targeted catalogued galaxies that were only then followed up \citep{Scolnic_2022}. Their selection thus depends on searches whose selection functions are not well known, so processing them with the same pipeline as DES does not produce a homogeneous sample. Moreover, the bias corrections are computed from simulations in a fiducial flat $\Lambda$CDM cosmology \citep{Brout_2022, DES_2024_SNe}, with $q_0 = \frac{3}{2}\Omega_{\rm m} - 1 \approx -0.5$. Near the limit of a legacy survey at $z \approx 0.1$, the corrections are largest while the survey constrains the shape of the expansion history only weakly, so the assumed shape can matter. The corrected magnitudes would then be pulled towards the fiducial $q_0$, so using them to argue against a late-time departure from $\Lambda$CDM would be partly circular. This matters less for DES, whose largest corrections lie near its magnitude limit at $z \ga 0.6$, where the expansion history we infer is close to that of $\Lambda$CDM (Figure~\ref{fig:dz_ratio}). The DES low redshift limit is set by the rarity of SNIa, since even faint SNIa lie well above the detection limit. We note that as $M_B$ is set by SNIa in Cepheid hosts at $z \la 0.01$ where the Hubble diagram is practically linear, it is barely affected by this issue, but it might affect the tilt of $D_{\rm L}/D_{\rm L}^\mathrm{SPA}$ with redshift at $z \la 0.1$;
    \item Light-curve modelling and host-galaxy subtraction: The host galaxy light must be subtracted in the same way for results to be comparable between surveys, but procedures have changed over the decades and legacy instruments typically had poorer angular resolution. Since SNIa often occur in crowded parts of their hosts, a different background estimate alters the inferred colour and light-curve shape, which then alters the standardised magnitude via the Tripp formula \citep{Tripp_1998}.
\end{enumerate}

We stress that our DES-only analysis is not completely independent of the legacy SNIa. Although we exclude the latter from our baseline cosmological fit, they are still used to train the algorithm that fits the DES light curves. To our knowledge, no analysis of DES has yet hidden the legacy magnitude--redshift relation from this training, so the size of the effect is currently unknown.

Importantly, systematics of this kind have little impact on the local distance ladder. This is because SH0ES does not need to recalibrate the legacy SNIa onto the DECam passbands, since many legacy surveys also found SNIa in hosts near enough for Cepheid, TRGB, or SBF distances, allowing the absolute magnitude to be calibrated on each photometric system \citep{Riess_2022_comprehensive}. We demonstrate this by refitting the public SH0ES data vector and covariance with a separate absolute magnitude $M_s$ for each SNIa survey $s$. Two surveys (SDSS and PS1MD) contain no SNIa in a Cepheid host and share none with any calibrated survey, so we remove their 57 Hubble-flow light curves. Since the SH0ES model is linear in its parameters, a generalised least-squares fit gives the exact posterior. This changes $H_0$ from the published $73.04 \pm 1.04\kmsMpc$ to $72.85 \pm 1.04\kmsMpc$, which is still $5.1\sigma$ above $H_0^\mathrm{SPA}$. Moreover, across the 297 remaining light curves (77 in Cepheid hosts and 220 in the Hubble flow), the 12 extra $M_s$ parameters improve $\chi^2$ by only 7.5, less than the improvement of about 12 expected from fitting noise alone. Thus, even our baseline approach of doubting cross-calibrations between surveys leaves the Hubble tension virtually unchanged, especially when we consider other ways of measuring distances that do not rely on SNIa \citep{Scolnic_2023}.

The cleanest test with existing data would be to reanalyse DES with a light-curve algorithm trained without the legacy SNIa apparent magnitudes, perhaps using them only for the light-curve shape to help with event classification. However, the definitive test requires a homogeneous SNIa sample that reaches below the DES floor, given that DES has little sensitivity at $z \la 0.2$. This will come from ongoing campaigns like the Dark Energy Bedrock All-Sky Supernova (DEBASS) program \citep{Sherman_2026}, the Asteroid Terrestrial-impact Last Alert System \citep[ATLAS; see the TITAN sample of][]{Marlin_2026_titan}, the Young Supernova Experiment \citep[YSE;][]{Jones_2021, Aleo_2023}, and the Zwicky Transient Facility \citep[ZTF;][]{Rigault_2025}. Of particular importance is DEBASS, which uses DECam, the same instrument as DES. This simplifies the treatment of any relative calibration offset, though selection effects for DEBASS differ from DES and must be handled with care.

Another route is to constrain the distance--redshift relation at $z \la 0.1$ without SNIa, for example with the fundamental plane (FP) distances of the DESI Peculiar Velocity Survey \citep{Said_2025, Scolnic_2025, Carr_2026}. At its outer limit, the FP distance ladder can be linked up with the homogeneous DES IDL. To bypass SNIa altogether, the FP results would have to reach $z \approx 0.2$ while BAO measurements are extended down to this redshift, which seems plausible in future DESI data releases.

\subsection{Comparison of Pantheon+ and DES}
\label{sec:discussion_pantheon_DES}

Even if there are systematics with the estimated apparent magnitudes of legacy SNIa had DES observed them, we still need to explain how $\Mdes$ and $M_B$ can differ by 0.13~mag given both are standardised peak magnitudes in rest-frame $B$. Our IDL uses DES, whereas the SH0ES $M_B$ is tied to Pantheon+ \citep{Riess_2022_comprehensive}. A detailed analysis of Pantheon+ is beyond the scope of this paper, but we briefly compare it to DES, starting with common features.

Both standardise their SNIa with SALT light-curve models to a peak magnitude in the rest-frame $B$ band. Pantheon+ contains many of the same low-redshift legacy surveys as our DES + legacy configuration (Section~\ref{sec:discussion-literature}), as well as a subset of the DES SNIa \citep{Vincenzi_2025}. Both correct for Malmquist bias and other selection effects using survey simulations \citep{Brout_2022, DES_2024_SNe}. While these are sometimes considered cosmology-independent, that is often taken to mean that $H_0$ and $\Omega_{\rm m}$ are free, not that the expansion history can depart sharply from $\Lambda$CDM at low redshift, as in our \texttt{exp} model or a local void \citep{Haslbauer_2020}. The bias corrections might thus implicitly assume a $\Lambda$CDM shape. This matters most near the limit of the legacy surveys at $z \approx 0.1$, where the corrections are large and the shape is weakly constrained by the data, making the prior more important (Section~\ref{sec:discussion_systematics_legacy}). The assumed shape of the expansion history is thus a more serious concern with Pantheon+, providing a further reason why we do not use it in this work.

Despite these similarities, Pantheon+ and DES also differ in several ways:
\begin{enumerate}
    \item Pantheon+ consists of spectroscopically classified SNIa and is weighted towards low redshift, whereas DES reaches higher redshift and classifies most of its SNIa photometrically \citep{Kunz_2007}. Their selection effects and light-curve fitting therefore differ;
    \item Host-galaxy subtraction might differ between the extended, well-resolved hosts of nearby SNIa -- especially the SH0ES calibrators -- and the small, faint hosts of DES SNIa (see also Section~\ref{sec:discussion_systematics_legacy});
    \item SNIa do not all have the same absolute magnitude even in the same filter, so the quoted $M_{\rm SNIa}$ refers to a standard SNIa with reference stretch, colour, and host properties. The conventions behind this choice can differ between analyses, for instance how the host-mass correction is split and whether the colour refers to the observed colour or the intrinsic colour after accounting for reddening by dust.
\end{enumerate}

It is difficult to estimate how much these effects would change $M_{\rm SNIa}$, but we can measure the net effect for SNIa common to both catalogues. An absolute magnitude is not a purely intrinsic property of SNIa because it depends on the photometric calibration, light-curve model, standardisation, and bias corrections of the pipeline that produces it. \citet{Vincenzi_2025} showed that DES and Pantheon+ assign different standardised magnitudes to the same SNIa. The Tripp formula \citep{Tripp_1998} corrects the magnitude using the observed colour, which combines the intrinsic colour of the SNIa with reddening by dust. Thus, how a pipeline separates the two contributions changes both the colour correction and the bias corrections. Matching the light curves present in both catalogues, we find that DES-Dovekie makes SNIa fainter than Pantheon+ by $0.011 \pm 0.006$~mag for 148 DES SNIa and by $0.046 \pm 0.006$~mag for 189 legacy SNIa. The individual terms differ by up to 0.11~mag, but they largely cancel. Differences between the pipelines are therefore comparable to $\Delta M_\mathrm{legacy}$, but an order of magnitude smaller than the gap between $\Mdes$ and $M_B$, and they would widen it rather than close it.

\citet{Camilleri_2026_Unite} recently put Pantheon+ and DES-SN5YR on a common calibration, light-curve model, and intrinsic scatter model. They find that the offset between low- and high-redshift SNIa changes by at most 0.02~mag relative to either parent catalogue. Even so, removing all SNIa at $z < 0.1$ is the only cut that shifts their cosmology substantially (by $\approx 2\sigma$), leading them to call for new low-redshift samples such as DEBASS. Since they marginalise over the absolute magnitude, their analysis does not directly address $M_B$ or $\Mdes$.

We find that the 0.13~mag gap is not specific to DES. Applying our $k_{\max} = 4$ IDL to Pantheon+ with the same BAO and $r_{\rm d}$ gives $M_B = -19.414 \pm 0.016$, which is 0.16~mag brighter than the SH0ES calibration within the same pipeline \citep[see also][]{Efstathiou_2026_late_dynamical}. The tension in the absolute magnitude is therefore between the nearby SNIa in Cepheid hosts and the more distant SNIa calibrated by the BAO, whichever catalogue is used. This makes it more difficult to solve the trilemma in Section~\ref{sec:discussion_implications}, especially since the 0.13~mag gap is too large to arise from photometric cross-calibration alone. However, the standardisation itself need not be the same for the few nearby SNIa that calibrate $M_B$ and the more distant SNIa that the BAO calibrate. We highlight four ways in which the two could differ:
\begin{enumerate}
    \item The light-curve models are not the same. DES-Dovekie uses a retrained SALT3 model \citep{Kenworthy_2021, Popovic_2026_Dovekie}, whereas Pantheon+ (and hence the SH0ES $M_B$) uses SALT2 \citep{Guy_2007, Brout_2022_supercal}. The spectral energy distribution (SED), the colour law, and the stretch--luminosity relation are all set by the training, so the standard SNIa to which each calibration refers need not be exactly the same. However, for the SNIa common to both catalogues, the net difference is only a few hundredths of a magnitude;
    \item The K-corrections differ. Nearby SNIa can be observed in a passband close to rest-frame $B$, so their K-corrections are small. In contrast, DES observes in fixed DECam passbands, so the rest-frame wavelengths it samples change with redshift. At $z = 0.51$--0.93 where the BAO set $\Mdes$, rest-frame $B$ falls in the $r$, $i$, or $z$ band or between them, while the bluer bands sample the rest-frame ultraviolet, where SNIa spectra are more diverse \citep{Foley_2016}. The conversion to rest-frame $B$ thus relies on the SED model. An error here would shift $\Mdes$ relative to $M_B$ without affecting the nearby SNIa;
    \item The SNIa populations might differ. In a volume-limited sample of about 1000 nearby SNIa, the SALT stretch $x_1$ is bimodal, with the low-stretch mode found only in old environments \citep{Ginolin_2025}. Since the cosmic star formation rate rises with redshift, this mode becomes rarer at higher redshift \citep{Nicolas_2021, Ginolin_2025}. The bimodality is indeed much less evident in DES, where a single asymmetric Gaussian describes the $x_1$ distribution almost as well as two components \citep{Wiseman_2022}. Importantly, \citet{Ginolin_2025} also find that the stretch--magnitude relation is not linear, so a linear standardisation \citep{Tripp_1998} calibrated on one mixture of the two modes would leave a residual offset for another. This holds whether the two modes are distinct subtypes of SNIa (Section~\ref{Introduction}) or the two ends of a single continuous but bimodal distribution, because nearby and distant surveys would still sample mostly different parts of it. We consider it unlikely that this alone explains the 0.13~mag, but it could contribute;
    \item The calibrators might not be typical of DES. The SH0ES $M_B$ rests on only 42~SNIa in 37 Cepheid hosts \citep{Riess_2022_comprehensive}, which are selected by the need for a Cepheid distance, and hence for a nearby star-forming host, rather than by their SNIa properties \citep[see also][]{Rigault_2015}. Since the standardisation depends on the SED of each SNIa, it is not clear that these few SNIa are representative of the DES SNIa at the BAO redshifts. In DES, one could test this by checking whether $\Mdes$ changes upon restricting the analysis to a particular subpopulation, for example by stretch or host environment. This is not possible for the calibrators as there are too few of them to split in this way while retaining precision, given the intrinsic scatter and measurement uncertainties.
\end{enumerate}
Since the pipelines account for only a few hundredths of a magnitude, this leaves items 3 and 4, or systematics in the local distance ladder. A population difference would have the right sign. Nearby SNIa are found almost regardless of their luminosity, so the calibrators should be dominated by the most common subtype, whereas a magnitude-limited survey at $z \approx 0.5$--0.9 preferentially finds the brighter SNIa, which might be a rarer subtype with a different progenitor channel (Section~\ref{Introduction}). Bias corrections model the selection of SNIa with given distributions of stretch and colour, so they would not remove a difference in standardised luminosity between subtypes whose mix changes with redshift. This would make $\Mdes$ brighter than $M_B$, as observed. Within DES, the same effect would make the nearest SNIa appear fainter and thus more distant, which is opposite to the deficit we find (Section~\ref{sec:discussion_DES}). We therefore argue that this feature is genuinely cosmological, while the difference between $M_B$ and our IDL estimate of $\Mdes$ need not be.

\subsection{Comparison with the literature}
\label{sec:discussion-literature}

The IDL has a record of giving a low $H_0$ similar to \citetalias{Planck_2020}, from BOSS BAO \citep{Aubourg_2015} and DES-SN3YR \citep[$H_0 = 67.8 \pm 1.3 \kmsMpc$;][]{Macaulay_2019} to DES-SN5YR combined with DESI~DR1 and DR2 BAO \citep{Camilleri_2025, Ling_2025}. Across three generations of SNIa samples and two BAO surveys, the IDL gives $H_0 \approx 67$--$68\kmsMpc$, in line with our DES + legacy result. \citet{Sabogal_2026} formalised this obstacle as a ``shape wall'', which forms one of the three arguments in the trilemma of Section~\ref{sec:discussion_implications}. If pre-recombination physics is standard, a late-time solution can raise $H_0$ only by changing the shape of $E(z)$. Relative SNIa distances and BAO constrain this tightly enough to restrict any increase in $H_0$ to only a few percent, well short of the $\approx 9\%$ needed to reach $H_0^\mathrm{SH0ES}$.

The closest analysis to ours is that of \citet{Popovic_2026_IDL}, who run a cosmographic IDL on the same DES-Dovekie catalogue, calibrated using DESI~DR2 and a standard CMB sound horizon. They conclude that the IDL leaves little room for a late-time solution. Adopting their methodology and likelihood, we recover their results to within $\approx 1\sigma$. Using a sample that includes the legacy SNIa, their nominal $H_0 = 66.82 \pm 0.74\kmsMpc$, which is consistent with our DES + legacy result at the $1.15\sigma$~C.L. However, there are two methodological differences. Firstly, their expansion is a polynomial in $z$ \citep[following][]{Visser_2004} with $k_{\max} = 4$ (their eq.~5), which does not converge over the redshift range of DES and DESI (Section~\ref{sec:setup}) and implies $H \propto z^4$ at high redshift, far steeper than the $z^{3/2}$ behaviour of $\Lambda$CDM. Their $H_0$ is thus tied to this polynomial ansatz rather than being a model-independent estimate. Secondly, they use separate polynomials for $H(z)$ and $D_{\rm L}(z)$, which is internally inconsistent because $D_{\rm L}$ follows from integrating $c/H$. This is a good approximation at low redshift, but not up to $z = 2.33$.

\citet{Popovic_2026_IDL} also argue that raising $H_0$ from 67 to $73\kmsMpc$ demands a magnitude gradient $\dd \Mdes/\dd z \approx 0.2$~mag, leading a flat $\Lambda$CDM fit to return $\Omega_{\rm m} \approx 0.23$, a value that is $10\sigma$ discrepant with other probes. However, this argument pre-supposes $\Lambda$CDM at late times. Since a late-time solution is by definition a departure from the $\Lambda$CDM background, a non-standard effective $\Omega_{\rm m}$ under an imposed $\Lambda$CDM fit is the expected signature of such a solution rather than a falsification of it. Most importantly, their nominal sample relies on the legacy SNIa, so their exclusion of late-time solutions holds only if the legacy SNIa are accurate.

The legacy rung has been questioned before. Our $\Delta M_\mathrm{legacy}$ reproduces the offset first reported by \citet{Efstathiou_2025} from a comparison between Pantheon+ and DES. \citet{Watkins_2026_consistency} find that although the high-redshift DES SNIa are well described by flat $\Lambda$CDM, the low-redshift sample is offset by $\approx 0.05$~mag, which they attribute to systematics rather than new physics. The interpretation of the offset is contested. \citet{Vincenzi_2025} reproduce the difference with Pantheon+, but attribute $\approx 43\%$ of it to improvements in the DES-SN5YR treatment of intrinsic scatter and host-galaxy properties, which mainly affect the legacy SNIa. They attribute a further $\approx 38\%$ to different selection functions for the DES subsets entering the two compilations, which changes the bias corrections \citep[see also][]{Efstathiou_2025}. For the IDL, the cause of the offset matters less than its size, since any offset of this kind has the same impact on $H_0$.

Our results in Figure~\ref{fig:dz_ratio} show that the legacy SNIa follow much the same distance--redshift relation as predicted by the SPA cosmology, but the lowest redshift DES SNIa are around 5\% closer than expected, a gap which closes as we get to higher redshift. We might then expect that if $H_0$ is estimated from Dovekie data at $z < z_{\rm ceil}$ without regard to the DES--legacy distinction, there would be a rising trend due to the transition from legacy data at low redshifts to the main DES sample further out. Just such a trend is indeed apparent in figure~2 of \citet{Roy_2026}, where the absolute calibration is not relevant because it would equally affect the results regardless of the adopted $z_{\rm ceil}$. They fix $q_0$ and $j_0$ to their $\Lambda$CDM values, which effectively fixes the shape of the expansion history to that in the SPA cosmology. Since $H_0$ should not have a trend with $z_{\rm ceil}$, the reported trend indicates that the expansion history has a different shape and/or that the data used is subject to systematics. We argue that both are correct. Once the legacy SNIa are removed, there is no appreciable trend in the recovered $H_0$ with the imposed redshift floor (Figure~\ref{fig:floorsweep-real-new}), suggesting that there is no internal consistency problem within the homogeneous DES dataset. Indeed, the lowest redshift DES data follow a smooth extrapolation of the BAO and SNIa distance--redshift relation at $z > 0.3$ (Figure~\ref{fig:dz_ratio}).

Further support for a legacy offset comes from the full DES-Dovekie catalogue showing some evidence for a violation of the Etherington relation \citep{Artola_2026}. This preference is unlikely to be genuine on theoretical grounds \citep{Etherington_1933}. Fortunately, the preference largely vanishes upon removing SNIa at $z < 0.1$ \citep{Artola_2026}. This corresponds closely to removing the legacy SNIa from the sample, though those authors did not phrase their results in these terms, sorting SNIa only in redshift rather than based on their source. We also note that independent reconstructions of the dark energy equation of state using DESI~BAO highlight the necessity of correcting the legacy rung to ensure consistency with DES \citep{Huang_2025_DESI, Ormondroyd_2026}. All these results make it likely that the legacy SNIa in the Dovekie compilation should not be taken at face value when combining them with the larger and more modern DES sample, which is necessary to reach high enough redshift to achieve a calibration via the IDL. Given the very different nature of how SNIa were discovered in DES compared to the legacy surveys, combining both into a single catalogue is inevitably challenging.

The legacy SNIa also explain why our conclusions differ from those of \citet{Bansal_2026}, who find that a purely late-time modification can fit $H_0^\mathrm{SH0ES}$, the DESI~BAO data, and relative SNIa distances only with a sharp step in $M_\mathrm{SNIa}$ at $z \la 0.01$ (Section~\ref{Introduction}). Their analysis rests on Pantheon+, which contains the same low-redshift legacy surveys as our DES + legacy configuration, in which we reach the same verdict. In contrast, the homogeneous DES sample gives $H_0^\mathrm{DES} = 76.25 \pm 2.87\kmsMpc$ using a single survey, filter set, and selection function to remove the need for cross-calibration between surveys. This is only $1.03\sigma$ above $H_0^\mathrm{SH0ES}$. Allowing the legacy SNIa to be recalibrated leaves $H_0$ only $0.99\sigma$ below $H_0^\mathrm{SH0ES}$. However, the good agreement with SH0ES in both cases despite assuming a standard $r_{\rm d}$ does not remove the 0.13~mag discrepancy between $\Mdes$ and $M_B$ (Section~\ref{sec:discussion}). With DES, a late-time solution still needs a step in $M_\mathrm{SNIa}$, but between the nearby and the DES SNIa rather than at $z \la 0.01$. Our results suggest that the legacy offset accounts for about half of this gap. We thus agree with \citet{Bansal_2026} that a step is needed, but differ on where it should be located.

\section{Conclusions}
\label{sec:conclusions}

We tested whether the IDL rules out late-time solutions to the Hubble tension by combining the homogeneous DES-Dovekie catalogue of uncalibrated SNIa \citep{Popovic_2026_Dovekie} with DESI~DR2 BAO \citep{DESI_2025}, which we calibrated with a sound horizon derived from the CMB assuming standard physics before recombination \citepalias{Planck_2020}. We used a cosmographic expansion in the lookback scale factor $\widetilde{a} \equiv z/(1 + z) = 1 - a$, selecting $k_{\max} = 4$ as the order preferred by the Bayesian evidence for DES. We checked our results against the BMA over $k_{\max} = 2$--6. Our baseline uses only the DES SNIa ($z \in [0.073, 1.14]$) and calibrates the BAO through the \citetalias{Planck_2020} prior on $r_{\rm d}$, which we replaced with CC in one analysis \citep{Moresco_2022}. We included the legacy SNIa ($z \in [0.025, 0.087]$) only as an extension so we could assess their impact on the IDL and test them for systematics. Our main results are as follows:
\begin{enumerate}
    \item Using SNIa only from DES and combining with the BAO assuming the \citetalias{Planck_2020} $r_{\rm d}$ predicts $H_0 = 76.25 \pm 2.87\kmsMpc$, consistent with SH0ES \citep{Breuval_2024} but $3.13\sigma$ above $H_0^\mathrm{SPA}$ \citep{SPT_2026}. This tension depends on the polynomial order used to fit the data and extrapolate below the DES floor at $z = 0.073$. Only 0.84\% of our $\Lambda$CDM mocks give such a high $H_0^\mathrm{DES}$ in a like-for-like comparison, while the BMA gives $74.77 \pm 4.83\kmsMpc$, only $1.56\sigma$ above SPA. Replacing the $r_{\rm d}$ prior with CC and fixing $k_{\max} = 4$ gives a similar but far less precise $H_0 = 75.7 \pm 5.3\kmsMpc$.
    \item Adding the legacy SNIa lowers the IDL-inferred $H_0$ dramatically to $67.98 \pm 0.69\kmsMpc$, which is $4.69\sigma$ below SH0ES ($66.61 \pm 1.18\kmsMpc$ and $4.50\sigma$ for the BMA). This leads to a decrease of $8.27\kmsMpc$ in the central estimate of $H_0$ when we add the legacy SNIa to the homogeneous DES sample.\footnote{The drop is $8.16\kmsMpc$ for the BMA, but our mocks only consider $k_{\max} = 4$, the preferred order with DES.} Only 0.85\% of $10^4$ mocks in the SPA cosmology show a larger drop ($\Delta H_0$) in a like-for-like comparison. We also find that when excluding all SNIa at $z < 0.3$ and then lowering the redshift floor to include more SNIa, the predicted $H_0$ remains fairly stable in the regime where only DES is relevant. However, a sharp drop occurs at $z \approx 0.08$, where the DES-Dovekie catalogue switches from DES to legacy SNIa. This is a strong hint that there is a systematic offset between the two.
    \item Allowing the legacy SNIa to be recalibrated relative to DES gives $\Delta M_\mathrm{legacy} = 0.065 \pm 0.039$~mag, in the same direction as several previous studies \citep{Efstathiou_2025, Huang_2025_DESI, Ormondroyd_2026, Watkins_2026_consistency}. This raises $H_0$ to $71.03 \pm 1.99\kmsMpc$, only $0.99\sigma$ below SH0ES ($69.50 \pm 2.43\kmsMpc$ and $1.43\sigma$ for the BMA). The offset shows no dependence on redshift and only weak evidence for being survey-specific within the legacy sample.
    \item With three SNIa-independent $H_0$ anchors \citep{Mukherjee_2021, Jensen_2025, Vogl_2025}, the \texttt{exp} model gives $H_0 = 70.00^{+4.78}_{-0.98}\kmsMpc$ and is preferred over $\Lambda$CDM for DES ($\ln B_{10} = +1.54$), but gives $H_0 = 67.66 \pm 1.01\kmsMpc$ for DES + legacy. Even for DES, SH0ES would measure only $H_0^{\rm win} = 70.23 \pm 0.99\kmsMpc$ in this model, $2.25\sigma$ below the published value. With the SH0ES anchor instead, the preference for the \texttt{exp} model with DES reaches 10.61~nats. In contrast, the CPL model cannot raise $H_0$ enough to be combined with local anchors. Its preference over $\Lambda$CDM without anchors relies on taking the legacy SNIa at face value (Figure~\ref{fig:w0wa_contours}).
    \item The IDL calibrates the DES SNIa to $\Mdes \approx -19.38$ for every order with appreciable weight, with or without the legacy SNIa. This is in $4.2$--$4.3\sigma$ tension with the SH0ES $M_B = -19.253 \pm 0.027$ \citep{Riess_2022_comprehensive} assuming the \citetalias{Planck_2020} prior on $r_{\rm d}$. Since both refer to the same rest-frame $B$ band, this is a new manifestation of the Hubble tension that a late-time solution cannot remove by itself. However, we note that the tension arises between the modern DES and legacy surveys operating in somewhat different ways, leading to a few possible non-cosmological explanations of the discrepancy (Section~\ref{sec:discussion_pantheon_DES}).
\end{enumerate}

These results suggest that the claimed exclusion of late-time solutions to the Hubble tension by the IDL depends crucially on the low-redshift legacy SNIa. We stress that the evidence for a calibration offset in them is modest and relies on extrapolating the DES relation below $z=0.073$. Furthermore, the DES-only analysis is not completely independent of the legacy SNIa, which still enter the training of the DES light-curve algorithm.

More generally, the Hubble tension, the shape wall, and the age of the Universe cannot all be as current observations indicate (Section~\ref{sec:discussion_implications}). A possible path out of this trilemma is a late-time solution enabled by the legacy offset, together with a slightly lower direct distance ladder value of $H_0 \approx 71 - 72\kmsMpc$ \citep{Anand_2022, Dhawan_2023, Uddin_2024, Hogas_2025, Li_2026, Stiskalek_2026_Cepheids, Stiskalek_2026_Cepheids_err, Stiskalek_2026_maser, Stiskalek_2026_TRGB}. This would avoid an early-time solution and hence a younger Universe \citep[see also][]{Vagnozzi_2022, Vagnozzi_2023, Banik_2025_cosmology}, naturally explaining the lack of any anomalies in the CMB anisotropies \citep{Calabrese_2025, SPT_2026, Lynch_2026} and their reasonable agreement with standard BBN \citep{Giovanetti_2026, Launders_2026, Poulin_2026}.

A full clarification of these issues awaits a homogeneous SNIa sample that extends from low redshift into the range covered by DES. Ongoing surveys like DEBASS \citep{Sherman_2026} will provide such a sample using the same instrument as DES. Preliminary DEBASS~DR0.5 distances do not show the $\approx 0.05$~mag offset between the legacy SNIa in the Dovekie catalogue and the extrapolated DES distance--redshift relation \citep{Watkins_2026_consistency}. This tentatively points to systematics in the legacy SNIa or in how they were analysed, in agreement with our results. This needs to be checked with a larger sample of SNIa and a careful treatment of survey selection effects, which differ somewhat from DES as SNIa in DEBASS were typically discovered elsewhere. It is also possible to obtain the distance--redshift relation out to the DES floor in other ways, though the shape should not be fixed to $\Lambda$CDM in such analyses \citep{Carr_2026}. In the meantime, we cannot confidently assert that the IDL rules out a late-time solution to the Hubble tension.

\section*{Acknowledgements}

All authors are supported by Royal Society University Research Fellowship grant 211046. Numerical computations were carried out on the SCIAMA High Performance Computing (HPC) cluster, which is supported by the Institute of Cosmology and Gravitation and the University of Portsmouth. The authors are grateful to Christopher Frohmaier for helpful discussions on supernova systematics. They also acknowledge comments from the anonymous referee.

\section*{Data Availability}

Our code is publicly available.\footnote{\url{https://github.com/antonionajeraq/inverse_distance_ladder_DES_DESI_Hubble}} The datasets used in this work are publicly available and come from the cited publications.

\bibliographystyle{mnras}
\bibliography{IDL_bbl}

\appendix

\section{Reconstructed expansion history}
\label{app:adot}

\begin{figure}
    \centering
    \includegraphics[width=1.0\linewidth]{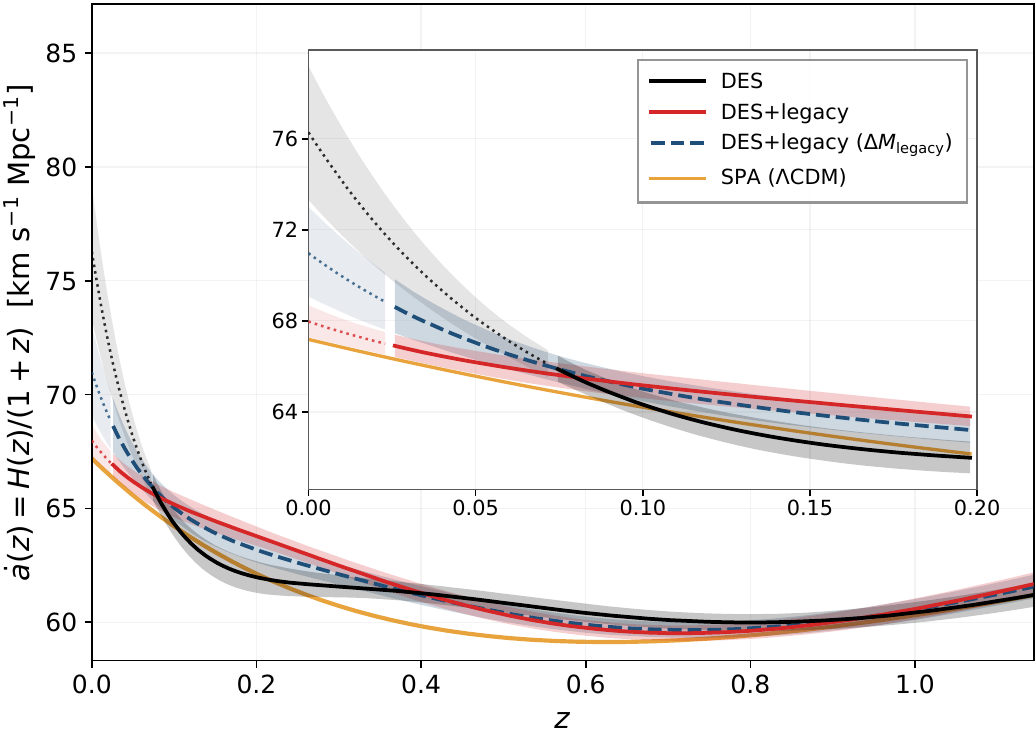}
    \caption{The median reconstructed $\dot{a}(z)$ in our cosmographic expansion with $k_{\max} = 4$ (Eq.~\ref{eqn:cosmographic-expansion}) constrained using SNIa from DES only (solid black line), DES + legacy (solid red line), or DES + legacy allowing the legacy SNIa to be recalibrated by $\Delta M_\mathrm{legacy}$ (dashed blue line at an intermediate level). The shaded bands show the $1\sigma$ uncertainty using Gaussian equivalent percentiles. The dotted lines in the same colour show the extrapolation below the lowest redshift SNIa in each case, which we mark with a small gap in the error band. The solid orange line without uncertainty shows the flat $\Lambda$CDM prediction using parameters that best fit the CMB \citep{SPT_2026}. The inset shows the $z < 0.2$ region in more detail.}
    \label{fig:adot-reconstruction}
\end{figure}

Figure~\ref{fig:adot-reconstruction} shows the median expansion history $\dot{a}(z) = H(z)/(1 + z)$ inferred by fitting our cosmographic expansion (Eq.~\ref{eqn:cosmographic-expansion}). As discussed further in Section~\ref{sec:results-params}, the reconstructed $\dot{a}$ is higher than the SPA expectation at $z \approx 0.4$. This discrepancy narrows towards lower redshift -- possibly even reversing sign. At the very lowest redshifts probed by the SNIa, $\dot{a}$ rises sharply towards lower $z$. This rise is most pronounced with DES alone and least pronounced with DES + legacy. Allowing recalibration of the legacy SNIa (Section~\ref{sec:systematics-legacy}) returns an intermediate result.

The inset to Figure~\ref{fig:adot-reconstruction} shows the $z < 0.2$ regime in greater detail. It is clear that if the legacy SNIa are taken at face value and assumed to be free of systematics, then the Hubble tension will be very difficult to solve at late times. But if the legacy SNIa are excluded or we assume that systematics with them can be rectified using a single recalibration parameter, then a late-time solution becomes much more plausible and may even be preferred.

\section{Alcock--Paczynski consistency}
\label{app:AP}

The AP test is an internal consistency test that is independent of both $r_{\rm d}$ and $\Mdes$ (Section~\ref{sec:AP_consistency}). We compare the SNIa predictions for $F_\mathrm{AP}$ with the DESI~DR2 measurements at effective redshifts of $z = 0.510$, 0.706, and 0.934, which lie within the range covered by the DES catalogue. We lack BAO measurements of $F_\mathrm{AP}$ at lower redshift, while DES lacks SNIa at higher redshift. Reaching the next DESI~DR2 BAO measurement at $z = 1.32$ would entail an extrapolation beyond the highest redshift DES SNIa at $z = 1.14$. This leads to much larger uncertainties, so we do not report the result.

\subsection{Methodology}
\label{app:AP-derivation}

Our predictions for $F_\mathrm{AP}$ are based on the same cosmographic expansion as used throughout this work (Eq.~\ref{eqn:cosmographic-expansion}). Both distances follow directly from it in a flat metric:
\begin{eqnarray}
    D_{\rm M} \left( z \right) ~=~ c \int_0^z \frac{\mathrm{d}\tilde{z}}{H(\tilde{z})} \, , \qquad D_{\rm H} \left( z \right) ~=~ \frac{c}{H \left( z \right)} \, .
    \label{eqn:AP-DM}
\end{eqnarray}
This implies $D_{\rm H} = D_{\rm M}'$, where a prime denotes differentiation with respect to $z$. In other words, the Hubble distance reported by DESI (based on the redshift depth of the BAO ruler) is exactly the derivative of the comoving distance (based on the angular scale of the BAO ruler). The SNIa reconstruction tells us the shape of $D_{\rm M}(z)$, which we can differentiate to obtain $D_{\rm H}$ and thus $F_\mathrm{AP}$. Cancelling out the speed of light, we obtain that
\begin{eqnarray}
    F_{\rm AP}(z) ~\equiv~ \frac{D_{\rm M}(z)}{D_{\rm H}(z)} ~=~ H(z) \int_0^z \frac{\mathrm{d}\tilde{z}}{H(\tilde{z})} \, .
    \label{eqn:AP-FAP}
\end{eqnarray}

To relate these results to our cosmographic expansion (Eq.~\ref{eqn:cosmographic-expansion}), we recall that $H(z) \equiv H_0 \, E(z)$, so $E(z)$ carries the shape of the expansion history and $H_0$ carries its normalisation. Substituting into Eq.~\ref{eqn:AP-FAP}, we obtain the following:
\begin{eqnarray}
    F_{\rm AP}(z) ~=~ H_0 \, E(z) \int_0^z \frac{\mathrm{d}\tilde z}{H_0 \, E(\tilde z)} ~=~ E(z) \int_0^z \frac{\mathrm{d}\tilde z}{E(\tilde z)} \, .
    \label{eqn:AP-shape}
\end{eqnarray}
This is the central equation for the AP test. Notice that since $H_0$ cancels out identically, $F_\mathrm{AP}$ is a function only of the shape coefficients $C_k$.

The SNIa absolute magnitude cancels for the same reason. SNIa constrain the apparent magnitude $m(z) = \Mdes + 5\log_{10}[(1+z) D_{\rm M}(z)/\rm{Mpc}] + 25$. Since $D_{\rm M} \propto 1/H_0$ at fixed shape, SNIa alone measure only the shape coefficients together with the single degenerate combination $\Mdes - 5\log_{10} H_0$, never $\Mdes$ or $H_0$ separately. Holding $\Mdes$ fixed in the fit is therefore a gauge choice that leaves $F_{\rm AP}$ untouched, so the calibration does not enter the prediction. This cancellation is also what makes the AP test precise despite the SNIa sample being uncalibrated. This leaves only the error in the shape of the expansion history, which is constrained quite well by SNIa given their large number and wide redshift range.

To predict BAO measurements of $F_\mathrm{AP}$, we must by definition not consider the BAO data we are trying to predict. We therefore fit the cosmographic parameters using only SNIa, assuming some fixed $\Mdes$. Since the SNIa only reach $z = 1.144$ ($\widetilde{a} = 0.534$), we impose the prior envelope of Eq.~\ref{eqn:envelope} up to $\widetilde{a}_{\max} = 0.534$ rather than $\widetilde{a}_{\max} = 0.70$ in this fit, so that the prior constrains only the range with available data. We then evaluate Eq.~\ref{eqn:AP-shape} by numerical quadrature on the actual posterior draws returned by the sampler. Every draw therefore carries the full covariance of the $C_k$ through to $F_\mathrm{AP}$.

\subsection{Results}
\label{app:AP-result}

\begin{table}
    \centering
        \begin{tabular}{lcccc}
        \hline
        Sample & $z$ & SNIa $F_{\rm AP}$ & DESI DR2 $F_{\rm AP}$ & Pull \\ \hline
        DES        & $0.510$ & $0.5853\pm0.0081$ & $0.6215\pm0.0171$ & $-1.91$ \\
                   & $0.706$ & $0.8948\pm0.0186$ & $0.8918\pm0.0208$ & $+0.11$ \\
                   & $0.934$ & $1.4150\pm0.0927$ & $1.2230\pm0.0192$ & $+2.03$ \\ \hline
        DES        & $0.510$ & $0.5847\pm0.0073$ & $0.6215\pm0.0171$ & $-1.98$ \\
        + legacy   & $0.706$ & $0.8935\pm0.0170$ & $0.8918\pm0.0208$ & $+0.06$ \\
                   & $0.934$ & $1.3677\pm0.0792$ & $1.2230\pm0.0192$ & $+1.78$ \\ \hline
    \end{tabular}
    \caption{Result of the AP consistency test between DESI~DR2 BAO and DES-Dovekie SNIa. Assuming only that the SNIa distance--redshift relation is smooth, we fit $H(z)$ to the SNIa alone and predict $F_{\rm AP} \equiv D_{\rm M}/D_{\rm H}$ at the three DESI~BAO redshifts, all of which lie inside the SNIa range. The test carries no absolute calibration and no sound horizon assumption.}
    \label{tab:AP-new}
\end{table}

Table~\ref{tab:AP-new} lists the SNIa-predicted and DESI-measured $F_{\rm AP}$ at the three redshifts suitable for a comparison. We find that the SNIa constraint on the shape of the expansion history predicts AP ratios that are in good agreement with BAO measurements, as quantified by $\chi^2 = 6.33$ for DES and 6.13 for DES + legacy. The corresponding $p$-values of a higher $\chi^2$ are 9.7\% and 10.5\%, respectively. These mild tensions indicate that DES SNIa and DESI BAO give consistent results for the shape of the expansion history. This agreement cannot be manufactured by a shared calibration, making it a genuine predictive success of the SNIa-derived shape.

Despite the overall agreement, the SNIa prediction at $z = 0.510$ falls $\approx 2\sigma$ below DESI (Table~\ref{tab:AP-new}). This feature is robust as it carries the same sign and size in both samples. It is also stable against the polynomial order and whether we bin the SNIa in redshift. Although a $2\sigma$ tension is not significant on its own, there is also a similar tension of the opposite sign at $z = 0.934$, while the pull at $z = 0.706$ is negligible. The tension is further mitigated by correlations between $F_\mathrm{AP}$ predictions at different redshifts, which we account for when calculating the total $\chi^2$.

\begin{figure}
    \centering
    \includegraphics[width=1.0\linewidth]{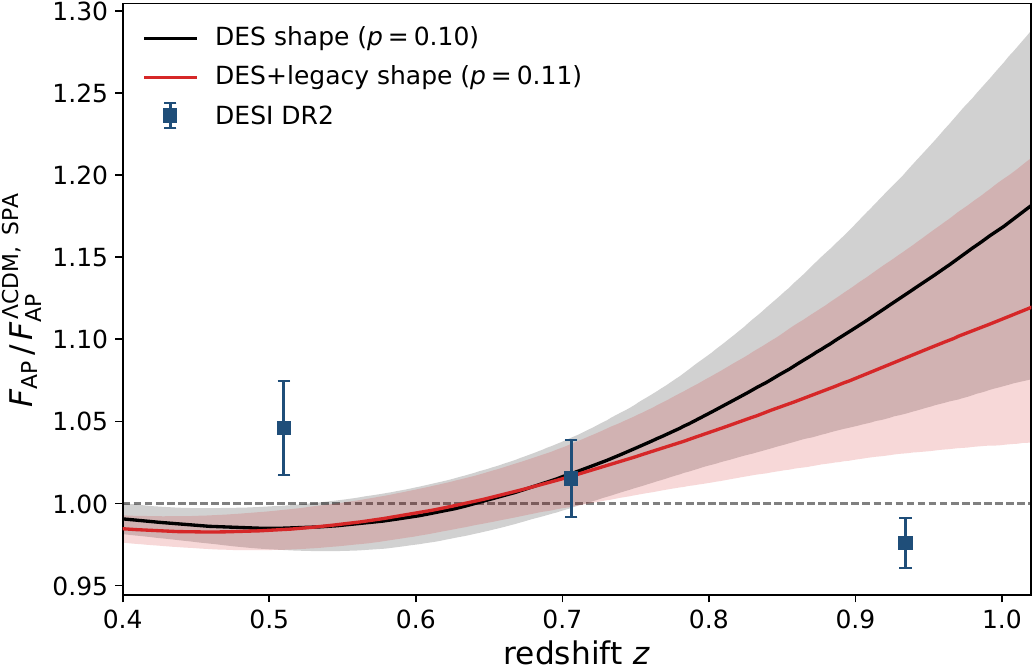}
    \caption{The shaded bands show the SNIa-shape mean prediction and $1\sigma$ uncertainty for the AP ratio $F_{\rm AP} \equiv D_{\rm M}/D_{\rm H}$
using either DES (black) or DES + legacy (red). The dark blue squares with error bars show DESI~DR2 measurements of $F_{\rm AP}$. The $F_{\rm AP}$ values shown here are relative to the predicted value at that redshift in the SPA cosmology \citep[$\Omega_{\rm m}=0.3175$;][]{SPT_2026}.}
    \label{fig:AP-new}
\end{figure}

Figure~\ref{fig:AP-new} shows the results relative to the fiducial SPA $\Lambda$CDM prediction. Given that BAO measurements are in reasonable agreement with the SPA cosmology \citep{Banik_2026_z0}, the AP consistency between SNIa and BAO at $z = 0.706$ also indicates consistency here with the CMB-derived expectation in $\Lambda$CDM. However, the BAO measurements of $F_\mathrm{AP}$ are not exactly in line with the
CMB-calibrated $\Lambda$CDM expectation \citep[cf. figure~3 of][which uses the \emph{Planck} cosmology]{Banik_2025_BAO}. This causes the SNIa-derived predictions to be in less tension with the SPA cosmology than they are with DESI~DR2 BAO -- though even here the tension is modest.

Despite the good overall agreement, SNIa and BAO appear to follow opposite trends in the ratio $F_\mathrm{AP}/F^{\rm{SPA}}_{\rm AP}$, which decreases with redshift when using BAO but increases with redshift when using SNIa. However, this is not too significant, especially given that the SNIa prediction becomes rather uncertain at $z > 0.8$, reducing the discrepancy with $\Lambda$CDM. Since the tension is largely in the slope rather than the normalisation, we can try to quantify this by fitting $F_{\rm SN}/F_{\rm BAO} = \alpha + \beta(z - z_{\rm p})$ over the three DESI redshifts, where $z_{\rm p} = 0.614$ is a suitable pivot redshift. This approach yields $\alpha - 1 = -0.019 \pm 0.021$ and $\beta = 0.41 \pm 0.18$ for DES, while $\alpha - 1 = -0.023 \pm 0.021$ and $\beta = 0.38 \pm 0.17$ for DES + legacy. The level is consistent with unity, but the shape tilt is around $2.2\sigma$ in both samples. This does not amount to a significant shape discrepancy. We cannot expect one because the total $\chi^2$ of the three residuals means that the tension cannot exceed $2.52\sigma$, even if it could be collapsed onto one parameter.

\section{Sweep over the cosmographic polynomial order}
\label{app:ordersweep}

\subsection{Results at a fixed order}
\label{app:fixedorder}

\renewcommand{\arraystretch}{1.2}
\begin{table*}
    \centering
    \setlength{\tabcolsep}{5pt}
    \begin{tabular}{cccccccccc}
    \hline
    $k_{\max}$ & $N_\mathrm{params}$ & $H_0 \; [\rm{km} \, \rm{s}^{-1} \, \rm{Mpc}^{-1}]$ & $\Delta M_\mathrm{legacy}$ [mag] & $\Mdes$ & $q_0$ & $\chi^2_{\min}$ & $\Delta \chi^2$ & $\Delta \ln Z$ & $w_k$ \\
    \hline
    \multicolumn{10}{c}{DES} \\ \hline
    $2$ & $5$ & $79.76 \pm 1.02$ & --- & $-19.413 \pm 0.012$ & $-2.60 \pm 0.07$ & $1634.24$ & $+209.74$ & $-97.05$ & $0.000$ \\
    $3$ & $6$ & $63.43 \pm 1.45$ & --- & $-19.381 \pm 0.012$ & $+1.31 \pm 0.38$ & $1454.47$ & $+29.96$ & $-10.33$ & $0.000$ \\
    $\mathbf{4}$ & $\mathbf{7}$ & $\mathbf{76.43 \pm 3.01}$ & --- & $\mathbf{-19.380 \pm 0.013}$ & $\mathbf{-3.01 \pm 0.73}$ & $\mathbf{1424.51}$ & $\mathbf{0.00}$ & $\mathbf{0.00}$ & $\mathbf{0.795}$ \\
    $5$ & $8$ & $68.52 \pm 4.99$ & --- & $-19.382 \pm 0.013$ & $+0.49 \pm 2.26$ & $1422.07$ & $-2.44$ & $-1.39$ & $0.198$ \\
    $6$ & $9$ & $63.30 \pm 7.75$ & --- & $-19.381 \pm 0.012$ & $+4.62 \pm 5.48$ & $1422.01$ & $-2.50$ & $-4.76$ & $0.007$ \\
    $\Lambda$CDM & $4$ & $68.98 \pm 0.48$ & --- & $-19.377 \pm 0.012$ & $-0.55 \pm 0.01$ & $1426.77$ & $+2.26$ & --- & --- \\
    BMA & --- & $74.77 \pm 4.83$ & --- & $-19.381 \pm 0.013$ & $-2.27 \pm 1.98$ & --- & --- & --- & --- \\
    \hline
    \multicolumn{10}{c}{DES + legacy} \\ \hline
    $2$ & $5$ & $69.43 \pm 0.59$ & --- & $-19.443 \pm 0.012$ & $-1.73 \pm 0.05$ & $2044.37$ & $+409.97$ & $-193.04$ & $0.000$ \\
    $3$ & $6$ & $66.18 \pm 0.60$ & --- & $-19.381 \pm 0.012$ & $+0.63 \pm 0.15$ & $1667.99$ & $+33.60$ & $-8.84$ & $0.000$ \\
    $4$ & $7$ & $67.99 \pm 0.71$ & --- & $-19.378 \pm 0.012$ & $-0.73 \pm 0.30$ & $1643.66$ & $+9.26$ & $-1.08$ & $0.250$ \\
    $\mathbf{5}$ & $\mathbf{8}$ & $\mathbf{66.14 \pm 0.90}$ & --- & $\mathbf{-19.380 \pm 0.012}$ & $\mathbf{+1.12 \pm 0.68}$ & $\mathbf{1634.40}$ & $\mathbf{0.00}$ & $\mathbf{0.00}$ & $\mathbf{0.734}$ \\
    $6$ & $9$ & $66.26 \pm 1.26$ & --- & $-19.381 \pm 0.012$ & $+0.94 \pm 1.30$ & $1634.33$ & $-0.07$ & $-3.80$ & $0.016$ \\
    $\Lambda$CDM & $4$ & $68.60 \pm 0.46$ & --- & $-19.379 \pm 0.012$ & $-0.54 \pm 0.01$ & $1644.95$ & $+10.56$ & --- & --- \\
    BMA & --- & $66.61 \pm 1.18$ & --- & $-19.379 \pm 0.012$ & \begin{tabular}[t]{@{}c@{}}$[-0.97, -0.48]$ \\ $\cup \; [+0.38, +1.84]$\end{tabular} & --- & --- & --- & --- \\
    \hline
    \multicolumn{10}{c}{DES + legacy ($\Delta M_\mathrm{legacy}$ free)} \\ \hline
    $2$ & $6$ & $79.44 \pm 0.99$ & $+0.282 \pm 0.021$ & $-19.412 \pm 0.012$ & $-2.57 \pm 0.07$ & $1862.92$ & $+230.25$ & $-103.14$ & $0.000$ \\
    $3$ & $7$ & $64.39 \pm 1.31$ & $-0.047 \pm 0.031$ & $-19.380 \pm 0.012$ & $+1.06 \pm 0.33$ & $1666.02$ & $+33.35$ & $-8.29$ & $0.000$ \\
    $4$ & $8$ & $70.92 \pm 1.99$ & $+0.063 \pm 0.039$ & $-19.378 \pm 0.012$ & $-1.51 \pm 0.55$ & $1640.66$ & $+7.98$ & $-0.14$ & $0.457$ \\
    $\mathbf{5}$ & $\mathbf{9}$ & $\mathbf{68.32 \pm 2.09}$ & $\mathbf{+0.045 \pm 0.038}$ & $\mathbf{-19.380 \pm 0.012}$ & $\mathbf{+0.41 \pm 0.91}$ & $\mathbf{1632.67}$ & $\mathbf{0.00}$ & $\mathbf{0.00}$ & $\mathbf{0.523}$ \\
    $6$ & $10$ & $68.02 \pm 2.13$ & $+0.043 \pm 0.040$ & $-19.380 \pm 0.012$ & $+0.65 \pm 1.30$ & $1632.66$ & $-0.01$ & $-3.23$ & $0.021$ \\
    $\Lambda$CDM & $5$ & $69.02 \pm 0.48$ & $+0.038 \pm 0.013$ & $-19.376 \pm 0.012$ & $-0.55 \pm 0.01$ & $1636.45$ & $+3.78$ & --- & --- \\
    BMA & --- & $69.50 \pm 2.43$ & $+0.053 \pm 0.040$ & $-19.379 \pm 0.012$ & \begin{tabular}[t]{@{}c@{}}$[-2.04, -0.64]$ \\ $\cup \; [-0.52, +0.63]$\end{tabular} & --- & --- & --- & --- \\
    \hline
    \end{tabular}
    \caption{Sweep over the truncation order $k_{\max}$ of our cosmographic expansion (Eq.~\ref{eqn:cosmographic-expansion}) in configuration~A. The column $w_k$ is the weighted posterior probability of order $k_{\max}$ (Eq.~\ref{eqn:order_weight}). The $\chi^2_{\min}$ column is the minimum $\chi^2$ at fixed $k_{\max}$ without any prior on the polynomial coefficients. $\Delta \chi^2$ is measured from the row with highest evidence (shown in bold), so the difference between $-\Delta \chi^2/2$ and $\Delta \ln Z$ is the change in the Occam penalty between the two orders. The third block repeats the DES + legacy analysis with free $\Delta M_\mathrm{legacy}$ (Section~\ref{sec:systematics-legacy}). The $\Lambda$CDM row of each block is our flat $\Lambda$CDM fit (Eq.~\ref{eqn:lcdm-expansion}), shown for reference only. Every row quotes the posterior mean and standard deviation, so a few entries differ slightly from the mode-based values in Table~\ref{tab:baseline}. The row labelled ``BMA'' is the Bayesian model average over $k_{\max} = 2$--6 (Eq.~\ref{eqn:bma}). At $k_{\max} = 4$, the nested sampling posteriors reproduce the NUTS results of Table~\ref{tab:baseline} to within 0.13 standard deviations in $H_0$. In the DES + legacy blocks, the BMA $q_0$ is bimodal, so we show its $1\sigma$ highest posterior density region as the union of two intervals.}
    \label{tab:ordersweep-ls-A}
\end{table*}
\renewcommand{\arraystretch}{1.0}

The results of our polynomial order sweep are shown in Tables~\ref{tab:ordersweep-ls-A} and \ref{tab:ordersweep-ls-B} for configurations~A and B, respectively. For DES alone, the second-order expansion gives a high $H_0$ similar to the fourth-order one, but $k_{\max} = 2$ does not fit the data (Section~\ref{sec:robustness_tests}). The $k_{\max} = 5$ result of $H_0 = 68.52 \pm 4.99\kmsMpc$, which is compatible with both SPA and SH0ES, is also consistent with $k_{\max} = 4$ at the $1.36\sigma$~C.L., while $k_{\max} = 6$ continues the trend towards a lower and less precise $H_0$. Since the evidence assigns 20\% of the probability to $k_{\max} = 5$, the fifth order cannot be dismissed, which is why we average over the orders in Appendix~\ref{app:blend}. We note that while the specific weights assigned to $k_{\max} = 4$ and 5 in the crucial DES-only case depend on the specific priors that we adopt (Table~\ref{tab:priors}), $\chi^2_{\min}$ improves by only 2.44 when going from the former to the latter. This is a very modest reduction given there are 1555 data points without the legacy SNIa, suggesting that any reasonable way of quantifying the Occam penalty would prefer $k_{\max} = 4$ when using DES alone. The situation is less clear when using DES + legacy because then the improvement in $\chi^2_{\min}$ is 9.26. Fortunately, the much better coverage of the legacy SNIa at low redshifts makes the choice of polynomial order far less relevant to the inferred cosmological parameters than for DES alone.

\renewcommand{\arraystretch}{1.2}
\begin{table*}
    \centering
    \setlength{\tabcolsep}{5pt}
    \begin{tabular}{cccccccccc}
    \hline
    $k_{\max}$ & $N_\mathrm{params}$ & $H_0 \; [\rm{km} \, \rm{s}^{-1} \, \rm{Mpc}^{-1}]$ & $r_{\rm d}$ [cMpc] & $\Mdes$ & $q_0$ & $\chi^2_{\min}$ & $\Delta \chi^2$ & $\Delta \ln Z$ & $w_k$ \\
    \hline
    \multicolumn{10}{c}{DES} \\ \hline
    $2$ & $5$ & $79.4 \pm 4.4$ & $148.25 \pm 8.13$ & $-19.427 \pm 0.119$ & $-2.60 \pm 0.07$ & $1641.13$ & $+209.60$ & $-97.00$ & $0.000$ \\
    $3$ & $6$ & $62.4 \pm 3.8$ & $149.96 \pm 8.42$ & $-19.421 \pm 0.121$ & $+1.28 \pm 0.38$ & $1460.19$ & $+28.65$ & $-10.17$ & $0.000$ \\
    $\mathbf{4}$ & $\mathbf{7}$ & $\mathbf{75.5 \pm 5.3}$ & $\mathbf{148.02 \pm 8.57}$ & $\mathbf{-19.391 \pm 0.125}$ & $\mathbf{-2.80 \pm 0.73}$ & $\mathbf{1431.54}$ & $\mathbf{0.00}$ & $\mathbf{0.00}$ & $\mathbf{0.834}$ \\
    $5$ & $8$ & $67.6 \pm 6.5$ & $148.38 \pm 8.40$ & $-19.398 \pm 0.123$ & $+0.74 \pm 2.38$ & $1428.82$ & $-2.71$ & $-1.67$ & $0.157$ \\
    $6$ & $9$ & $62.0 \pm 8.0$ & $148.03 \pm 7.93$ & $-19.393 \pm 0.115$ & $+5.23 \pm 5.73$ & $1428.80$ & $-2.74$ & $-4.54$ & $0.009$ \\
    $\Lambda$CDM & $4$ & $68.5 \pm 3.9$ & $148.48 \pm 8.39$ & $-19.395 \pm 0.122$ & $-0.55 \pm 0.01$ & $1433.06$ & $+1.52$ & --- & --- \\
    BMA & --- & $74.1 \pm 6.3$ & $148.08 \pm 8.54$ & $-19.392 \pm 0.125$ & $-2.18 \pm 1.94$ & --- & --- & --- & --- \\
    \hline
    \multicolumn{10}{c}{DES + legacy} \\ \hline
    $2$ & $5$ & $67.7 \pm 3.8$ & $151.50 \pm 8.49$ & $-19.503 \pm 0.122$ & $-1.73 \pm 0.05$ & $2054.26$ & $+413.29$ & $-193.93$ & $0.000$ \\
    $3$ & $6$ & $65.4 \pm 3.8$ & $149.24 \pm 8.58$ & $-19.410 \pm 0.124$ & $+0.62 \pm 0.14$ & $1673.44$ & $+32.47$ & $-7.89$ & $0.000$ \\
    $4$ & $7$ & $67.1 \pm 3.9$ & $149.43 \pm 8.59$ & $-19.410 \pm 0.125$ & $-0.70 \pm 0.30$ & $1650.02$ & $+9.05$ & $-0.26$ & $0.427$ \\
    $\mathbf{5}$ & $\mathbf{8}$ & $\mathbf{65.6 \pm 3.9}$ & $\mathbf{148.55 \pm 8.54}$ & $\mathbf{-19.399 \pm 0.125}$ & $\mathbf{+1.15 \pm 0.70}$ & $\mathbf{1640.97}$ & $\mathbf{0.00}$ & $\mathbf{0.00}$ & $\mathbf{0.553}$ \\
    $6$ & $9$ & $65.4 \pm 4.1$ & $149.40 \pm 8.87$ & $-19.412 \pm 0.129$ & $+0.91 \pm 1.30$ & $1640.84$ & $-0.13$ & $-3.34$ & $0.020$ \\
    $\Lambda$CDM & $4$ & $68.1 \pm 4.0$ & $148.51 \pm 8.68$ & $-19.397 \pm 0.127$ & $-0.54 \pm 0.01$ & $1651.18$ & $+10.21$ & --- & --- \\
    BMA & --- & $66.2 \pm 4.0$ & $148.94 \pm 8.58$ & $-19.404 \pm 0.125$ & \begin{tabular}[t]{@{}c@{}}$[-1.10, -0.26]$ \\ $\cup \; [+0.59, +1.67]$\end{tabular} & --- & --- & --- & --- \\
    \hline
    \end{tabular}
    \caption{As Table~\ref{tab:ordersweep-ls-A}, but for configuration~B in which the sound horizon is inferred from the CC data without CMB input.}
    \label{tab:ordersweep-ls-B}
\end{table*}
\renewcommand{\arraystretch}{1.0}

If we replace our model-independent cosmographic expansion with flat $\Lambda$CDM, the resulting $\chi^2_{\min}$ is only 2.3 (1.3) above that at $k_{\max} = 4$ for DES (DES + legacy). Since $\Lambda$CDM has three fewer shape parameters than a quartic, this might suggest that $\Lambda$CDM is the preferred description of the data. However, it is important to stress that the cosmographic expansion should not be seen as a rival model to $\Lambda$CDM, but rather as a Taylor series which can describe any expansion history over a given redshift range. For this reason, we do not include $\Lambda$CDM in the average over orders, none of which can exactly match $\Lambda$CDM for finite $k_{\max}$. We instead compare $\Lambda$CDM to specific alternatives using the Bayesian evidence in Section~\ref{sec:Model_comparison}. An important caveat to our results is that the intrinsic scatter term in the SNIa covariance is calibrated so that $\Lambda$CDM provides an acceptable fit to the Hubble diagram, since the intrinsic scatter is difficult to obtain any other way. This means a reduced $\chi^2$ close to unity is not necessarily a success of the $\Lambda$CDM model, justifying our focus on relative rather than absolute likelihoods.

\subsection{Bayesian model average over the polynomial order}
\label{app:blend}

Appendix~\ref{app:fixedorder} shows that we cannot confidently confirm the Hubble tension with the DES IDL because the predicted $H_0$ depends on the order. Rather than selecting one order, we treat $k_{\max}$ as a discrete parameter and marginalise over it. The posterior of any quantity $\theta$ is then the BMA, defined as follows:
\begin{eqnarray}
    P \left( \theta \, | \, D \right) ~=~ \sum_{k_{\max} = 2}^{6} w \left( k_{\max} \right) \, P \left( \theta \, | \, D, k_{\max} \right),
    \label{eqn:bma}
\end{eqnarray}
where $P(\theta \, | \, D, k_{\max})$ is the posterior at order $k_{\max}$ and $w(k_{\max})$ is the weighted posterior probability of that order (Eq.~\ref{eqn:order_weight}). We quote the mean and standard deviation of Eq.~\ref{eqn:bma} in the BMA rows in Tables~\ref{tab:ordersweep-ls-A} and \ref{tab:ordersweep-ls-B}. Its variance is the probability-weighted average of the variances at each order plus the variance of their central values, so the average carries the uncertainty in the order as well as the uncertainty at each order. The only exception is $q_0$ for DES + legacy, whose posterior is bimodal because $k_{\max} = 4$ and 5 place it on opposite sides of zero. For these cases, we report the two intervals that together comprise the $1\sigma$ highest posterior density region.

The first result to highlight is that the BMA is dominated by $k_{\max} = 4$ and 5 in every case. The lower orders are excluded by at least 7.9~nats, while $k_{\max} = 6$ never carries more than 2.1\% of the probability. Configuration~B ranks the orders in the same way, with $k_{\max} = 4$ preferred for DES and $k_{\max} = 5$ marginally preferred for DES + legacy (Table~\ref{tab:ordersweep-ls-B}).

The BMA changes the interpretation of the DES result more than that of any other sample. For DES alone, the tension with SPA falls to $1.56\sigma$, whereas DES + legacy remains $4.50\sigma$ below SH0ES (Section~\ref{sec:robustness_tests}). The DES average remains consistent with SH0ES at the $0.33\sigma$~C.L. Once the legacy SNIa are recalibrated, the average is consistent with SPA at the $0.94\sigma$~C.L. and with SH0ES at the $1.43\sigma$~C.L. The recalibration also raises the averaged $H_0$ by $2.9\kmsMpc$, which is similar to the $\approx 3\kmsMpc$ shift at $k_{\max} = 4$ (Section~\ref{sec:systematics-results}). Thus, the BMA confirms that a late-time solution to the Hubble tension remains possible insofar as we exclude or recalibrate the legacy SNIa.

Unlike a least-squares fit, the evidence depends on the prior, so it is important to check how much the weights rely on it. Our prior is uniform over all expansion histories with
$0 \leq E(\widetilde{a}) \leq 2 + 7\widetilde{a}$ (Section~\ref{sec:setup}). Halving the width of this band about the line
$E = 1 + 3.5\widetilde{a}$ is a linear contraction in the $C_k$, so it divides the prior volume at order $k$ by exactly $2^k$. Provided the posterior lies inside the narrower band, $\ln Z$ then rises by $k \ln 2$, so the evidence ratio between consecutive orders shifts by $\ln 2 \approx 0.69$ in favour of the higher order (the Occam penalty is lower due to the narrower prior range). We find that at least 99\% of the posterior lies inside the halved width band for every sample at $k_{\max} \geq 3$, so this shift applies almost exactly. Halving the band moves the DES weights at $k_{\max} = 4$ and 5 from 0.80 and 0.20 to 0.65 and 0.33, respectively, while the DES + legacy weights shift from 0.25 and 0.73 to 0.14 and 0.82. However, the averaged $H_0$ moves by at most $1.2\kmsMpc$ and is now $73.56 \pm 5.60 \kmsMpc$ for DES, $66.41 \pm 1.10 \kmsMpc$ for DES + legacy, and $69.05 \pm 2.38 \kmsMpc$ with the legacy SNIa recalibrated. These lie $1.13\sigma$ above SPA, $4.85\sigma$ below SH0ES, and $1.63\sigma$ below SH0ES, respectively, so our conclusions are not driven by the width of the prior. This highlights the advantage of a prior on the normalised expansion history itself rather than independent priors on each $C_k$, whose widths are difficult to set but would nonetheless set the Occam penalty of every order separately.

\subsection{Marginal priors on the cosmographic coefficients}
\label{app:prior}

Although our prior is uniform in the $C_k$ over the allowed region, the envelope of Eq.~\ref{eqn:envelope} couples the different orders,
making the marginal prior on each $C_k$ bounded and unimodal rather than uniform. However, it is very broad, with $\sigma_{C_1} = \sigma_{q_0} = 17.3$, $\sigma_{C_2} = 148$, $\sigma_{C_3} = 378$, and $\sigma_{C_4} = 293$, where $\sigma_x$ is the standard deviation of the prior on $x$. This is $24$--$73\times$ the width of the corresponding posterior in all four of our cosmographic analyses (DES and DES + legacy in configurations~A and B). As a result, the marginal prior density changes by less than 4\% across the $\pm 2\sigma$ range of any posterior. This is because every posterior sample stays at least 0.8 in $E$ from both limits of Eq.~\ref{eqn:envelope}. Thus, our prior neither truncates the posteriors nor appreciably shapes them.

We obtain virtually the same results with other priors. In configuration~A, using independent uniform priors on the dimensional coefficients $H_0 C_k$ over ranges much wider than the posteriors gives $H_0 = 76.83 \pm 2.88 \kmsMpc$ for DES and $68.02 \pm 0.70 \kmsMpc$ for DES + legacy, against $76.25 \pm 2.87$ and $67.98 \pm 0.69 \kmsMpc$, respectively, with our baseline prior. Moreover, a least-squares fit in the standard cosmographic parameters $q_0$, $j_0$, $s_0$, and $l_0$, which places no limits on them, returns the same $H_0$ as the least-squares fit in the $C_k$ (Section~\ref{sec:robustness_tests}). Even so, we adopt the envelope because it is simpler to implement and its edges are known in advance, since they are set by the expansion history itself rather than by a separate range on each $C_k$ that would have to be tuned for every $k_{\max}$ and then checked against the posteriors.

\section{The compressed CMB likelihood}
\label{app:compCMB}

Here we give the details of the compressed CMB likelihood summarised in Section~\ref{sec:comp_CMB}. The mean $\bm{\mu}_{\rm CMB}$ and covariance $\matr{C}_{\rm CMB}$ in Eq.~\ref{eqn:CMB_compressed} were obtained from \citet{Banik_2026_z0} from the chains of \citet{SPT_2026}. The methodology involves computing the auto-correlation time $\tau_i$ of each parameter $i$ in $\bm{x}$, finding $\max(\tau_i)$ across these parameters, and removing $3\max(\tau_i)$ steps from the start of the chains to account for burn-in \citep{Najera_2026}.

We follow \citet{Banik_2026_z0} in assuming three neutrino species with masses of $(50, 10, 0)$~meV$/c^2$ and $N_\mathrm{eff} = 3.044$, with the two massive species changing from relativistic to non-relativistic behaviour with a sharpness of $k_\nu = 2$ (see their appendix~B). The photon density follows from blackbody physics at the CMB temperature of $T_{\rm CMB} = 2.72548$~K \citep{Fixsen_2009}. The matter density relevant at late times is thus $\Omega_{\rm m} = \Omega_{\rm bc} + \Omega_\nu$, which includes the contribution $\Omega_\nu$ of massive neutrinos. We note that radiation changes $H(z)$ by $<0.04\%$ over the redshift range of the SNIa, BAO, and CC, so it matters only for the CMB, where its impact is substantial.

We compute $r_\star$ using the analytic result of \citet{Hu_1996}:
\begin{eqnarray}
    r_\star ~=~ \frac{2c}{3} \sqrt{\frac{3 \, a_{\rm eq}}{\Omega_{\rm bc} H_0^2 \mathcal{R}_{\rm eq}}} \ln \left( \frac{\sqrt{1 + \mathcal{R}_\star} + \sqrt{\mathcal{R}_\star + \mathcal{R}_{\rm eq}}}{1 + \sqrt{\mathcal{R}_{\rm eq}}} \right),
    \label{eqn:r_star}
\end{eqnarray}
where the subscript eq denotes matter--radiation equality (treating neutrinos as radiation) and $\mathcal{R} \equiv 3\rho_{\rm b}/(4\rho_\gamma)$ is the ratio between the baryon and photon densities, which sets the sound speed of the photon--baryon fluid. The BAO sound horizon $r_{\rm d}$ follows from the same expression evaluated at the end of the drag epoch, i.e. using $\mathcal{R}_{\rm d}$ instead of $\mathcal{R}_\star$. This requires the scale factors at recombination and at the end of the drag epoch, which we obtain from $a_\star = T_{\rm CMB}/T_\star$ and $a_{\rm d} = T_{\rm CMB}/T_{\rm d}$, respectively, where $T$ denotes the photon temperature at the epoch indicated by the subscript. We find $T_\star$ and $T_{\rm d}$ using quadratic fits to a grid of \textsc{class} runs \citep{CLASS}, as described in section~2.6 and appendix~A of \citet{Banik_2026_z0}:
\begin{eqnarray}
    T_\star &=& T_\star^{\rm ref} + \boldsymbol{d}^{\rm T} \boldsymbol{g}_\star + \frac{1}{2} \boldsymbol{d}^{\rm T} \matr{H}_\star \boldsymbol{d}, \nonumber \\
    \boldsymbol{d} &\equiv& \left( \omega_{\rm b}, \, \omega_{\rm bc} \right)^{\rm T} - \left( 0.0225, \, 0.1450 \right)^{\rm T},
    \label{eqn:T_star}
\end{eqnarray}
where $T_\star^{\rm ref} = 2970.29$~K. The temperature $T_{\rm d}$ at the end of the drag epoch is found in the same way, with $T_{\rm d}^{\rm ref} = 2892.74$~K. The gradients and Hessians are as follows, with temperatures in K:
\begin{eqnarray}
    \boldsymbol{g}_\star &=& \begin{pmatrix} -3188.92 \\ 194.09 \end{pmatrix}, ~~ \matr{H}_\star = \begin{bmatrix} 265960.27 & -7860.93 \\ -7860.93 & -488.36 \end{bmatrix}, \nonumber \\
    \boldsymbol{g}_{\rm d} &=& \begin{pmatrix} 5925.64 \\ 203.34 \end{pmatrix}, ~~ \matr{H}_{\rm d} = \begin{bmatrix} -189076.80 & -4767.87 \\ -4767.87 & -626.49 \end{bmatrix}.
    \label{eqn:T_fits}
\end{eqnarray}
These fits reproduce the \textsc{class} temperatures to within 0.8~mK across the grid, a negligible fraction of $T_\star \approx 2970$~K.

We stress that Eq.~\ref{eqn:r_star} neglects dark energy before recombination. This is an excellent approximation for $\Lambda$CDM and the \texttt{exp} model, but not for CPL models with $w_0 + w_a > 0$ as the dark energy density in such models grows faster than the matter density towards high redshift. However, the posterior assigns these models a probability below $10^{-34}$ in every case we consider, so this approximation has no impact on our results.

\section{Full posteriors}
\label{app:triangles}

\begin{figure}
    \centering
    \includegraphics[width=1.0\linewidth]{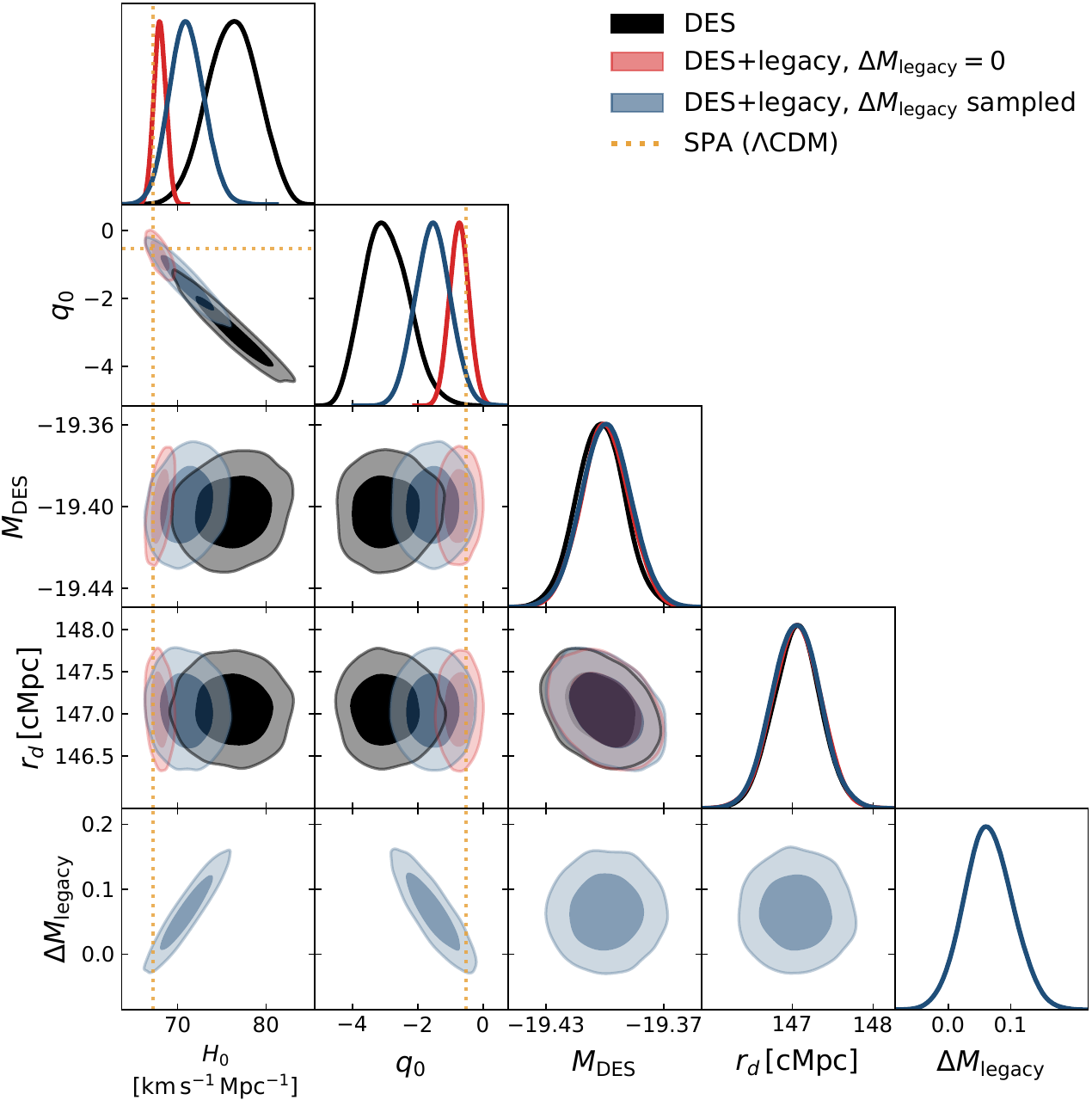}
    \caption{Triangle plot showing the main parameters in our baseline analysis, which uses a fourth-order cosmographic expansion with the \citetalias{Planck_2020} prior on $r_{\rm d}$ (configuration~A). The IDL is constrained using SNIa from DES (black), DES + legacy (red), or DES + legacy with free $\Delta M_\mathrm{legacy}$ (blue; Eq.~\ref{eqn:legacy_offset}). The results in the latter case are intermediate between the other two. The dotted orange lines mark $H_0$ and $q_0$ in flat $\Lambda$CDM with the SPA parameters, whose small uncertainties are neglected here \citep{SPT_2026}.}
    \label{fig:tri-legacy}
\end{figure}

Figure~\ref{fig:tri-legacy} shows the main parameters in our baseline analysis. If we include the legacy SNIa but allow them to be recalibrated relative to the main DES sample by $\Delta M_\mathrm{legacy}$, we obtain results intermediate between the DES-only and DES + legacy cases. The bottom right panel shows that this is driven by a preference for $\Delta M_\mathrm{legacy} > 0$, though it is probably $\la 0.1$~mag.

\begin{figure*}
    \vspace{0.5cm}
    \centering
    \begin{overpic}[width=0.49\linewidth]{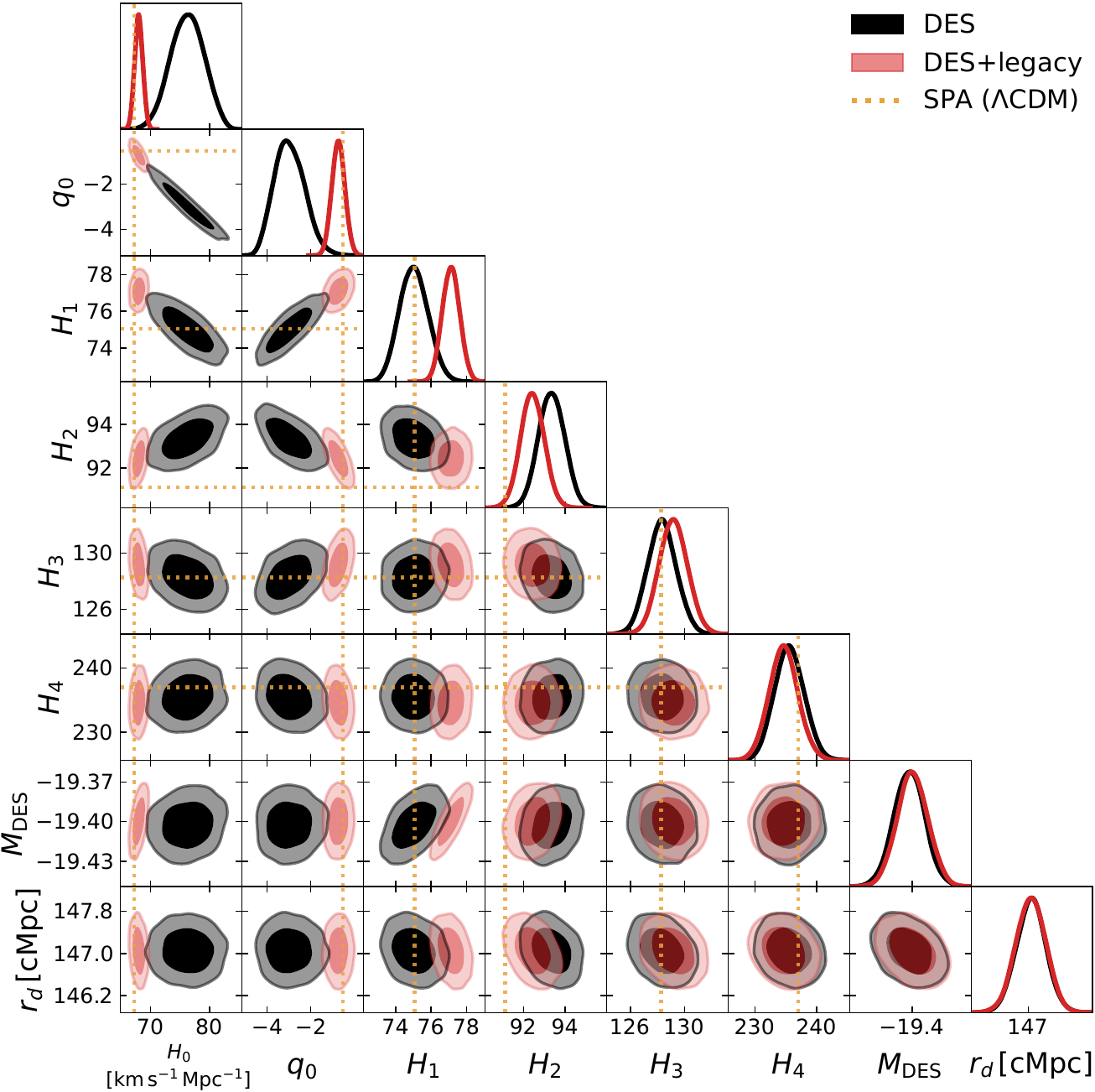}
        \put(15,101){\textbf{Cosmography with $k_{\max} = 4$, configuration A}}
    \end{overpic}
    \hfill
    \begin{overpic}[width=0.49\linewidth]{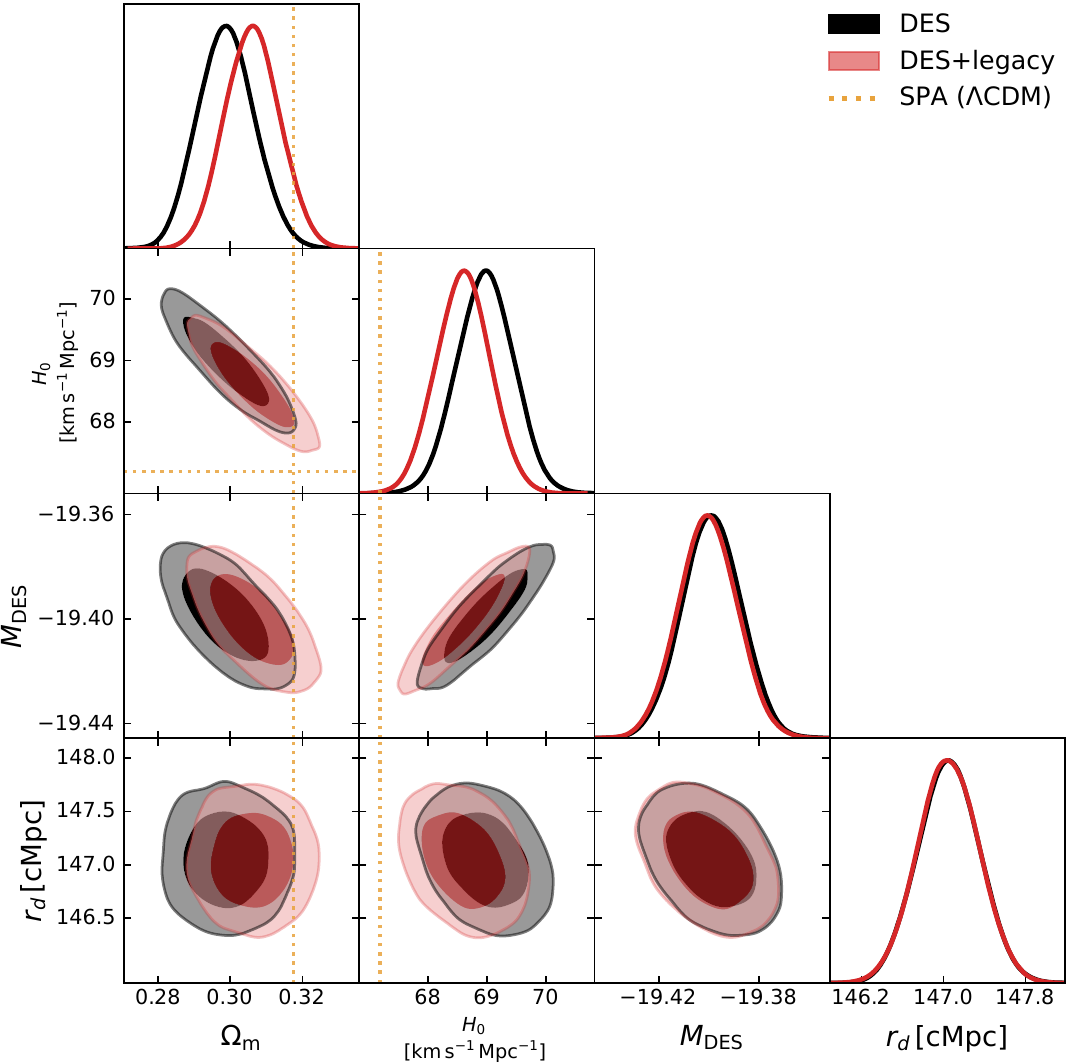}
        \put(35,101){\textbf{Flat $\Lambda$CDM, configuration A}}
    \end{overpic}

    \par\vspace{5ex}

    \begin{overpic}[width=0.49\linewidth]{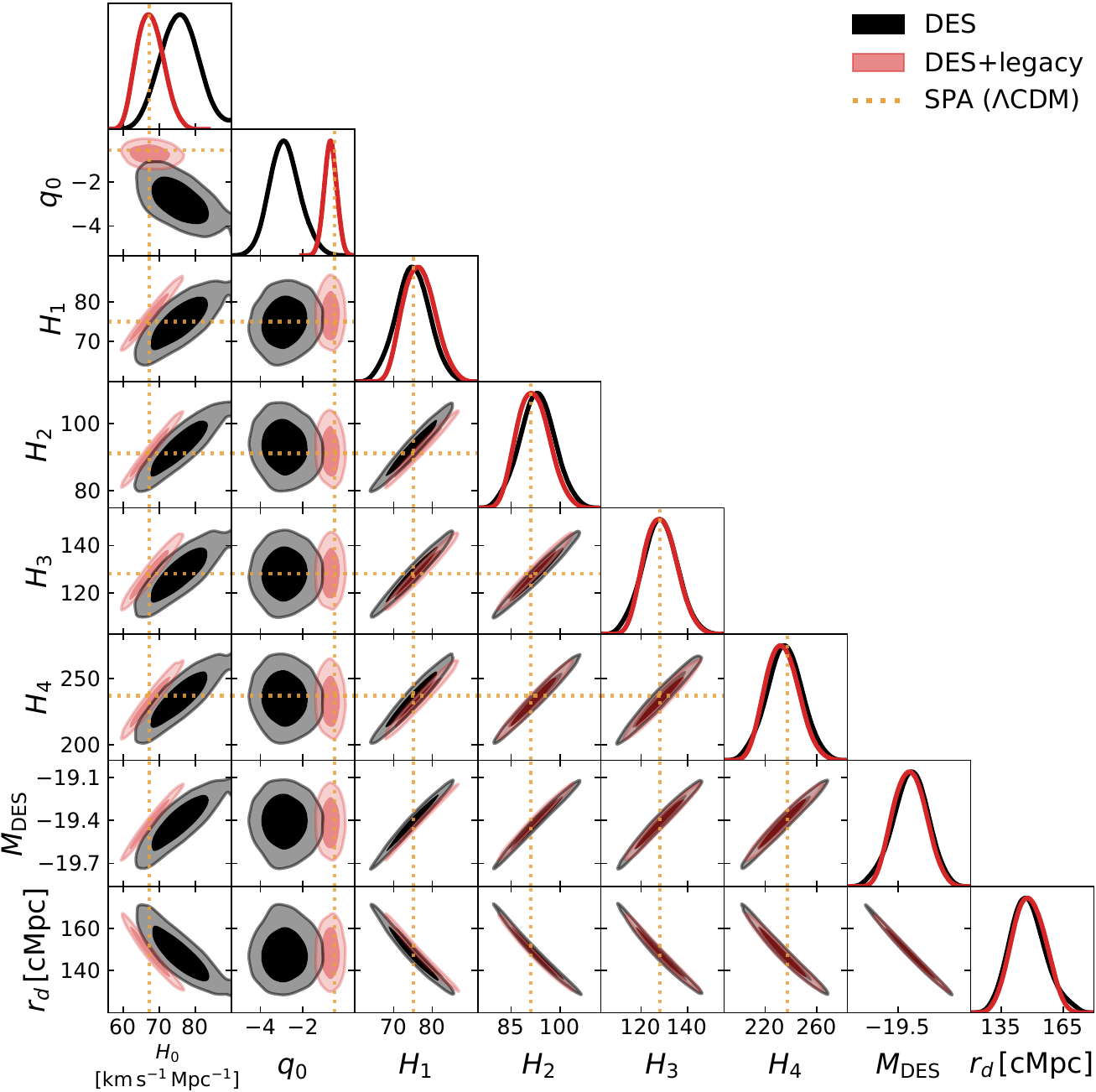}
        \put(15,101){\textbf{Cosmography with $k_{\max} = 4$, configuration B}}
    \end{overpic}
    \hfill
    \begin{overpic}[width=0.49\linewidth]{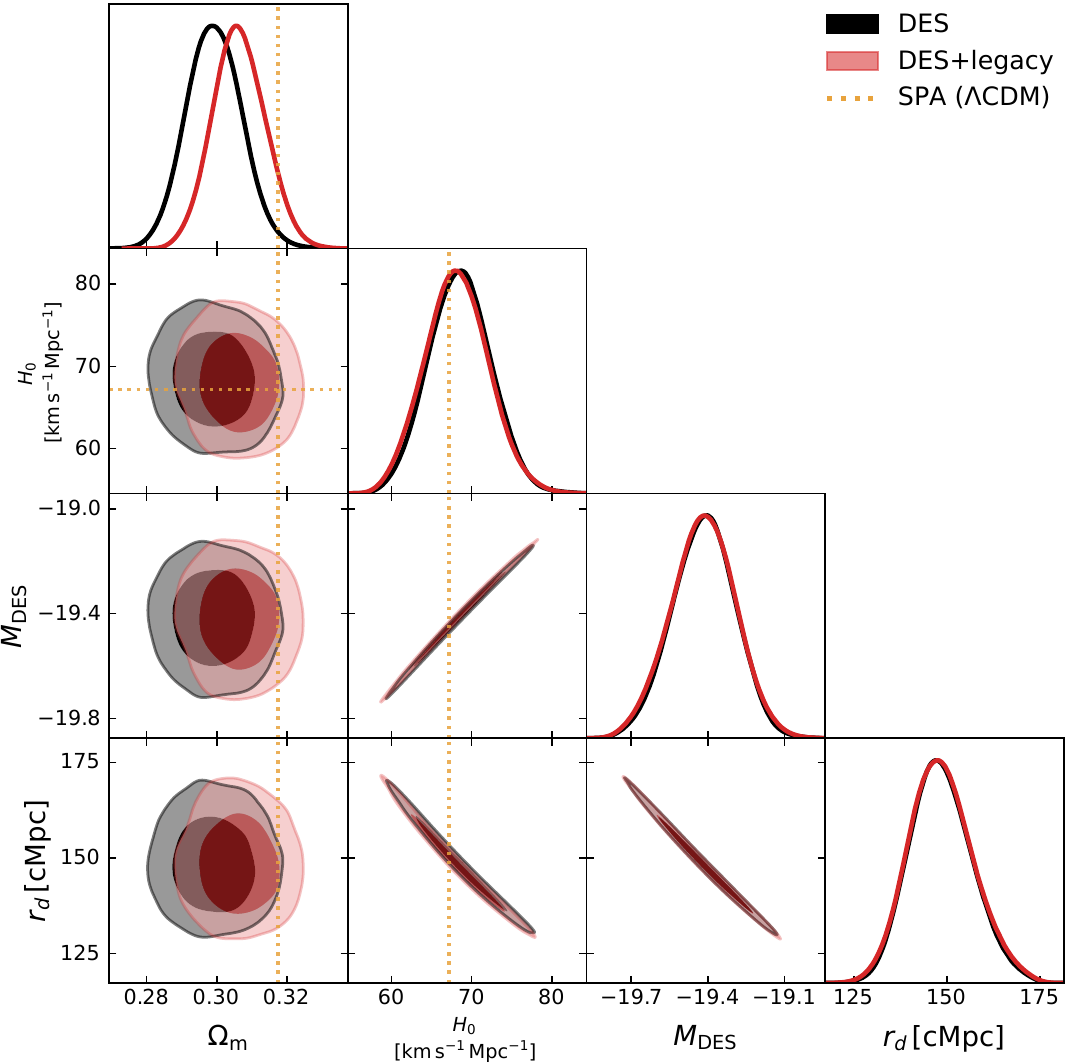}
        \put(35,101){\textbf{Flat $\Lambda$CDM, configuration B}}
    \end{overpic}
    \caption{Triangle plots for the main analyses presented in this paper, showing the $1\sigma$ and $2\sigma$ contours for DES (black) and DES + legacy (red). The left panels use our fourth-order cosmographic expansion (Eq.~\ref{eqn:cosmographic-expansion}), which we express through the expansion rate $H_k \equiv H(\widetilde{a}_{\max} k/k_{\max})$ at four points equally spaced in $\widetilde{a}$, which lie at $z = 0.21$, 0.54, 1.11, and 2.33 for $k = 1$--4. All $H_k$ are given in km~s$^{-1}$~Mpc$^{-1}$. We also show $q_0 = C_1 - 1$, though it is fully determined by the $H_k$. Over the $H_k$ ranges shown here, the priors are essentially flat. The right panels assume flat $\Lambda$CDM (Eq.~\ref{eqn:lcdm-expansion}). The dotted orange lines on some panels show predictions for the corresponding quantities in flat $\Lambda$CDM with the SPA parameters, neglecting their small uncertainties \citep{SPT_2026}. The top panels show results for configuration~A with a tight prior on $r_{\rm d}$ \citepalias{Planck_2020}. This is relaxed in the bottom panels, which show our configuration~B results where $r_{\rm d}$ is inferred from CC, making it far less precise. As a result, the $r_{\rm d}$ axis range differs between the top and bottom panels. Notice how the DES + legacy parameters in the cosmographic configuration~A analysis largely fall outside the range expected from DES alone.}
    \label{fig:tri-main}
\end{figure*}

Figure~\ref{fig:tri-main} shows the triangle plots for the main analyses in this paper, namely the cosmographic ($k_{\max} = 4$) and $\Lambda$CDM versions of the IDL constrained by DES and DES + legacy in configurations~A and B. The legacy SNIa do not much shrink the $\Lambda$CDM contours because $\Lambda$CDM already imposes a rigid shape to the expansion history, minimising the uncertainty associated with extrapolating across the DES coverage gap at $z \la 0.2$. In our more flexible cosmographic expansion, the legacy SNIa substantially shrink the parameter uncertainties. However, the preferred region is not expected on the basis of DES alone.
This emphasises that the legacy SNIa are likely subject to systematics, at least when recalibrated onto the DES filter system in order to enable a joint analysis.

\subsection{Late-time solutions}
\label{app:LTS}

\begin{figure}
    \centering
    \includegraphics[width=1.0\linewidth]{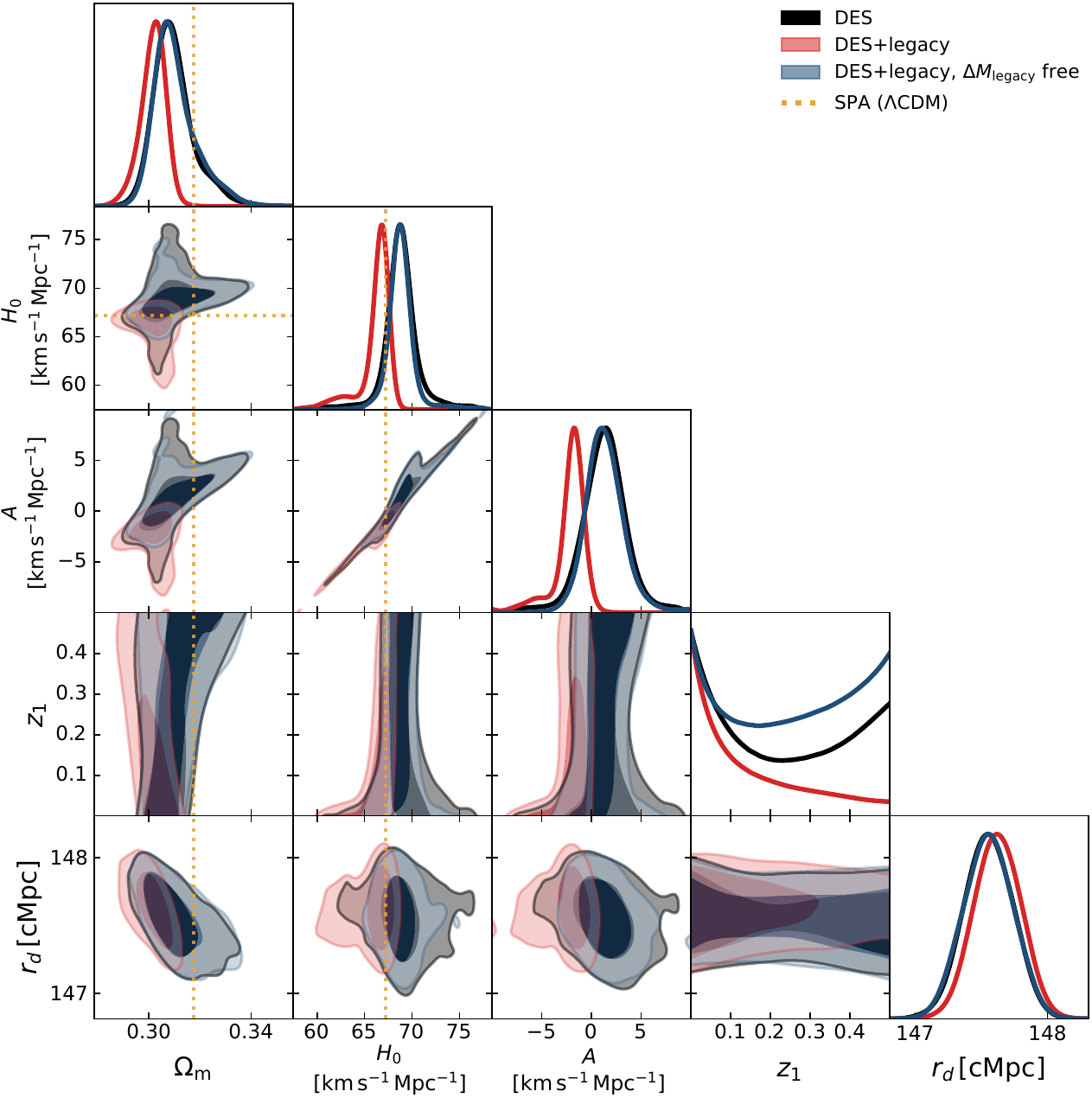}
    \caption{Triangle plot for the \texttt{exp} model (Eq.~\ref{eqn:H_z_extra}), proposed as a late-time solution to the Hubble tension \citep{Najera_2026}. Results are shown using SNIa from DES (black), DES + legacy (red), and DES + legacy with free $\Delta M_\mathrm{legacy}$ (blue). No $H_0$ anchors are considered here. As a result, the excess amplitude $A$ is consistent with zero and the redshift scale $z_1$ is unconstrained. The dotted orange lines mark $\Omega_{\rm m}$ and $H_0$ in flat $\Lambda$CDM with the SPA parameters, whose small uncertainties we neglect \citep{SPT_2026}.}
    \label{fig:tri-evidence-exp}
\end{figure}

In what follows, we only consider the \texttt{exp} model (Eq.~\ref{eqn:H_z_extra}) because the CPL model predicts a low $H_0$ and thus cannot be combined with local $H_0$ anchors (Table~\ref{tab:evidence}). We first show the triangle plots for the analyses unconstrained by any local $H_0$ anchor (Figure~\ref{fig:tri-evidence-exp}). As might be expected without any local anchor, the amplitude term $A$ is consistent with zero in all cases, though there is an interesting $1.3\sigma$ preference for $A < 0$ using DES + legacy. This is associated with a preference for lower $H_0$, making it even more difficult to solve the Hubble tension. If we allow recalibration of the legacy SNIa, the result is very similar to that using DES alone. Given the lack of preference for any excess $H_0$ at all, the redshift scale of the excess ($z_1$) is largely unconstrained.

\begin{figure}
    \centering
    \includegraphics[width=1.0\linewidth]{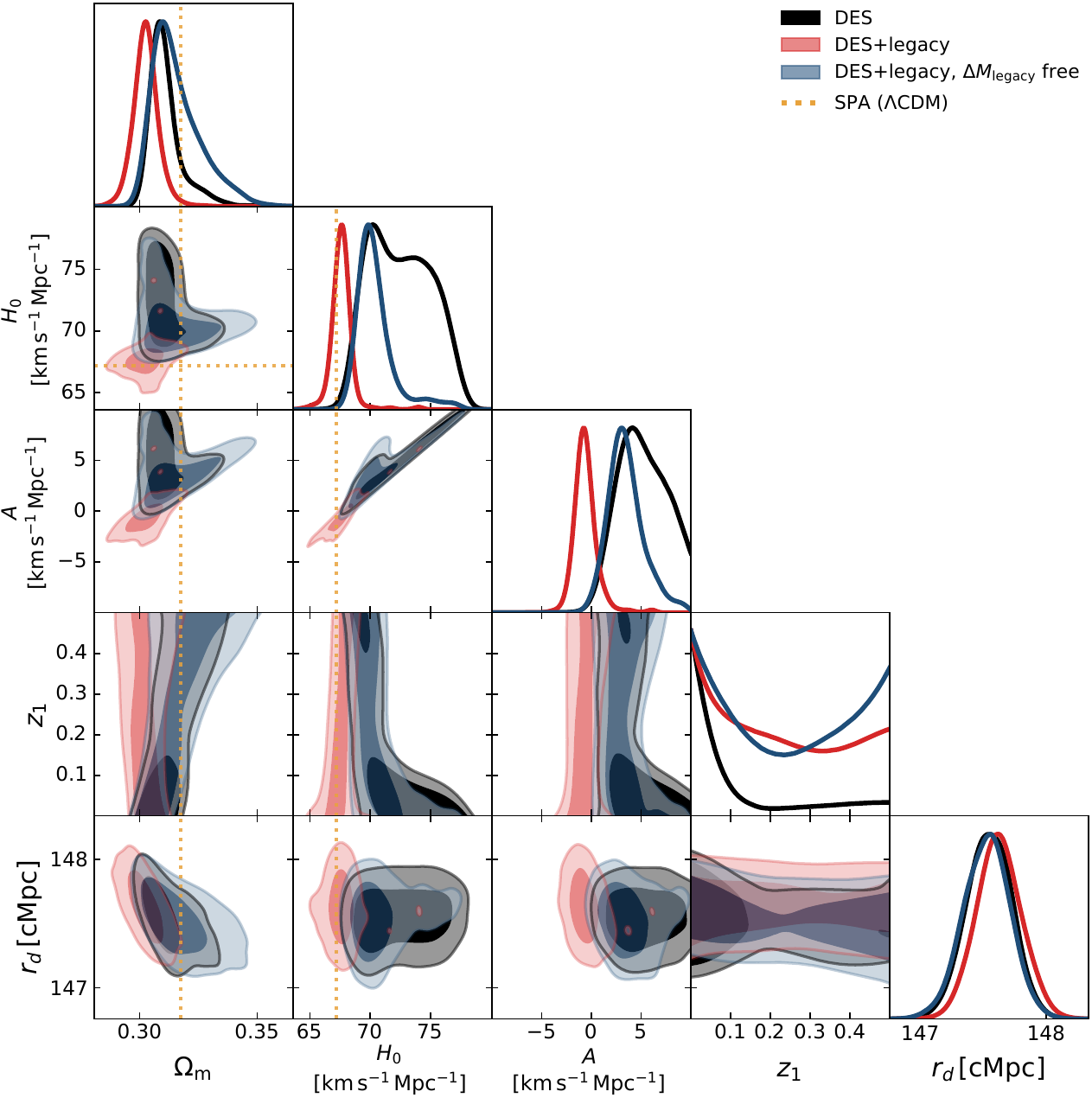}\\
    \includegraphics[width=1.0\linewidth]{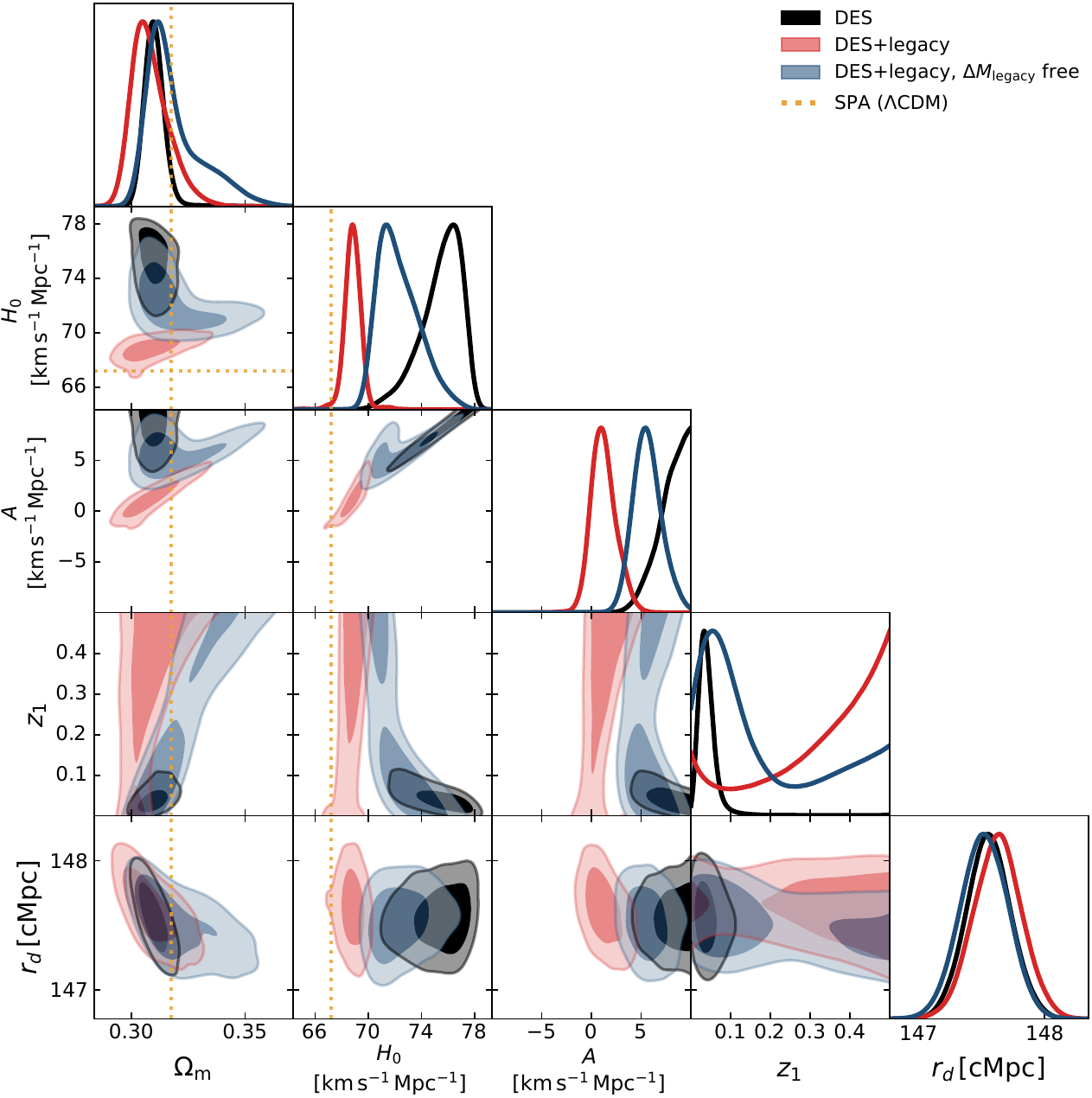}
    \caption{Similar to Figure~\ref{fig:tri-evidence-exp}, but including a local $H_0$ anchor (Section~\ref{sec:H0_anchors}). The \textit{top} panel jointly uses the three anchors not including SNIa \citep{Mukherjee_2021, Jensen_2025, Vogl_2025}. The \textit{bottom} panel uses the SH0ES anchor \citep{Breuval_2024}.}
    \label{fig:tri-evidence-exp-anchored}
\end{figure}

We next show the triangle plots when adding a local $H_0$ anchor (Figure~\ref{fig:tri-evidence-exp-anchored}). The top panel considers three anchors that do not involve SNIa, while the bottom panel uses the SH0ES anchor \citep{Breuval_2024}. These anchors are in principle independent but are consistent with each other, allowing us to combine all four anchors. We refrain from doing so because the statistical power would be dominated by SH0ES. Our results show that $z_1$ is not well constrained if we include the legacy SNIa, even if we allow them to be recalibrated. Excluding them leads to a tight upper limit on $z_1$. With the non-SNIa anchors, the $z_1$ posterior is truncated at low values only by our prior floor of 0.001. However, when using DES + SH0ES without the legacy SNIa, the $z_1$ posterior becomes sharply peaked and is not truncated by the prior on $z_1$. As we discuss in Section~\ref{sec:logz1}, this peak is set by the upper edge of our prior on $A$, so it is not a measurement of $z_1$.

\section{Log-uniform prior on the \texttt{exp} model $z_1$}
\label{app:logz1}

Table~\ref{tab:logz1} repeats each \texttt{exp} model analysis of Table~\ref{tab:evidence} with a uniform prior on $z_1$ and on $\ln z_1$, the latter being motivated in Section~\ref{sec:logz1}. Since several of these posteriors are markedly non-Gaussian, we compare the two priors using the posterior mean and standard deviation of $A$ and $H_0$, which can differ from the mode and highest posterior density interval quoted in Table~\ref{tab:evidence}. We note that a model with $z_1 < 0.001$ cannot realistically be expected to solve the Hubble tension because this would imply essentially no change to the $\Lambda$CDM distance--redshift relation in the SH0ES range \citep[$z = 0.023$--0.15;][]{Riess_2022_comprehensive}. The upper limit is less important because we will see that the results imply a low $z_1$.

\renewcommand{\arraystretch}{1.2}
\begin{table*}
    \centering
    \begin{tabular}{llcccccc}
        \hline
        Sample & Anchor & $z_1$ prior & $\ln B_{10}$ & $A \, [\rm{km} \, \rm{s}^{-1} \, \rm{Mpc}^{-1}]$ & $H_0 \, [\rm{km} \, \rm{s}^{-1} \, \rm{Mpc}^{-1}]$ & $H_0^{\rm win} \, [\rm{km} \, \rm{s}^{-1} \, \rm{Mpc}^{-1}]$ & $z_1$ \\
        \hline
        DES & \multirow{2}{*}{None} & $\mathcal{U}(z_1)$ & $-1.08$ & $+1.26 \pm 2.40$ & $68.83 \pm 2.02$ & $68.64 \pm 1.00$ & Unc. \\
        & & $\mathcal{U}(\ln z_1)$ & $-0.28$ & $+0.95 \pm 5.11$ & $68.93 \pm 5.02$ & $68.40 \pm 1.09$ & $<0.015$ \\
        \noalign{\smallskip}
        & \multirow{2}{*}{Excl.\ SNIa} & $\mathcal{U}(z_1)$ & $+1.54$ & $+5.10 \pm 2.41$ & $72.53 \pm 2.58$ & $70.23 \pm 0.99$ & $<0.086$ \\
         & & $\mathcal{U}(\ln z_1)$ & $+2.53$ & $+6.47 \pm 2.30$ & $74.23 \pm 2.31$ & $70.21 \pm 0.97$ & $0.013^{+0.018}_{-0.010}$ \\
        \noalign{\smallskip}
         & \multirow{2}{*}{SH0ES} & $\mathcal{U}(z_1)$ & $+10.61$ & $+8.00 \pm 1.46$ & $75.49 \pm 1.57$ & $71.93 \pm 0.67$ & $0.031^{+0.017}_{-0.011}$ \\
         & & $\mathcal{U}(\ln z_1)$ & $+11.26$ & $+8.25 \pm 1.32$ & $75.83 \pm 1.36$ & $71.87 \pm 0.64$ & $0.027^{+0.015}_{-0.009}$ \\
        \hline
        DES + legacy & \multirow{2}{*}{None} & $\mathcal{U}(z_1)$ & $+0.07$ & $-2.17 \pm 1.65$ & $66.21 \pm 1.70$ & $67.18 \pm 0.46$ & $<0.194$ \\
         & & $\mathcal{U}(\ln z_1)$ & $+1.21$ & $-4.20 \pm 2.89$ & $63.92 \pm 2.92$ & $67.27 \pm 0.43$ & $<0.019$ \\
        \noalign{\smallskip}
         & \multirow{2}{*}{Excl.\ SNIa} & $\mathcal{U}(z_1)$ & $-1.67$ & $-0.61 \pm 1.26$ & $67.66 \pm 1.01$ & $67.76 \pm 0.46$ & Unc. \\
         & & $\mathcal{U}(\ln z_1)$ & $-1.20$ & $-1.37 \pm 3.56$ & $66.77 \pm 3.53$ & $67.86 \pm 0.40$ & $<0.017$ \\
        \noalign{\smallskip}
         & \multirow{2}{*}{SH0ES} & $\mathcal{U}(z_1)$ & $-1.78$ & $+1.22 \pm 1.28$ & $68.84 \pm 0.70$ & $68.74 \pm 0.48$ & $>0.288$ \\
         & & $\mathcal{U}(\ln z_1)$ & $-1.16$ & $+0.14 \pm 3.85$ & $68.28 \pm 3.75$ & $68.41 \pm 0.40$ & $<0.021$ \\
        \hline
        DES + legacy & \multirow{2}{*}{None} & $\mathcal{U}(z_1)$ & $-1.22$ & $+1.35 \pm 1.99$ & $68.86 \pm 1.56$ & $68.59 \pm 0.81$ & Unc. \\
        ($\Delta M_\mathrm{legacy}$ free) & & $\mathcal{U}(\ln z_1)$ & $-0.65$ & $+0.32 \pm 4.38$ & $68.25 \pm 4.29$ & $68.18 \pm 0.72$ & $<0.016$ \\
        \noalign{\smallskip}
         & \multirow{2}{*}{Excl.\ SNIa} & $\mathcal{U}(z_1)$ & $+0.54$ & $+3.47 \pm 1.86$ & $70.44 \pm 1.72$ & $69.61 \pm 0.76$ & Unc. \\
         & & $\mathcal{U}(\ln z_1)$ & $+1.03$ & $+4.52 \pm 2.62$ & $72.29 \pm 2.72$ & $69.34 \pm 0.82$ & $<0.047$ \\
        \noalign{\smallskip}
         & \multirow{2}{*}{SH0ES} & $\mathcal{U}(z_1)$ & $+7.87$ & $+5.61 \pm 1.43$ & $72.30 \pm 1.58$ & $71.05 \pm 0.75$ & $<0.225$ \\
         & & $\mathcal{U}(\ln z_1)$ & $+7.81$ & $+5.91 \pm 1.42$ & $73.22 \pm 1.57$ & $71.25 \pm 0.71$ & $0.049^{+0.040}_{-0.029}$ \\
        \hline
    \end{tabular}
    \caption{Impact of the $z_1$ prior on the \texttt{exp} model results. We repeat each analysis in Table~\ref{tab:evidence} using a uniform prior on $z_1$ or on $\ln z_1$, the latter being more representative of a local structure due to its unknown size (see the text). All results shown here are relative to the $\Lambda$CDM control. For clarity, we omit the numerical uncertainty on $\ln B_{10}$ of $\pm 0.19$. To make the results with the two $z_1$ priors more comparable, we show the mean and standard deviation of $A$ and $H_0 = H_{0, \Lambda \mathrm{CDM}} + A$. $H_0^{\rm win}$ is the $H_0$ that SH0ES would infer if the Universe followed each posterior, i.e. the kinematic fit of Section~\ref{sec:H0_anchors} applied over the redshifts of the SH0ES Hubble-flow SNIa ($z = 0.023$--0.149).}
    \label{tab:logz1}
\end{table*}
\renewcommand{\arraystretch}{1.0}

Table~\ref{tab:logz1} shows the impact of changing the $z_1$ prior from $\mathcal{U}(z_1) \to \mathcal{U}(\ln z_1)$. We begin with the case without a local $H_0$ anchor, so the inferred $H_0$ can be considered a genuine prediction of the IDL approach constrained by SNIa + BAO + CC + CMB + BBN. With SNIa from DES alone, $A$ remains very consistent with zero for both priors, though the uncertainty rises from 2 to $5\kmsMpc$ when going from the uniform prior to the log-uniform prior. While these results are more in line with $A = 0$ and $H_0 = H_0^\mathrm{SPA}$, they cannot exclude the possibility that $A \approx 6\kmsMpc$ and $H_0 = H_0^\mathrm{SH0ES}$ -- especially with the log-uniform prior. The picture remains similar upon including the legacy SNIa with free $\Delta M_\mathrm{legacy}$, which again shows no preference for $A > 0$. The Bayesian evidences are also not appreciably different from that of $\Lambda$CDM.

The results are quite different with DES + legacy regardless of the prior choice. We infer $A<0$ with either $z_1$ prior. Although the Bayesian evidence no longer disfavours the \texttt{exp} model and actually prefers it with the log-uniform prior, this reflects a preference for $A < 0$, which is strongly incompatible with the $A \approx +6\kmsMpc$ required to solve the Hubble tension. This is evident in that the inferred $H_0$ is $66.21 \pm 1.70\kmsMpc$ with the uniform $z_1$ prior, which is lower than the $67.94 \pm 0.24\kmsMpc$ we obtain for flat $\Lambda$CDM from the same data. With the log-uniform prior on $z_1$, the inferred $H_0$ drops to just $63.92 \pm 2.92\kmsMpc$. In both cases, the results are considerably below the SH0ES measurement. We note that a sharp reduction in predicted $H_0$ upon including the legacy SNIa is also evident from our sweep of the redshift floor using a model-independent cosmographic expansion (Figure~\ref{fig:floorsweep-real-new}). We are able to recover this feature in the context of our \texttt{exp} model due to it allowing a sharp feature in the expansion history at $z \la z_1$. Despite the preference for $A < 0$, we are still not able to meaningfully constrain $z_1$ with DES + legacy if we adopt a uniform prior. With a log-uniform prior, we obtain $z_1 < 0.019$ at the $1\sigma$~C.L., which is considerably tighter than the $z_1 < 0.070$ given by the prior alone (Table~\ref{tab:logz1}). Since $A$ is still plausibly consistent with zero, we suggest that there is no clear preference for the \texttt{exp} model when we include the legacy SNIa at face value.

Once we impose a local $H_0$ anchor, the inferred $H_0$ is no longer a prediction, so the important quantity is now the Bayesian evidence. In five of the six anchored cases, the log-uniform prior raises $\ln B_{10}$ by 0.47--0.99 (Section~\ref{sec:logz1}), which exceeds the numerical uncertainty of $\pm 0.19$. The exception is DES + legacy with free $\Delta M_{\rm legacy}$ and the SH0ES anchor, for which $\ln B_{10}$ barely changes ($+7.87$ vs $+7.81$). The biggest preference for \texttt{exp} over $\Lambda$CDM is still with DES + SH0ES, which prefers the \texttt{exp} model by 11~nats. There is an almost 8~nat preference with DES + legacy (free $\Delta M_{\rm legacy}$) if we use the SH0ES anchor. With an anchor, the only preference for $\Lambda$CDM occurs with DES + legacy, which prefers $\Lambda$CDM by 1.2--1.8~nats with either anchor. Allowing free $\Delta M_{\rm legacy}$ reverses this, so that the SNIa-independent anchors then prefer the \texttt{exp} model by 0.5--1.0~nats. In the DES + legacy case with the SH0ES anchor, changing from a uniform prior to a log-uniform prior reduces $H_0$ from $68.84\kmsMpc$ to $68.28\kmsMpc$, demonstrating that even the tight SH0ES anchor cannot pull the legacy-dominated fit up to $H_0^\mathrm{SH0ES}$ regardless of the $z_1$ prior. In the DES-only case with SNIa-independent anchors, the clear preference for the \texttt{exp} model with the log-uniform prior is related to a rise in the posterior mean $H_0$ from $72.53\kmsMpc$ to $74.23\kmsMpc$. The impact of using a uniform prior on $\ln z_1$ rather than on $z_1$ is quite modest if we use the SH0ES anchor.

\section{Redshift-floor sweep in \texorpdfstring{$\Lambda$CDM}{LCDM} mocks}
\label{app:floor-sweep-mock}

To determine how plausible it is that $H_0^\mathrm{DES}$ exceeds $H_0^\mathrm{DES+legacy}$ by the observed $\Delta H_0 = 8.27 \kmsMpc$, we set up mock datasets assuming flat $\Lambda$CDM with the SPA parameters based on the latest CMB observations \citep[$H_0 = 67.19\kmsMpc$ and $\Omega_{\rm m} = 0.3175$;][]{SPT_2026} and the \citetalias{Planck_2020} sound horizon of $r_{\rm d} = 147.05$~cMpc. We neglect uncertainties in these parameters because we expect larger uncertainties from the SNIa, especially for DES alone due to the need to extrapolate to the present epoch. We generate mock BAO and SNIa data at the same redshifts as the real data, ensuring that every mock realisation of $\Lambda$CDM with the SPA parameters has the same redshift coverage as the real data.

We calculate `true' SNIa apparent magnitudes in the assumed cosmology as follows:
\begin{eqnarray}
    m_{\rm SNIa}^{\Lambda\mathrm{CDM}} = \Mdes + 5\log_{10}\left[ \left( 1 + z_{\rm hel} \right) \, D_{\mathrm{M}, \Lambda {\rm CDM}}(z_{\rm HD}) \right] + 25,
    \label{eqn:mock-mag}
\end{eqnarray}
where $\Lambda$CDM superscripts indicate that the quantity does not include observational errors, the redshifts are the same as observed, and $D_\mathrm{M}$ is the comoving distance to $z_{\rm HD}$ in the SPA cosmology. Since the absolute calibration of the SNIa is irrelevant to the inferred cosmological parameters, we adopt $\Mdes = -19.40$ and do not report $\Mdes$ results from our mocks. We add observational scatter based on the total covariance matrix $\matr{C}_{\rm SNIa}$ of the real data in the DES-Dovekie catalogue \citep{Popovic_2026_Dovekie}. The apparent magnitude is therefore as follows:
\begin{eqnarray}
    m_{\rm SNIa} ~\sim~ \mathcal{N} \left( m_{\rm SNIa}^{\Lambda\mathrm{CDM}}, ~\matr{C}_{\rm SNIa} \right).
\end{eqnarray}

We follow a similar approach with the BAO. We generate mock BAO data at the same redshifts as those with real data from DESI~DR2, propagating errors using their published covariance matrix $\matr{C}_{\rm BAO}$ at each redshift \citep[except at the lowest redshift where only $D_{\rm V}$ is available; see][]{DESI_2025}. The mock BAO measurements of $D_{\rm H, M, V}/r_{\rm d}$ are therefore as follows:
\begin{eqnarray}
    \frac{D_{\rm M,H,V}}{r_{\rm d}} ~\sim~ \mathcal{N} \left( \frac{D_{\rm M,H,V}^{\Lambda\mathrm{CDM}}}{r_{\rm d}}, ~\matr{C}_{\rm BAO} \right).
\end{eqnarray}
For both SNIa and BAO, we set up a single noiseless mock in which we do not add observational errors, so $m_{\rm SNIa} = m_{\rm SNIa}^{\Lambda\mathrm{CDM}}$ and $D_{\rm M,H,V}/r_{\rm d} = D_{\rm M,H,V}^{\Lambda\mathrm{CDM}}/r_{\rm d}$. We use the noiseless mock to separate the impact of observational errors and deficiencies in our fourth-order cosmographic expansion.

\begin{figure}
    \centering
    \includegraphics[width=1.0\linewidth]{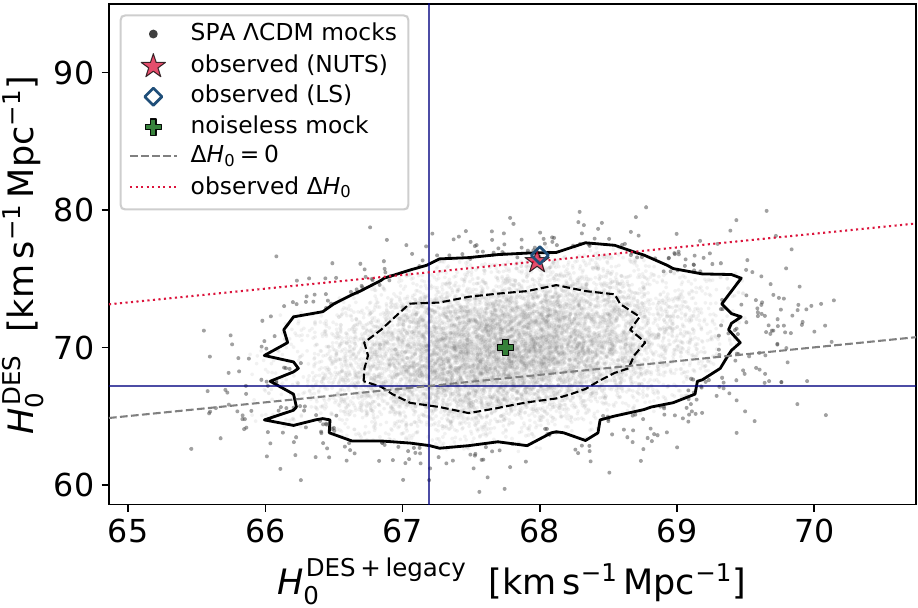}\\
    \vspace{0.3cm}
    \includegraphics[width=1.0\linewidth]{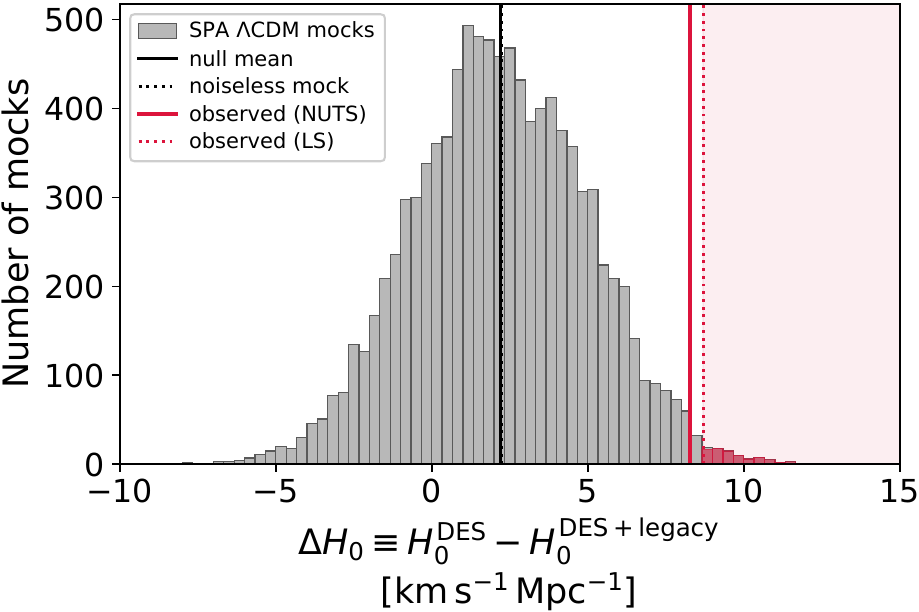}
    \caption{\textit{Top}: Joint distribution of $H_0^\mathrm{DES}$ and $H_0^\mathrm{DES+legacy}$, inferred from applying our fourth-order cosmographic IDL analysis to $10^4$ mock datasets assuming $\Lambda$CDM with the SPA parameters \citep{SPT_2026}. The true $H_0$ is indicated using solid lines. The contours show the $1\sigma$ and $2\sigma$ regions, outside which individual points are shown with higher opacity. The dashed grey line is the line of equality. The dotted red line is offset upwards by the observed $\Delta H_0 = 8.27\kmsMpc$ so it intersects the observed point, shown here as the red star. This is obtained using NUTS, but the open blue diamond shows that the corresponding least-squares result (labelled `LS') is almost identical, so we use least-squares for the mocks. The green plus shows the result of our noiseless mock. \textit{Bottom}: The distribution of $\Delta H_0 \equiv H_0^\mathrm{DES} - H_0^\mathrm{DES+legacy}$ from the same mocks. The solid black line shows the mean result, while the dotted black line shows the result in the noiseless mock. The solid red line shows the observed $\Delta H_0$ using NUTS, while the dotted red line shows the almost identical result using least-squares. The red shading at the right shows the region where $\Delta H_0$ exceeds the observed least-squares value of $8.70\kmsMpc$.}
    \label{fig:dh0-hist}
\end{figure}

We conduct $10^4$ realisations of the above setup. In each case, we extract $H_0^\mathrm{DES}$ and $H_0^\mathrm{DES+legacy}$ using the least-squares approach to reduce computational costs. The top panel of Figure~\ref{fig:dh0-hist} shows the joint distribution of these two $H_0$ estimates. It is immediately apparent that the observed combination is unlikely in the SPA cosmology. In our mocks, both recovered $H_0$ values are somewhat higher than the truth. This remains the case even in the noiseless mock. The bias is thus an inherent property of our cosmographic IDL, though it is small compared to the random scatter caused by observational errors. In particular, $\langle H_0^\mathrm{DES} \rangle$ is biased by $+2.76 \kmsMpc$ compared to the true value, but the standard deviation of the best-fitting $H_0^\mathrm{DES}$ between different mocks is $2.86 \kmsMpc$. $H_0^\mathrm{DES}$ is thus biased upwards by only $+0.96\sigma$. Due to the better redshift coverage of DES + legacy, it achieves a smaller bias: $\langle H_0^\mathrm{DES+legacy} \rangle$ exceeds the true value by only $+0.57 \kmsMpc$. Since the standard deviation of the best-fitting $H_0^\mathrm{DES+legacy}$ values is $0.69 \kmsMpc$, this represents a bias of $+0.82\sigma$. The results also show a weak positive correlation between $H_0^\mathrm{DES}$ and $H_0^\mathrm{DES+legacy}$,
with a correlation coefficient of $+0.25$. Higher $H_0^\mathrm{DES+legacy}$ is thus generally associated with a correspondingly higher $H_0^\mathrm{DES}$ (the regression of $H_0^\mathrm{DES}$ on $H_0^\mathrm{DES+legacy}$ has a slope of 1.04), albeit with large scatter. This correlation mitigates the impact on $\Delta H_0$.

The bottom panel of Figure~\ref{fig:dh0-hist} shows the distribution of $\Delta H_0$. Our noiseless mock sits near the mean of the mocks with noise, both of which indicate a small positive bias of around $+2.2\kmsMpc$ that must be due to truncation error. Since the mocks are fitted by least squares, the like-for-like statistic is the least-squares $\Delta H_0 = 8.70\kmsMpc$ of the real data, which only 85 of the $10^4$ mocks exceed (0.85\%). This is the result we quote in Section~\ref{sec:robustness_tests} and in the abstract. Comparing the NUTS value of $8.27\kmsMpc$ with the least-squares mocks would mix two estimators, modestly raising the fraction to 1.24\%. These results indicate that $H_0^\mathrm{DES}$ cannot easily be reconciled with the SPA cosmology, despite assuming standard pre-recombination physics to set $r_{\rm d}$ and not using the standard distance ladder.

\begin{figure*}
    \centering
    \includegraphics[width=0.48\linewidth]{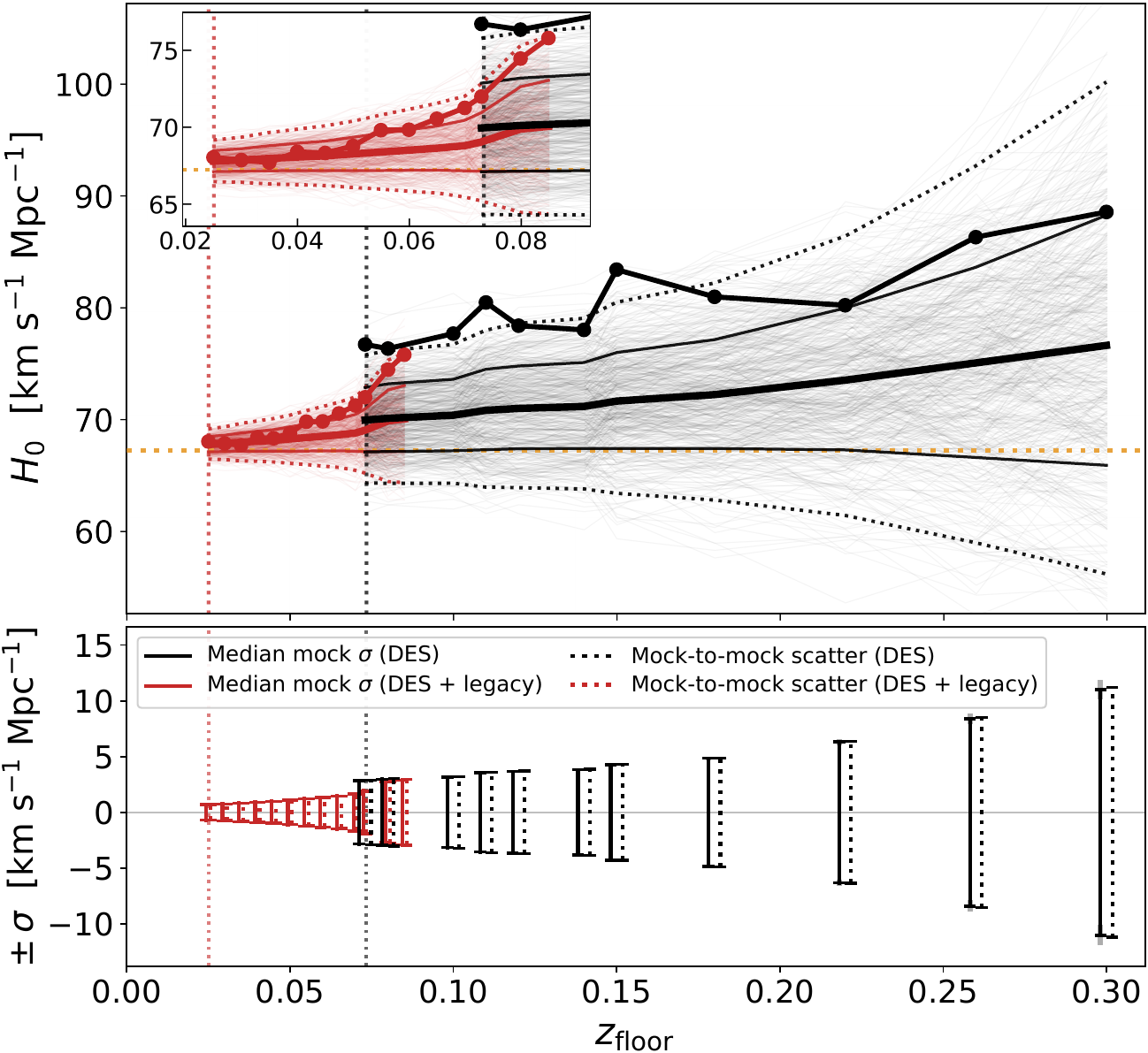}
    \includegraphics[width=0.5\linewidth]{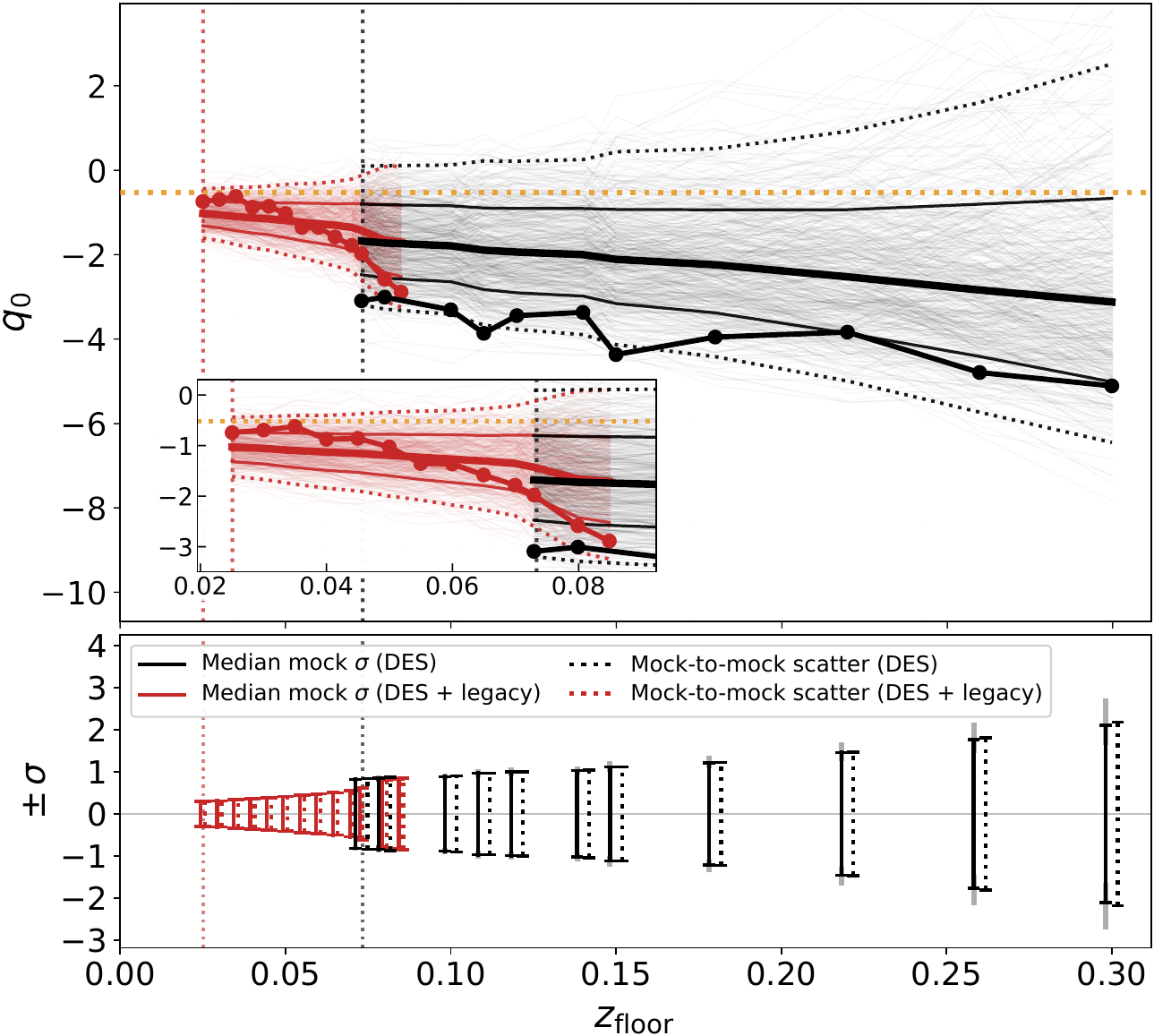}
    \caption{\textit{Top}: The results of our redshift-floor sweep for $H_0$ (\textit{left panels}) and $q_0$ (\textit{right panels}) in our $\Lambda$CDM mocks where the SNIa redshift distribution matches DES (black) and DES + legacy (red). The true values are shown using horizontal dotted orange lines. The thin lines show 500 of the mocks in each sample's own colour, using a fixed stride through the ensemble. The median of all $10^4$ mocks is shown as a thick solid line without points. The thin solid lines around it in the same colour show the $1\sigma$ range using Gaussian equivalent percentiles, while the dotted lines show the $2\sigma$ range. The result with the real data, obtained by least squares as for the mocks, is shown using a thick solid line with points. The inset focuses on the region covered by the legacy SNIa. Their redshift floor is shown with a vertical dotted red line, while the DES redshift floor is shown with a vertical dotted black line. We do not show the DES + legacy results beyond the redshift range of the legacy SNIa, as only DES SNIa are available there. \textit{Bottom}: The uncertainty is shown in two ways, both plotted as $\pm \sigma$ error bars around a central value of zero. The vertical scale is the same as the top panels. For clarity, the redshift is dithered symmetrically relative to the actually adopted $z_\mathrm{floor}$. The dotted bars show half the 16th--84th percentile range of the best-fitting value between mocks. The solid bars show the median value across the mocks of the uncertainty returned by our analysis from each mock. Since not all mocks yield the same uncertainty, we use shaded rectangles in the colour of each sample to show the $1\sigma$ Gaussian equivalent range in the uncertainty returned by each mock. These are only visible at high $z_\mathrm{floor}$ because different mocks give results with a very similar precision.}
    \label{fig:floorsweep-mock}
\end{figure*}

Our mocks show that the recovered $H_0$ becomes less precise and less accurate (more biased) if we remove the legacy SNIa due to the loss of coverage at low redshifts. To better understand this process, we repeat the redshift-floor sweep of Section~\ref{sec:z_floor_result} in the $\Lambda$CDM mock datasets mentioned above. The results are shown in Figure~\ref{fig:floorsweep-mock}, with the left panel showing $H_0$ and the right panel showing $q_0$ in every 20\textsuperscript{th} mock, so that the same 500 realisations appear in both panels. The recovered $H_0$ and $q_0$ are not much biased at low $z_\mathrm{floor}$ values due to the minimal extrapolation required to reach the present epoch. As we increase $z_\mathrm{floor}$, the recovered $H_0$ is systematically biased high, while the opposite occurs for $q_0$. However, this bias is not sufficient to explain the high observed $H_0^\mathrm{DES}$ within the SPA cosmology. Indeed, very few mocks have a curve that rises steeply to match the observed $H_0^\mathrm{DES}$ at its redshift floor of 0.073.

The bottom panels of Figure~\ref{fig:floorsweep-mock} show that the bias of our analysis is largely within its statistical uncertainty. The uncertainty our analysis returns for a single mock closely matches the scatter of the best-fitting values between mocks, both of which grow steadily with $z_\mathrm{floor}$. Across a broad range of redshift floors, including that of DES, the true $H_0$ lies close to the $1\sigma$ lower limit of the median $H_0^\mathrm{DES}$ from the mocks, while the true $q_0$ lies close to the $1\sigma$ upper limit. The two mirror each other because a lower $q_0$ implies a faster late-time acceleration, and hence a higher $H_0$. Our mocks cannot tell us the truncation error for the real cosmology, but in a $\Lambda$CDM-like one it is comparable to our reported uncertainty, even if DES had no SNIa below $z = 0.3$. Since DES reaches $z = 0.073$ and has good coverage down to $z \approx 0.2$, we expect our cosmographic expansion to perform well on the real data.

We caution that in a redshift-floor sweep, the different points are all strongly correlated because every point uses a dataset that is strictly a subset of the data used at the preceding point. Even so, it is interesting that the DES-only result remains quite stable to the progressive addition of SNIa at low redshifts by decreasing $z_\mathrm{floor}$. This suggests a lack of redshift-dependent internal tension within DES, making it all the more surprising that adding the legacy SNIa causes such a sharp change in behaviour for both $H_0$ and $q_0$. The change is so sharp that in the narrow redshift range where DES and legacy samples overlap, there is a visible discrepancy between the results using DES and DES + legacy. This is especially evident in the insets to the top panels of Figure~\ref{fig:floorsweep-mock} at $z \approx 0.08$. The smoother behaviour in the DES alone case is much more in line with what we might expect, but in this case the values of $H_0$ and $q_0$ differ from those in the SPA cosmology despite assuming a standard $r_{\rm d}$. This issue could presumably be rectified if $H_0 r_{\rm d}$ and $q_0$ had values different from those in the SPA cosmology. However, the sharp feature in the DES + legacy case at precisely the redshift where legacy SNIa start being included seems unlikely to arise in any realistic cosmology.

\bsp
\label{lastpage}
\end{document}